\documentclass[12pt]{report}
\usepackage{setspace}
\usepackage[utf8]{inputenc}
\usepackage[titletoc]{appendix}
\usepackage{graphicx}
\usepackage{listings}
\usepackage{dirtytalk}
\graphicspath{ {images/} }
\usepackage{float}
\usepackage{multirow}
\usepackage{booktabs}
\usepackage{amsmath}
\usepackage{multicol}
\usepackage{amsmath,amsthm,amssymb}
\usepackage[english,greek]{babel} 
\usepackage{csquotes}
\usepackage[bookmarksnumbered,pdfpagelabels=true,plainpages=false,colorlinks=true,
           linkcolor=black,citecolor=blue,urlcolor=blue, pdftitle={SubatomicKallitsopoulouMasterThesis}]{hyperref}
\hypersetup{pdfauthor={AlexandraKallitsopoulou}} 
\usepackage{afterpage}
\usepackage{titlesec}
\titleformat*{\subsection}{\Large\bfseries}
\titleformat*{\subsubsection}{\large\bfseries}
\titleformat{\chapter}[display]
{\normalfont\huge\bfseries\color{black}}
{\chaptertitlename\thechapter}{20pt}{\LARGE}

\usepackage{pgf}
\usepackage{pgfpages}

\pgfpagesdeclarelayout{boxed}
{
 \edef\pgfpageoptionborder{0pt}
}
{
 \pgfpagesphysicalpageoptions
  {%
  logical pages=1,%
  }
\pgfpageslogicalpageoptions{1}
  {
border code=\pgfsetlinewidth{0.5pt}\pgfstroke,%
border shrink=\pgfpageoptionborder,%
resized width=.95\pgfphysicalwidth,%
resized height=.95\pgfphysicalheight,%
center=\pgfpoint{.5\pgfphysicalwidth}{.5\pgfphysicalheight}%
  }%
}
\pgfpagesuselayout{boxed}

\usepackage{xcolor}
\usepackage[a4paper,top=2cm, bottom=3.5cm, left=2.54cm, right=2.54cm, headheight=14pt]{geometry}
\definecolor{myred}{HTML}{581000}
\usepackage{fancyhdr}
\renewcommand{\footrulewidth}{3pt}
\renewcommand{\footrule}{\hbox to\headwidth{\color{myred}\leaders\hrule height \footrulewidth\hfill}}
\usepackage{caption}
\usepackage{subcaption}
\fancypagestyle{plain}{%
\fancyhf{}
\fancyfoot{}
\fancyfoot[R]{\thepage}
\fancyfoot[L]{\fontsize{10}{14}\selectlanguage{english}SubAtomic Physics and Technological Applications-Thessaloniki 2021}

\renewcommand{\footrulewidth}{3pt}
\renewcommand{\footrule}{\hbox to\headwidth{\color{myred}\leaders\hrule height \footrulewidth\hfill}}
}

\usepackage{array}
\newcolumntype{Y}{>{\centering\arraybackslash}m{4.4cm}}
\newcolumntype{X}{>{\centering\arraybackslash}m{7cm}}
\newcolumntype{P}{>{\centering\arraybackslash}m{2.5cm}}
\newcolumntype{L}{>{\raggedright\arraybackslash}m{8cm}}
\newcolumntype{M}{>{\raggedright\arraybackslash}m{3.7cm}}
\newcolumntype{Z}{>{\justify}m{15cm}}

\usepackage{biblatex}
\title{
{Design and Implementation of Vehicle Control Unit in Formula Student Electric race-car}}
\date{2021}

\usepackage{nomencl}
\makenomenclature

\begin{document}
\selectlanguage{english}
\renewcommand{\thepage}{\roman{page}}
\begin{sloppypar}

\begin{titlepage}
    \begin{center}
        \vspace*{0.1cm}
          
        \LARGE
        \textbf{Aristotle University of Thessaloniki}
        \vspace{0.5cm}
        
        \LARGE
        \textbf{Development of a Simulation Model \\
        and Precise Timing Techniques \\
        for PICOSEC-Micromegas Detectors.}
         \vspace{0.5cm}
        \par 
        \large
        Submitted in partial fulfillment of requirements\\
        for the degree of\\
        \Large
        \par
        \textbf{Master of Science}
        \\
        \large
        \par
         \vspace{0.5cm}
        by
        
        \Large
        \par
        \textbf{Alexandra Kallitsopoulou}\\
        
        \Large
        \par
        University ID:  1119133664838179\\

        \large
        \vspace{0.5cm}
        Supervisor:\\
        \Large
        \textbf{Prof. Spyros Eust. Tzamarias}
            
        \vfill
            
        \vspace{0.5cm}
            
        \includegraphics[width=0.5\textwidth]{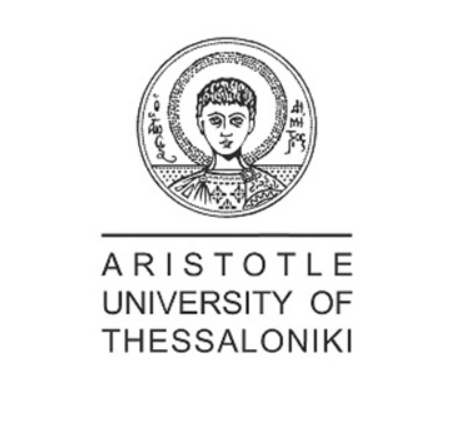}\\
        \vspace{0.5cm}   
        \large
        \textbf{Physics Department }\\
        \textbf{SubAtomic Physics and Technological Applications}\\
        \normalsize
        \textbf{}\\
        \vspace{0.5cm}
        \large
        \textbf{Thessaloniki October 2021}\\
        \vspace{0.8cm}

    \end{center}
\end{titlepage}

\setstretch{1.5}

\newpage
    \begin{center}
    
        \vspace*{0.1cm}
        \large
        {Aristotle University of Thessaloniki}\\
        
        \vspace{0.8cm}
        
        \large
        \textbf{Development of a Simulation Model and Precise Timing Techniques for PICOSEC-Micromegas Detectors}\\
        \vspace{0.5cm}
        \large  Master thesis \\
        \vspace{0.8cm}
        \large \textbf{Alexandra Kallitsopoulou}\\
        
\vspace{5cm}
        
    \end{center}

\begin{center}
    \textbf{Examination Committee}
\begin{multicols}{3}
			\textbf{}\\[1cm]
			\hrulefill \\
			 	\textbf{Spyros Eust. Tzamarias}\\
				\textbf{Professor, AUTH}
	\columnbreak
			
			\textbf{}\\[1cm]
			\hrulefill  \\
			 	\textbf{Dimitrios Sampsonidis}\\
				\textbf{Professor, AUTH}
	\columnbreak

		\textbf{}\\[1cm]
		\hrulefill \\
			\textbf{Konstandinos Kordas}\\
		    \textbf{Assistant Professor, AUTH}
	
\end{multicols}
\end{center}

\vspace{2cm}

\vspace{1cm}
\begin{center}
\noindent
{\large Thessaloniki, 15 October 2021}
\end{center}

\addcontentsline{toc}{chapter}{Acknowledgements}
\newpage
\thispagestyle{plain}
    \begin{center}
    
        \vspace*{0.1cm}
        \Large
        \textbf{Acknowledgements}\\
        \vspace{0.5cm}
    \end{center}

Despite the strict approach, this thesis would not have been implemented without the immense support of my supervisors, my family, and my friends. For this reason, I would like to express my sincere gratitude to everyone who stood by my side.  First and foremost I would like to thank, my supervisor Prof. Spyros Eust. Tzamarias. During our collaboration I acquired a great deal of knowledge, I've been involved to real experimental research and scientific process, and last but not least it was the most suitable closure period for the master program I have attended, because we have used techniques and methods, which were well presented in the courses, and with their real application, they reveal their all majesty. Furthermore, the least I can offer is a big thanks also to Prof. Dimitrios Sampsonidis for his guidance and support to the early stages of this thesis, during these hard times, which turned to be very important, and who also supported me throughout the whole period. Of course and I could not leave out Dr. Lampoudis who has become my mentor, he believed in me when I was losing hope and he encouraged me to pursue the research path I wanted. Dr. Ioannis Manthos, thank you not only for proof-reading my thesis but also for all the scientific, and not, support from the first time we met 3 years now.  Last but not least to my colleagues, Ph.D. candidates Ioannis Maniatis, and Aggelos Tsiamis, for providing me useful analysis tools and methods, for their patience, our endless discussions, who were always there for me, to answer my questions,  and the infinite hours of work we all spent together. Especially Aggelos, for his output data of the Neural Network, and all the needed information helping me to learn more about Machine Learning, with the additional information provided also by the Ph.D. candidate Ioannis Karkanias. Ioannis M., thank you for being the bright example and motivational power for me. I had not come this far without you all, I offer you great gratitude. \newline
To my parents for their warmth and loving tender, as they have been those who have been supporting me throughout my life, I offer you the biggest gratitude for who I have become and what I have achieved.

\addcontentsline{toc}{chapter}{Abstract}
\newpage
    \begin{center}
    
        \vspace*{0.05cm}
        \Large
        \textbf{Abstract}\\
        \vspace{0.5cm}
        
    \end{center}

This work aims at the development of signal processing algorithms that can explore the potential of a PICOSEC-MicroMegas detector and offers the ability for online, precise timing. The PICOSEC-MicroMegas detector is a novel gaseous detector offering precise timing resolution in experimental measurements. The performance of the proposed signal processing algorithms performance is evaluated by a variety of tests using experimental data. The data used in the present research, were accumulated during a Laser Beam Test at CEA/IRAMIS-SACLAY. We use two sets of experimental events: the SPE-set comprises single photoelectron waveforms, and the EXP-set contains waveforms corresponding to an average of 7.8 photoelectrons. The EXP-set waveforms, when fully analysed, result to a timing resolution of 18.3 $\pm$ \,0.2 ps. Alternative algorithms, based on Constant Threshold Discrimination, but using multi-Charge above threshold to correct for systematic timing effects, are described and evaluated in detail. Furthermore, a timing technique using Artificial Neural Networks(ANN) has been advanced and evaluated. To generate an appropriate learning set for the ANN, a novel simulation model has been developed. We have used the EXP-set waveforms, to quantify the accuracy of the developed techniques and we found that the same timing precision, as in the full signal processing analysis, is achieved. We have also proved that the estimation of the Signal Arrival Time using the developed techniques is  consistent and unbiased. Finally, another attractive feature of the new developed techniques is the fact that they can be used during data acquisition to provide fast timing of the particles arrival for triggering and event selection. 
\selectlanguage{greek}
\addcontentsline{toc}{chapter}{\texorpdfstring{$\Pi\epsilon\rho\iota\lambda\eta\psi\eta$}{Lg}}
\newpage
    
    \begin{center}
    
        \vspace*{0.05cm}
        \Large
        \textbf{Περίληψη}\\
        \vspace{0.5cm}
        
    \end{center}

Η παρούσα εργασία στοχεύει στην ανάπτυξη αλγορίθμων επεξεργασίας σήματος που μπορούν να διερευνήσουν τις δυνατότητες ενός ανιχνευτή \selectlanguage{english}PICOSEC-MicroMegas \selectlanguage{greek}και προσφέρει τη δυνατότητα για \selectlanguage{english}online, \selectlanguage{greek}ακριβή χρονισμό. Ο ανιχνευτής \selectlanguage{english}PICOSEC-MicroMegas\selectlanguage{greek} είναι ένας νέος ανιχνευτής αερίου που προσφέρει ακριβή ανάλυση χρονισμού σε πειραματικές μετρήσεις. Η απόδοση των προτεινόμενων αλγορίθμων επεξεργασίας σήματος αξιολογείται από μια ποικιλία δοκιμών χρησιμοποιώντας πειραματικά δεδομένα. Τα δεδομένα που χρησιμοποιήθηκαν στην παρούσα έρευνα, συγκεντρώθηκαν κατά τη διάρκεια \selectlanguage{english} Laser Test Beams, \selectlanguage{greek}στο\selectlanguage{english} CEA/IRAMIS-SACLAY.\selectlanguage{greek} Χρησιμοποιούμε δύο σετ πειραματικών δεδομένων: το \selectlanguage{english}SPE-set \selectlanguage{greek}περιλαμβάνει μεμονωμένες κυματομορφές φωτοηλεκτρονίων και το σύνολο \selectlanguage{english}EXP \selectlanguage{greek}περιέχει κυματομορφές που αντιστοιχούν σε μέσο όρο 7,8 φωτοηλεκτρόνων. Οι \selectlanguage{english}EXP-set \selectlanguage{greek}κυματομορφές, όταν αναλυθούν πλήρως, καταλήγουν σε χρονική διακριτική ικανότητα  \selectlanguage{english}18,3 $\pm$ \,0,2 ps.\selectlanguage{greek} Εναλλακτικοί αλγόριθμοι, βασισμένοι στην \selectlanguage{english}Constant Fraction Discrimination \selectlanguage{greek}τεχνική, αλλά χρησιμοποιώντας το φορτίο πάνω από κάποιο κατώφλι για τη διόρθωση συστηματικών σφαλμάτων χρονισμού, περιγράφονται και αξιολογούνται λεπτομερώς. Επιπλέον, έχει αναπτυχθεί και αξιολογηθεί μια τεχνική χρονισμού με χρήση \selectlanguage{english} Artificial Neural Networks (ANN).\selectlanguage{greek} Για να δημιουργηθεί ένα κατάλληλο σετ μάθησης για το ANN, έχει αναπτυχθεί ένα νέο μοντέλο προσομοίωσης. Χρησιμοποιήσαμε τις κυματομορφές του \selectlanguage{english}EXP-set,\selectlanguage{greek} για να ποσοτικοποιήσουμε την ακρίβεια των τεχνικών που αναπτύχθηκαν και διαπιστώσαμε ότι επιτυγχάνεται η ίδια ακρίβεια χρονισμού, όπως στην πλήρη ανάλυση επεξεργασίας σήματος. Έχουμε επίσης αποδείξει ότι η εκτίμηση του \selectlanguage{english} Signal Arrival Time \selectlanguage{greek}χρησιμοποιώντας τις αναπτυγμένες τεχνικές είναι συνεπής και αμερόληπτη. Τέλος, ένα άλλο ελκυστικό χαρακτηριστικό των νέων αναπτυγμένων τεχνικών είναι το γεγονός ότι μπορούν να χρησιμοποιηθούν κατά τη λήψη δεδομένων για να παρέχουν γρήγορο χρονισμό των σωματιδίων για σκανδαλισμό και επιλογή γεγονότων. 

\selectlanguage{english}
\setlength{\parskip}{0em}
\tableofcontents
\addcontentsline{toc}{chapter}{List of Figures}
\listoffigures
\addcontentsline{toc}{chapter}{List of Tables}
\listoftables


\clearpage
\setcounter{page}{1}
\setcounter{chapter}{0}
\renewcommand{\thepage}{\arabic{page}}	
\setlength{\parindent}{2em}
\setlength{\parskip}{0em}

\chapter{Introduction}\label{cha:introduction}
\paragraph{}Finding ways to detect electromagnetic radiation, was a fundamental question from the early stages of its discovery. Because of the agility, it offers for electrons and ions, gas is the right medium to use. Many inventions have been made, and which split the history of gas-filled detectors into periods. Overcoming the first period with the three original devices, the ionization chamber, the analog counter, and the Geiger-Muller counter, the historical background starts from a very important moment, that of the Nobel Prize (1992), and the discovery of the MultiWire Proportional Chamber(hereafter called MWPC) by G.Charpak in 1968\cite{Charpak:1968kd}. The introduction of the MWPC sparked other ideas, such as the use of electronic drifting time to obtain spatial information, a drift chamber. To cover large areas with MWPC, many traffic channels and electronics are valuable. A MultiWire Drift Chamber is a device built to reconstruct the track of particles in high-intensity experiments.

As successful as these chambers seemed, they posed serious limitations when key questions were raised about improving spatial resolution and better rate capability. To overcome these limitations, Oed proposed the Micro-Strip Gas Chambers (MSGC)\cite{OED1988351}. These detectors are known to have high radiation resistance, which makes them suitable for use in high radiation environments. Another advantage of these detectors is the good spatial resolution they provide. Detectors with a spatial resolution of 20-30μm have been developed and used in high-energy experiments. Despite their use in high-energy experiments, their application is not limited at all in this area. For example, they have applications in medical imaging, albeit with a modified design. The MSGC era was followed by a series of discoveries that led to the design of Micropattern detectors, which take advantage of new technology in microelectronics and photolithography and have high granularity, such as the Micromegas detector. 

The purpose of modern detector technologies is to make precise measurements for the position, the time and the energy of the passing particle through the detector. Energy can be measured with the detector suited in a calorimeter, using materials that after interaction, they absorb an amount of energy that can be measured. Position is measured by reconstructing the track after the passing particle hits several position sensitive detectors. 

When it comes to timing, new physics discoveries in high-intensity experiments, such as the High Luminosity of the Large Hadron Collider, demand a high ability to tag multiple events, each comprising many very paricles. In such a dense topology, i.e. the environment of the H-LHC where 140 interactions take place on every beam crossing, it is very difficult to identify them all at the same time. For this reason, all these experiments, have an intense need for detectors with precise timing properties. As it is known, Micro Channel Plates are good candidates providing timing resolution in the order of a few ps per MIP, the only drawback of their use in experiments like HL-LHC is the need for large area coverage, their construction seems very costly. Fortunately, there are cheaper solutions called Micropattern gaseous detectors and silicon structures, which provide excellent timing properties as well. The distinctive ability for these events is of the order of 20-30ps, or better, which means that our technologies need some modifications to reach this limitation. Apart from the need of large area coverage, there is also a need of resistance to radiation damage, in order for these applications to be used in different domains, as for example in Cosmology for energy or speed measurements and correlations, or in accelerators for 4D tracking.    

Summing up the upcoming study, in the Chapter \ref{cha:interaction},  is given a brief introduction of the interactions of charged particles with matter. Subsequently, Chapter \ref{cha:introgaseous} gives a short overview of the general consideration of gaseous detectors, resulting to the special case of the Micromegas Detector and even more for the novel technology of the PICOSEC Micromegas Detector. Moving on to the Chapter \ref{cha:design} , the experimental setup and the precise timing techniques are being presented in detail. In the Chapter \ref{cha:implementation} a detailed analysis of the PICOSEC-MicroMegas signal is implemented, while on Chapter \ref{cha:simulation} the assiduous study of the development of simulated pulses as an input to an Artificial Neural Network is described, resulting to the Chapter \ref{cha:conclusion} were there is a complete summary of the study.

\chapter{Interaction of Particles with Matter \label{cha:interaction}}\label{chap:interaction}
\vspace{-0.8cm}
\noindent\fbox{%
    \parbox{\textwidth}{%
        A requisite step in our way of designing a detector is to understand the interaction mechanisms of the radiation in the interactive medium.  This chapter presents the basic properties of radiation's interaction with matter, as a summary in order to refresh useful principles, necessary for the following Chapters.
}%
}
\section{Interaction of Heavy Charged Particles }\label{section:InteractionOfHCP}

There are various of nuclear end electronic mechanisms through which charged particles can interact with the molecules of a medium. However, the bottom line result of all different interaction mechanisms is a reduction in the energy of the incoming particle, which consists the radiation, as it passes through the medium. The main process of interaction for the charged particles is the electromagnetic processes, which conclude to two different branches, called \emph{excitation} and \emph{ionization} of the molecules of the medium, for particles heavier than the electron\cite{BLUM}. Moreover, besides these processes electrons can also lose energy through radiating photons, or via other electromagnetic processes such as Cherenkov and transition radiation. 
\subsection{Energy Loss}\label{subsection:EnergyLoss} 
\begin{figure}[hbt!]
 \centering 
 \includegraphics[width=0.7\textwidth]{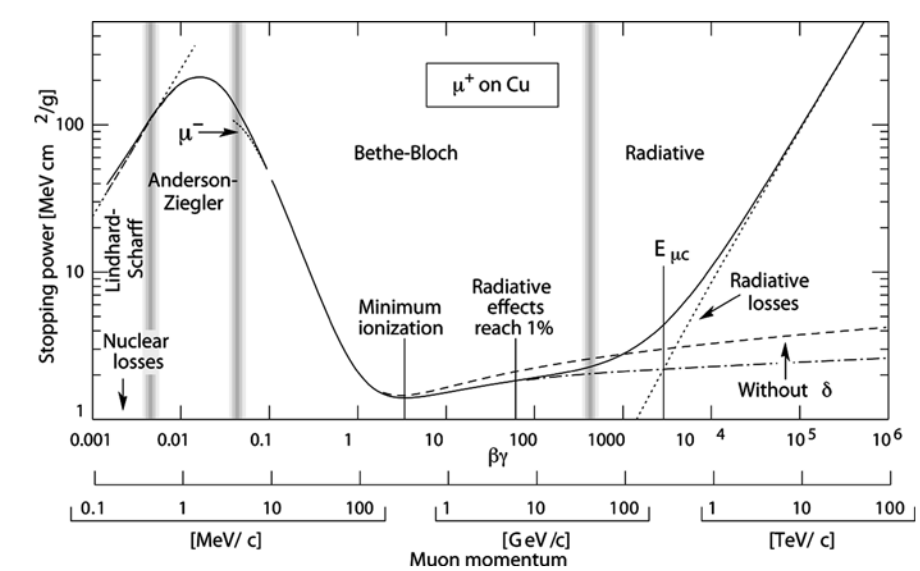}
\caption{Energy Loss of a Muon in Copper \cite{TAVERN}.}\label{fig:bethe}
\end{figure}

\paragraph{} Although the phenomenology of these processes referred above implies very complicated interaction mechanisms, in fact, the energy loss can be predicted by a number of semi-empirical relations that have been developed through the years\cite{AHMED}.  The rate of losing energy, for a particle that passes through a material, depends not only on the nature of the incident particle but also on the nature of the target particles, i.e the type of interactive medium. This quantity that describes the energy loss, is referred to in the literature as \emph{stopping power} of the material\cite{AHMED}. One piece of information to notice here is that the stopping power does not imply the energy loss per unit time, but the energy that the particle loses per unit length of the interactive medium while traversing it. The well-known formula related to the computation of the energy loss, using quantum mechanics, derived by Bethe and Bloch is given in Eq.\ref{eq:bethe} 
\begin{equation}\label{eq:bethe}
    \Big[-\frac{dE}{dx}\Big]\Big|_{Bethe-Bloch} = \frac{4\pi N_A r^2_e m_e c^2 \rho Z q^2}{A\beta^2}\Big[ln\Big(\frac{W_{max}}{I}\Big)- \beta^2\Big]
\end{equation}
where $N_A = 6.022\times10^{23}mole^{-1}$ is the Avogadro number, $r_e = 2.818\times10^{-15}m$ is the radius of the electron, $m_e = 9.109\times10^{-31}kg$ is the rest mass of the electron, q is the electrical charge of the ion, $\rho$ is the density of the medium, A is the mass number of the medium, I is the ionization potential of the medium and $\beta$ is a correction factor.
One significant information about Eq.\ref{eq:bethe} is that the energy loss decreases rapidly due to the term $1/\beta^2$ , as $\beta \gamma$ increases. For relativistic energies, i.e close to speed of light, the energy loss graph reaches a minimum, which is called energy-loss minimum, which is defined for particles within the relative energy range as minimum ionising particles, near the value $\beta\gamma = 3$. After its minimum the curve follows a logarithmic rise, due to the homonym term in the brackets\cite{TAVERN}. As a matter of fact, the Bethe-Bloch formula has to be corrected and has been, for two other factors which seem to be significant at very high and low energies. The first correction is necessary because of the polarization of electrons due to the electric field caused by moving ions, created in the medium, and the term is called \emph{shielding} \cite{AHMED}. It is obvious that the density of electrons affects the importance of this effect, and it becomes even more significant in the case of the higher energy of the incident particle. The second term that needs to be inserted as a correction is defined because of the orbital velocities of the electrons in the medium and it becomes significant at low energies. Both  terms are subtractive in the formula and presented by symbols $\delta$ and $C$ respectively, so that the Bethe-Bloch Eq.\ref{eq:bethe} for stopping power, becomes :
\begin{equation}\label{eq:bethenew}
     \Big[-\frac{dE}{dx}\Big]\Big|_{Bethe-Bloch} = \frac{4\pi N_A r^2_e m_e c^2 \rho Z q^2}{A\beta^2}\Big[ln\Big(\frac{W_{max}}{I}\Big)- \beta^2 -\frac{\delta}{2} - \frac{C}{Z} \Big]
\end{equation}
\paragraph{} Both Eq.\ref{eq:bethe} and Eq.\ref{eq:bethenew} can be used to find the integrated energy up to the maximum of the energy loss that is transferred to the atoms of the medium. In the case of electrons as incoming particles, because their energy loss in a wide range of energies, even at low ones, is mainly due to bremsstrahlung processes and due to the fact its mass and the target's electron mass is the same, that causes extra problems to calculate the energy transfer, and the calculation is more complicated. A schematic example representation of the Energy Loss as a fraction of the momentum is given in Figure \ref{fig:bethe} \cite{TAVERN}   
Bethe-Bloch formula is a semi-empirical formula that gives an idea of the average energy loss due to ionization and excitation, for charged particles. 
\section{Interaction of Electrons}\label{section:InteractionOfElectrons}

\begin{figure}[hbt!]
 \centering 
 \includegraphics[width=0.7\textwidth]{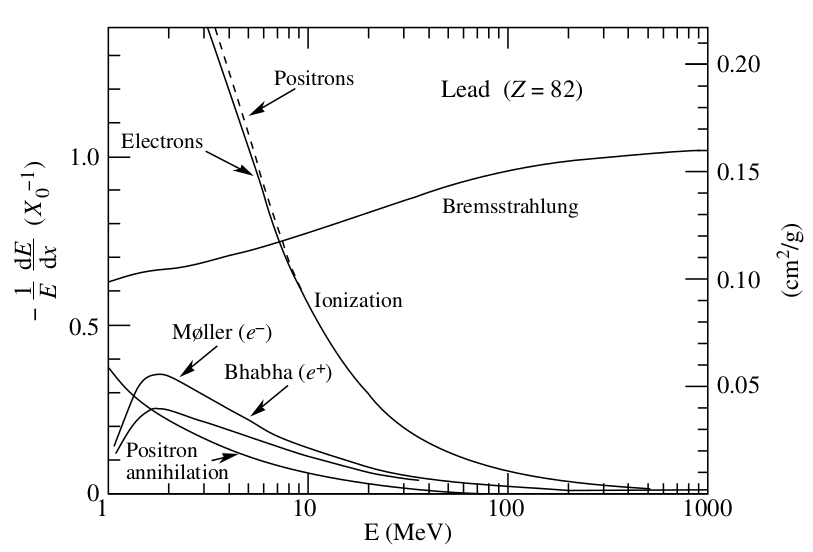}
\caption{Fractional energy loss of electrons and positrons  as a function of energy \cite{AHMED}}\label{fig:electrons}
\end{figure}

The way an electron interacts with the molecules of interactive material depends on its energy. Between low to moderate energies, electrons interact via ionization, scattering and electron-pair annihilation. At higher energies the Bremsstrahlung radiation dominates as it is referred before, as shown in Figure \ref{fig:electrons}.
\subsection{Cherenkov radiation}\label{subsection:CherenkovRad}
\begin{figure}[hbt!]
 \centering 
 \includegraphics[width=0.5\textwidth]{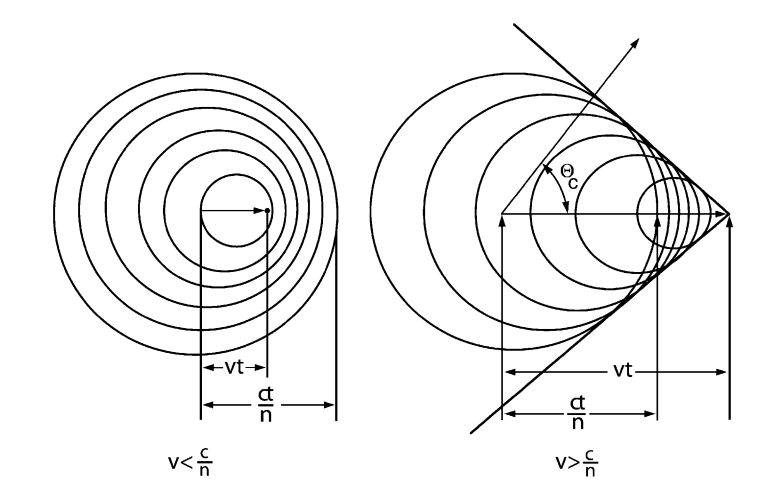}
\caption{On the left: Particle moving with lower velocity than the speed of light in th medium. On the right the opposite process, when Cherenkov light is emitted\cite{TAVERN}.}\label{fig:cherenkov}
\end{figure}

\paragraph{} As it is mentioned earlier in this chapter, an electron with high energy in a medium can emit Cherenkov radiation. What we mean with high energy is that the energy to be high enough so that electron velocity in the medium can become higher than the relative speed of light in it. The optical property that help us to define the velocity of light in a medium is the optical index of refraction $n$, so that let us define the relative velocity as $c/n$\cite{TAVERN}. The Cherenkov effect can be visualized by the Figure \ref{fig:cherenkov}. As we can see for the Cherenkov radiation to be emitted, the velocity of the incoming radiation(i.e travelling charged particles) in the medium has to be greater than the relative velocity of the light in the medium.

Cherenkov's effect uses Huygens' principle as a way to be understood. When charged particles are moving in a medium, their electric field will polarise the charges of the medium. After the effect of this phenomenon, the medium returns to its first state (unpolarized). The result of this change in polarization state represents an \emph{electromagnetic perturbation} that will travel with the speed of light \cite{TAVERN}. On the left side of Figure \ref{fig:cherenkov} the slow-moving particle case is represented, slower than the speed of light in the medium. The result of this movement is a small electromagnetic perturbation due to polarization and depolarization of the medium and it propagates faster than the particles themselves. Far from the particle's trajectory, all these perturbations arise in a random way and annihilate each other. On the right side of Figure \ref{fig:cherenkov} the particle travels faster than the speed of light in the medium. In that case, the small electromagnetic perturbations propagate slower than the particles. As a result, all these perturbations form in one \emph{wavefront}. The perturbation represents a wave that moves in a direction defined by both the speed of the particle and the speed of the light in the medium. From the geometry of this phenomenon, using necessary trigonometric relations, we can derive the expression of the angle between the particle and the wave to be as in Eq.\ref{eq:cherenkov}.
\begin{equation}\label{eq:cherenkov}
    \cos{\theta_c} = \frac{(c/n)t}{vt} = \frac{c}{nv} =\frac{1}{\beta n}
\end{equation}

When it comes to energy loss, the amount of this energy is smaller than the energy loss due to ionization. Nevertheless the interesting about this effect is that depends only on the velocity of the incoming particle so that if we know its energy or its momentum by other means, a measurement of the Cherenkov effect lets us know the mass of the particle and as a result its nature.

\chapter{Properties of Gaseous Detectors\label{cha:introgaseous}}\label{chap:gas}
\vspace{-0.8cm}
\noindent\fbox{%
    \parbox{\textwidth}{%
        The idea of detecting and measuring radiation is based on the interactions that are made within a material and then studying the resulting waveform created in the system configuration. In this Chapter a brief overview of the general design considerations of gaseous detectors is given in detail. Moreover, the basic principles of their operation starting with the production of electron-ion pairs, the energy loss process, the diffusion and drift of charges in the gas, are also being discussed. 
    }%
}
\section{Transportation of Electrons and Ions }\label{section:TransportOfElectrons}
\paragraph{} The operating principles of the gas detectors to be discussed in the next section are based on the formation of ions and electrons as the particle interacts with the gas of the detector until it reaches the readout electronics. The thermal motion of electrons and ions by sliding and interacting with atoms/molecules under the influence of an electric field will be analyzed below. Unless a strong electric field is applied, the ion-electron pairs can reconnect, engage in charge transfer collisions, or lose much of their energy due to diffusion.

\subsection{Drift of Ions}\label{subsection:DriftOfIons}
\paragraph{} In gas-filled detectors, the shape of the received pulse and its height is based not only on the motion of the electrons but also on that of the ions. Ions, as positively charged and heavier than electrons, move more slowly. In most filling gas detectors, and especially in ionization chambers, the output signal can be measured from either the positive or negative electrodes (cathode). In both cases, the measured result is the change in the electric field in the active gas region. Therefore the slippage of electrons and ions contributes to the total outgoing pulse.

In the presence of an external electric field in the volume of the gas, some of the positive ions move along the cathode field. In the case of gas detectors, the average kinetic energy of the ions is comparable to the thermal velocity of the gas atoms. When the mass ion m interacts with the limits of the mass gas M, the constant f expresses the energy loss of the ion and is given by the following equation \cite{AHMED}: 
\begin{equation}\label{eq:drift}
    f = \frac{2mM}{\sqrt{(m+M)^2}}
\end{equation}
\paragraph{} The mean drift velocity of the ions depends linearly on the external electric field and can be expressed as: 
\begin{equation}\label{eq:iondrift}
    u^{ion}_d \simeq \Big(\frac{1}{m+M}\Big)^{1/2}\Big(\frac{1}{3kT}\Big)^{1/2}\Big(\frac{1}{m+M}\Big)^{eE/N\sigma} = \mu E
\end{equation}
where $\mu$ is the mobility factor of the ion inside the gas and $\sigma$ is the cross section of scattering for the ions in the gas molecules, k is the Bolzmann's constant, T the temperature and N the density. 

The \emph{drift velocity} of ions is essentially the velocity of the cloud that is created around the ions as they move along the dynamic lines of the field. This velocity is much slower than the instantaneous velocity of the ions. The drift velocity is a very important parameter, as it expresses how fast we wait for the ions to reach the cathode to be collected. It has been calculated that as long as there is no breakdown in the gas, this velocity remains proportional to the ratio of the electric field to the gas pressure. The drift velocity of ions is two to three orders of magnitude smaller than that of electrons.
The value of gas mobility is characteristic of a particular ion in a gaseous medium while in the case of stable environmental conditions it can be considered constant. The following table Table\ref{tab:mobility} shows the mobility of the different ions for the different gas mediums of the detector\cite{AHMED}.

\begin{table}[hbt!]
\begin{center}
\resizebox{5cm}{!}{
\begin{tabular}{|c|c|c|}
 \hline
\textbf{Gas}&\textbf{Ion}&\textbf{\begin{tabular}{@{}c@{}}Mobility\\ ($cm^2V^{-1}s^{-1}$) \end{tabular}}\\
 \hline
 Ar & $Ar^+$ & 1.54  \\
 \hline
 He & $He^+$ & 10.4 \\
 \hline
$CO_2$ & $CO_2^+$ & 1.09  \\
 \hline
Ar & $CH_4^+$ & 1.87 \\
 \hline
Ar & $CO_2$ & 1.72  \\
 \hline
\end{tabular}}
\caption{\label{tab:mobility}Experimental Values for ion mobility in a few gases\cite{TAVERN}.}
\end{center}
\end{table}

In the presence of an electric field, a specific distribution of thermal ions diffuses symmetrically through multiple scattering with the material of medium. Their distribution in space follows a Gaussian form, as shown in the following Eq.\ref{eq:gaussian}.
\begin{equation}\label{eq:gaussian}
    \frac{dN}{N}(x,t) = \frac{1}{\sqrt{4\pi D t}}exp\Big(-\frac{x^2}{\sqrt{4Dt}}\Big) dx
\end{equation}
The standard deviation of the above equation gives an estimate of the diffusion efficiency of ions equal to $\sigma^{ion}_x = \sqrt{2Dt}$ for Cartesian coordinates. Under the influence of an electric field, the ions move along the length of the field at an average diffusion velocity $u_d$. In this case, the diffusion constant D is related to the mobility of the m ions according to the following Einstein relation \cite{AHMED}:
\begin{equation}\label{eq:einstein}
    \frac{D}{\mu} = \frac{kT}{e}
\end{equation}
Substituting the constant diffusion from the above relation, the linear standard deviation for distance x will be equal to:
\begin{equation}\label{eq:std}
    \sigma_x^{ion} = \sqrt{\frac{2kTx}{eE}}
\end{equation}

\subsection{Drift of Electrons}\label{subsection:DriftOfElectrons}
\paragraph{} In case of electrons the situation turns to be very different, due to their small mass in addition to ions, electrons can transfer only a small amount of energy to their neighbor atoms and molecules during their interaction.   

\begin{figure}[hbt!]
 \centering 
 \includegraphics[width=0.5\textwidth]{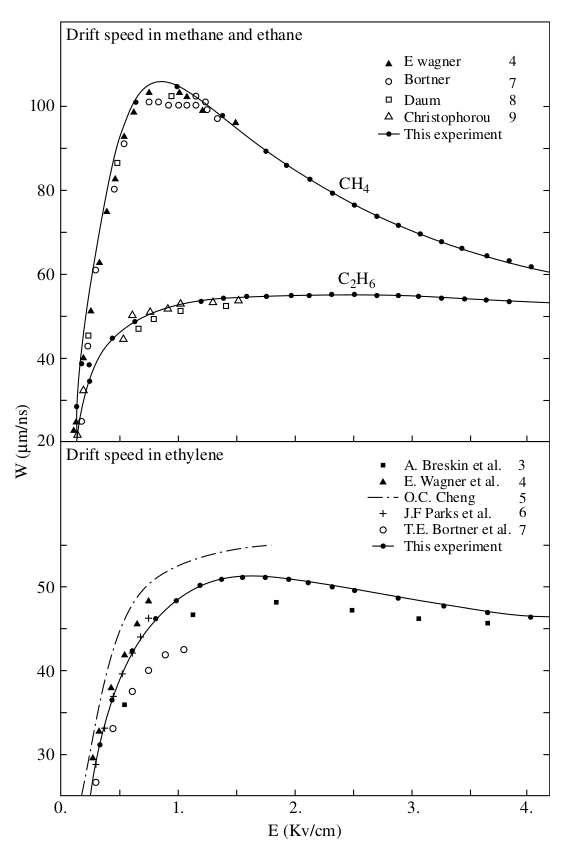}
\caption{Variation pf drift velocity of electrons in methane, ethane and ethylene\cite{AHMED}}\label{fig:drift}
\end{figure}

In absence of electric field, a free electron in a gas has a thermal kinetic energy equal to $3/2k_BT$. In the presence of electric field, the electron begins to collide with gas molecules and if the mean free path between two collisions is $\Delta t$ then the drift velocity can be expressed from Townsend Eq.\ref{eq:townsed} that follows\cite{AHMED}: 
\begin{equation}\label{eq:townsed}
    u_d^e = k\frac{eE}{m}\tau
\end{equation}
where $\tau$ is the mean free path between two collisions. However, the correlation between the drift velocity and the electric field is more than complicated, due to the fact that the mean free path is influenced not only from the electric field but also from the properties of the detector.  
\paragraph{} By applying a constant electric field between the electrodes, the electrons, due to their small mass, accelerate rapidly between the scatterings, gaining energy. The energy they lose during the scattering of gas molecules is much less than that they gain, due to acceleration. A natural consequence of this is that their energy distribution is no longer described by a Maxwell curve. Along the dynamic lines of the field, electrons slide at a speed one order of magnitude faster than their thermal motion. However, the magnitude of the drift velocity depends on the applied electric field and is limited to the moment of breakdown in the gas (breakdown). Drift speed has been shown to increase with field energy, only at low electric field values(Figure \ref{fig:drift}).

Ionization electrons can easily come up to thermal equilibrium with gas molecules, in absence of an electric field, but because of their smaller mass, electrons can diffuse with a greater diffusion coefficient compared with the ions. In presence of an electric field, electron's mobility is significantly changing with the strength of the electric field, due to the fact that their kinetic energy can grow during the collisions with atoms/molecules and the standard deviation $\sigma_x^e$ starts to deviate from the value $1/\sqrt{E}$ (same formula as the Eq.\ref{eq:std})that has been observed in ions.  On the contrary, with the classical approach has been observed that for various gases the diffusion coefficient is not uniform in all directions of the electric field. In these cases, two coefficients of variation must be taken into account, one perpendicular to the displacement direction and one perpendicular to the direction transverse to it. 

Nowadays, for the development of charge gas detectors, many experimental studies have been carried out in order to determine the drift velocity of electrons in the gases that are widely used in the composition of detectors.

\section{Principle of Operation of Gas-Filled Detectors}\label{section:OperationGas}
\paragraph{}As charged particles pass through the detector's filling gas, it can ionize the gas molecules, provided that the energy it carries is greater than the gas ionization potential. By applying an external electric field, the charged pairs generated can be pushed in opposite directions, thus creating a measurable electrical pulse. This application was used in the process of creating gas-filled detectors. A typical gaseous detector consists of the gas region along with positive and negative electrodes. The electrodes are at a high potential difference, which can start at less than 100V and reach up to several thousand volt, which depends on the design and operation of the detector \cite{AHMED}.  

While radiation passing through the gas, the generation and movement of charged pairs disrupt the applied external electric field, resulting in the generation of an electrical signal at the electrodes. The final charge, current, or pulse at one of the electrodes can be measured. In combination with proper calibration, sufficient information can be given about the ionizing particle, which caused this effect, such as energy, or in the case of a beam of particles, in addition to the energy, we can also obtain information about its intensity \cite{KNOLL}. 

It is obvious that such a system will work efficiently only when the largest percentage of the generated pairs can be collected from the electrodes, before they are reconnected in neutral molecules, with the gas ions. The choice of gas, the geometry of the detector, and the applied potential difference allow us to control the process of generating pairs of charges and their movement in the gas. 

Each time radiation passes through a gas, it interacts with gas molecules in different ways. In addition to the active cross-section of the interaction of each type of radiation with gas molecules, an important quantity that determines the probability of this interaction is the average energy required to create an electron-ion pair in the gas. This action is referred to as \textbf{\textit{W-value}}. With this assumption, it can be assumed that since the interaction mechanisms depend on the energy of the radiation and the type of gas, the W-value will also depend on these parameters. In fact this value is weakly dependent on these parameters and ranges between 25-45 eV per pair for most gases and types of radiation, as it can be seen in Table \ref{tab:W-value}\cite{AHMED}.

\begin{table}[hbt!]
\begin{center}
\resizebox{16cm}{!}{
\begin{tabular}{|c|c|c|c|c|c|c|c|}
 \hline
\textbf{Gas}&\textbf{Z}&\textbf{\begin{tabular}{@{}c@{}}First Ion. \\ Potential[eV]\end{tabular}} &\textbf{\begin{tabular}{@{}c@{}}
Density\\$\times 10^{-4}g/cm^{3}$\end{tabular}}&\textbf{\begin{tabular}{@{}c@{}}W\\$eV/pair$\end{tabular}}&\textbf{\begin{tabular}{@{}c@{}}$dE/dx$\\keV/cm\end{tabular}}&\textbf{\begin{tabular}{@{}c@{}}Primary Ion.\\ $ion pairs/cm$\end{tabular}}&\textbf{\begin{tabular}{@{}c@{}}Total Ion.\\ $ion pairs/cm$\end{tabular}}\\
 \hline
 $H_2$ & 2 & 0.8 & 15.4 & 37 & 0.34 & 5.2 & 9.2  \\ 
 \hline
 $O_2$ & 16 & 13.3 & 1.2 & 31 & 2.26 & 22 & 73 \\
 \hline
Ar & 18 & 17.8 & 15.8 & 26 & 2.44 & 29 & 94 \\
 \hline
$CO_2$ & 22 & 18.6 & 13.7 & 33 &3.01 & 34 & 91 \\
 \hline
 $CH_4$ & 10 & 6.7 & 10.8 & 28 & 1.48 & 46 & 53 \\
 \hline
\end{tabular}}
\caption{\label{tab:W-value}Gases' properties at Standard Pressure and Temperature (STP) for MIP\cite{BUGAGOV}.}
\end{center}
\end{table}

An interesting point to be underlined, as it can be seen in the table above, is that W-value is significantly higher than the First Ionization Potential in gases, suggesting that not all of the energy of the incoming radiation to the gas goes to the formation of electron-ion pairs. This is understandable as we know that the transmitting radiation can, in addition to ionizing atoms, also cause their stimulation. 

The charges generated by the incoming radiation are called \emph{primary} charges so that we can distinguish them from those produced indirectly in the active volume of the gas (\emph{secondary} charges). The mechanisms of production of these secondary charges are similar to those of the primary ones except that they are produced by ionization caused by the primary charges and not by the incoming radiation. W-value represents all these indirect ionizations that occur in the medium \cite{AHMED}. 

Both electrons and ions are produced as a result of the rapid energy loss of the transmitting radiation, due to multiple scatterings with the gas molecules. How these charges move through the gas depends on the type and net force they face. In the absence of an external electric field, electrons and ions have energy \emph{E} characterized by the Maxwell Energy Distribution. The average energy of the loads as shown by the distribution is \cite{KNOLL}:
\begin{equation}\label{eq:maxwell}
\bar{E} = \frac{3}{2}kT
\end{equation}

At room temperature, this energy is equal to 0.04eV. As there is no external electric field, there is no preferred direction for the motion of the charges in a homogeneous gas mixture, so their diffusion is isotropic. Electrons, due to their small mass, diffuse faster. This involves comparing the thermal velocities of electrons and ions, which usually differ by two or three orders of magnitude. Therefore the diffusion coefficient of electrons is very different from that of ions for the same medium. As the diffusion coefficient is directly dependent on mass and charge, we conclude that it takes different values for different ions. With the presence of an electric field, the propagation ceases to be isotropic and thus the diffusion coefficient cannot be expressed by a gradient. In this case, it will be reproduced by a tensor with two non-zero components: one horizontal and one vertical in the direction of diffusion \cite{KNOLL}.

In gas-filled detectors, the Maxwell form of the energy distribution of charges cannot be considered standard. The reason is the creation of polarization, due to the movement of charges, which create an electric field inside the active area of the gas. Electrons, due to their small mass, face strong electrical force, and therefore their energy distribution deviates from the pure Maxwell form. On the other hand, the energy distribution of electrons is not significantly affected if the applied field is not high enough to cause discharge in the gas \cite{AHMED}.

\section{Regions of Operation of Gas-filled detectors}\label{section:RegionsOperation}
\begin{figure}[hbt!]
 \centering 
 \includegraphics[width=0.7\textwidth]{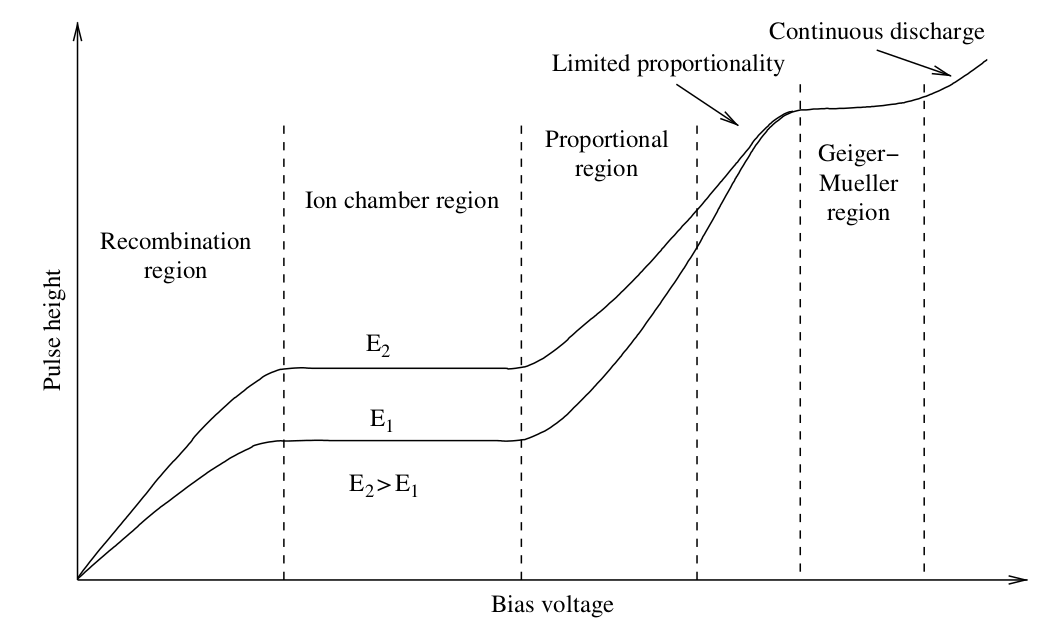}
\caption{Different regions of operation for gas-filled detectors, with respect to the applied voltage. The two curves correspond to particles with different energy deposition within the gas. \cite{AHMED}}\label{fig:region}
\end{figure}

\paragraph{} Figure \ref{fig:region} gives an overview of the different regions of operation for gas-filled detectors. As it can be seen, a detector can be operated in many modes, with the only difference between them being the number of charges produced and how they travel inside the active volume of the detector. The choice of mode of operation depends on the application, but in general, is it used for the detectors to work in a range of applied voltage. These different regions are going to be discussed next. \cite{AHMED}

\subsection{Recombination region}
\paragraph{} In the absence of an electric field, the generated charges reconnect very quickly creating neutral molecules. By applying a polarizing(bias) voltage some of the charges begin to drift towards the oppositely charged electrodes. As this voltage increases, the degree of reconnection decreases, and the current flowing through the detector increases. The reconnection region (recombination region) shown in Figure \ref{fig:region} comprises a range of potential difference values, above which the reconnection is negligibly small. Due to the calculated rate of reconnection in this area, the current which is measured at the output of the detector does not exactly represent the energy deposited in the gas, due to the passing radiation. Therefore, to study the properties of radiation, such as its energy, we will not use this operating range in the detector.

\subsection{Ion chamber region}
\paragraph{} The collection efficiency of the electron-ion pairs in the reconnection region increases with the applied voltage until all the charges produced can be collected. This point is the beginning of the area called \emph{the ionization chamber} area. In this region, a larger increase in voltage does not affect the measured current, as long as all the charges produced continue to be collected efficiently by the electrodes. The current measured by the electronics in this area is called \emph{the saturation current} and is proportional to the energy deposited due to the passing radiation. Detectors operating in this area are called ionization chambers. 

\subsection{Proportional region}
\paragraph{} Previously, we referred to the process of production of electron-ion pairs due to the passage of radiation into the gas volume. This type of ionization is referred to as \emph{primary ionization} \cite{AHMED}. If the charges produced during primary ionization have enough energy, they can produce additional pairs of electrons - ions, a process called \emph{secondary ionization} \cite{AHMED}. Additional ionization from these charges can occur as long as they have sufficient energy for this purpose. Obviously, this process can only take place if a sufficiently high potential is applied between the electrodes, at which point the charges can reach high speeds. Although the energy gained by ions increases with increasing voltage, electrons, due to their small mass, are the ones that cause the most of the extra ionizations. 

Those extra ionizations cause a process called \emph{multiplication}. This multiplication of the charges in high fields takes advantage of the proportional chambers to increase the height of the output pulse. In this type of detector, the charge is multiplied in such a way that the output pulse remains proportional to the energy deposited. For this reason, they are also called proportional detectors. From Figure \ref{fig:region}, we can conclude that in analog counters the height of the output pulse is proportional to the applied voltage. This assumption is correct only as an approach. The real reason they are called proportional counters is that the total number of charges produced after multiplication is proportional to the original number of charges of the transmitted radiation. At this point, it is useful to see how the process of multiplication takes place.
\subsubsection{Multiplication of Charges}
\begin{figure}[hbt!]
 \centering 
 \includegraphics[width=0.6\textwidth]{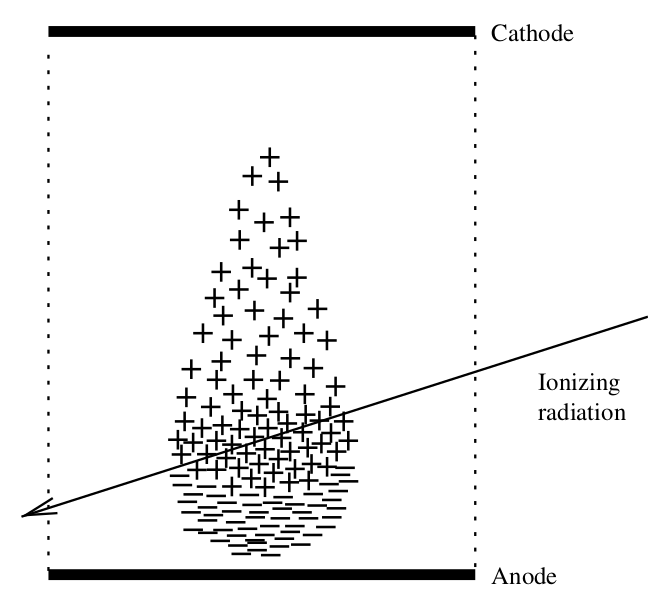}
\caption{Geometrical representation of avalanche multiplication in the active volume of gas-filled detector. Known as droplet like shape \cite{AHMED}.}\label{fig:droplet}
\end{figure}

\paragraph{} For a detector operating in proportionality, an electric field of a few kV / cm is not uncommon. This high electric field does not only reduce the charge collection time but also leads to a process called avalanche multiplication, which is a rapid multiplication of charges from the main charge produced due to radiation emission. The result of this particle storm is an increase in the height of the outgoing pulse. An interesting aspect of the avalanche is its geometric evolution, which resembles a drop, due to the large difference between the slip velocities of electrons and ions, as shown in Figure \ref{fig:droplet}.
Because of the existing high electric filed between the electrodes, all the charges between collisions gain energy. The total amount of energy gain needs to be higher than the ionization potential of the chosen gas, so that it can create another electron-ion pair. Assuming that the electric field is uniform and pressure and temperature of the surrounding area remain constant, then the change in the number of pairs created per unit length is proportional too the total number of pairs \cite{AHMED}.
\begin{equation}\label{eq:npairs}
    \frac{dN}{dx} = \alpha N
\end{equation}
where N represents the total numver of pairs and $\alpha$ is the so called \emph{first Townsend coefficient}. This value refers to the number of collisions that a particle can suffer before causing the first ionization per unit length of the particle track, an it is related to the \emph{mean free path} $\lambda$, for ionization as: 
\begin{equation}\label{eq:Townsend}
    \alpha = \frac{1}{\lambda}
\end{equation}
where $\alpha$ depends on the energy gaining in a $\lambda$, and the ionization potential of the gas. So the total number of pairs created is from \ref{eq:npairs}: 
\begin{equation}
    N = N_0 e^{\alpha x}
\end{equation}
while if $\alpha >0$ this number follows an exponential growth with distance .  

In reality, however, the electric field is not uniform, but has the form $\alpha = \alpha(x)$.
An initial population of electrons $n_e$ will be multiplied by a distance path $dx$ into $dn_e$ electrons and the multiplied number of electrons for a given path $x_1 \rightarrow x_2$ is given by Eq.\ref{eq:nonuniform}, where $n_0$ is the number of electrons at $x = 0$ and $n$ is the total number of secondary electron ions produced by the length of the detector volume. Then, the multiplication factor, also known as Gain is given by the ratio between the final and the initial population of electrons as $G = n/n_0$. 
\begin{equation}\label{eq:nonuniform}
N = N_0 exp\Big( \int_{x_1}^{x_2}\alpha(x)dx\Big)
\end{equation}
where the above expression describes the mean evolution of the avalanche, but in reality, the interactions between electrons and atoms are dominated by statistical fluctuations and as a result, the size of the avalanche has to follow a probability density function. 
\begin{figure}[hbt!]
 \centering 
 \includegraphics[width=0.5\textwidth]{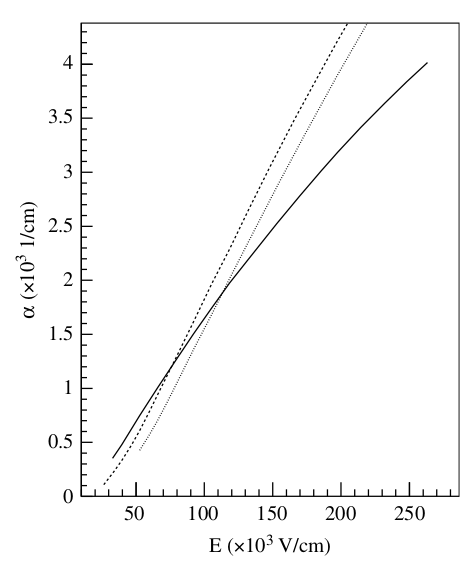}
\caption{Dependence of Townsend coefficient on electric field strength for different gas mixtures: $80\% Ne + 20\% CO_2$ (solid line), $90\% Ar + 10\% CH_4$ (dashed line), $70\% Ar + 30\% CO_2$ (dotted line).  \cite{AHMED}.}\label{fig:Townsend}
\end{figure}
Depending on the Townsend coefficient and the electric field, the probability to have $n$ electrons after a distance $x$, for an avalanche that starts from a single electron, is expressed via Furry's law, Eq. \ref{eq:furry}, where the probability decreases exponentially for greater n, with maximum value $n=1$.
\begin{equation}\label{eq:furry}
    P(n,x) \cong \frac{e^{-n/\bar{n}}}{\bar{n}}
\end{equation}

Moreover, it has been observed experimentally that electrons, under the influence of the electric field, use a significant part of their distance in order to reach the energy needed to cause the ionization process, and thus, the exponential function of Furry's distribution become the well known \emph{Polya distribution} \cite{BORTFELDT2021165049,Polya}.  As the field gets stronger, this probability of ionization is correlated by the path the electron traveled inside the active volume of the gas, so that the Polya distribution, which describes this phenomenon to be expressed like: 
\begin{equation}\label{eq:Polya}
    P(n,\bar{n};\theta) = \frac{(1+\theta)^{1+\theta}}{\bar{n} \Gamma(1+\theta)}\Big(\frac{n}{\bar{n}}\Big)^\theta exp\Big[-(1+\theta)\frac{n}{\bar{n}}\Big]
\end{equation}
where $\theta$ is a parameter related to the relative gain variance, called the shape parameter. Its properties, as expectation value and variance are :
\begin{equation}\label{eq:expectation}
    E[n] = \bar{n}
\end{equation}
\begin{equation}\label{eq:variance_polya}
V[n] = \frac{\bar{n}^2}{1+\theta}
\end{equation}
In the case we assume N electrons having the same starting point, and cause the creation of an avalanche, the number of its electrons will be given by the N-Polya distribution, being:
\begin{equation}\label{eq:N-polya}
    P_N(n;\bar{n},\theta,N)  = \frac{(1+\theta)^{N(1+\theta)}}{\bar{n} \Gamma(N(1+\theta))}\Big(\frac{n}{\bar{n}}\Big)^{N(\theta+1)-1} exp\Big[-(1+\theta)\frac{n}{\bar{n}}\Big]
\end{equation}
having its respective expected value and variance to be:
\begin{equation}\label{eq:Npolyaexpect}
     E[n] = N\bar{n}
\end{equation}
\begin{equation}\label{eq:Npolyavar}
    V[n] = \frac{N\bar{n}^2}{1+\theta}
\end{equation}
The fact of this generalization of Polya distribution when it comes to N electrons , is that it gives us a useful way to predict the distribution of any number of primary electrons, given the calibration of a single primary electron avalanche.

On the other hand, during the multiplication process, it is likely that the electrons can cause excited states of the atoms, because of gaining enough energy. Their decay to the ground state through cause the emission of a photon with a wavelength in and around the ultraviolet region of spectrum. These photons in turn, can cause ionization processes anywhere in the gas randomly. That applies a limit in the multiplication process, which cannot increase beyond that, otherwise it will lead to a breakdown of the detector supplied voltage. 
\begin{figure}[hbt!]
 \centering 
 \includegraphics[width=0.8\textwidth]{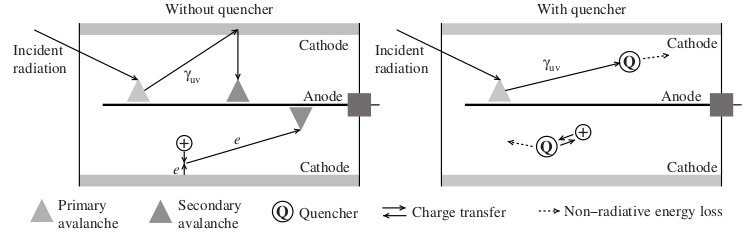}
\caption{Representation of a gas with and without quenchers \cite{AHMED}.}\label{fig:quenching}
\end{figure}
Another drawback of these photons is that when they strike the cathode, they may export an electron from the metal, and if this electron enters the gas, it can approach the anode as well as the initial electrons from the first avalanche, and create another one. This process is represented in the above Figure \ref{fig:quenching}. The way that it is found to solve this problem is by adding an agent in the gas, with the optical property of the high absorption coefficient to those photons. Polyatomic gases (i.e $CH_4$) fulfill this criterion. This mechanism of decreasing the probability of creation o a second avalanche in the active volume of the detector is called \emph{quenching}, and the agent gas is named as a \emph{quencher} \cite{AHMED}.

\subsubsection{Penning Effect} 
\paragraph{} As far as we have mention the benefits of gas mixtures, it is convenient to mention another aspect. Let's assume a gas mixture with two components A and B, if the excited state of component A is higher than the ionization potential of the component B, then a further ionization can occur through the process: 
\begin{equation}\label{eq:penning}
    A^* + B \rightarrow A + B^+ + e^-
\end{equation}
This is called \emph{Penning Effect} studied by F. M. Penning \cite{PENNING}, and is has to be taken into account, because it contributes to the multiplication process, when it comes to the gas gain calculation.
\subsection{Region of limited proportionality}
\paragraph{} As the polarization potential increases, more and more loads are generated within the active volume of the detector. So as heavy positive charges move significantly slower than electrons, they tend to create a cloud of positive charges between the electrodes. This cloud acts as a shield in the electric field reducing its efficiency. Consequently, the proportionality of the total number of charges produced by the primary charges of the incoming radiation is not determined. This area is referred to as the limited proportionality region. As the loss of proportionality also means a reduction in linearity, the radiation detectors do not work in this area.
\subsection{Geiger-Muller region}
\paragraph{} By further increasing the value of the local electric field, reaching high values(above 700V), where an extensive avalanche is created in the gas, many charged pairs are produced. Therefore, a high pulse is recorded in the reading electronics. This is the beginning of an area called Geiger - Mueller. In this region, it is possible to measure the incoming particles individually, as each particle causes a breakdown in the detector and a pulse higher than the background height. Since the output pulse is not proportional to either the energy deposited or the type of radiation, detectors operating in this area are not suitable for spectroscopy.
\section{MicroMegas Detector}\label{section:Micromegas}
The name of MicroMegas detector comes from the acromym MICROMEsh GAseous Structure. It belongs to a specific category of gas-filled detectors, the basic principle of which is the ionization of the gas, as a particle passes through and ionizes the gas, producing an electron-ion pair. The technology of these detectors was developed in accelerator devices and particle physics experiments. The MicroMegas detector, which we use in the present work, is a mature detector technology that was developed by Ioannis Giomataris - G. Charpak (1996)\cite{GIOMATARIS}. The main advantages of the MicroMegas detector are fast response, precise timing resolution of the order of tens of \,ns, exact energy resolution, accurate spatial resolution, high efficiency, and radiation resistance.
\subsection{Principle of Operation}
\paragraph{} MicroMegas is a two-phase detector with two areas of different field. What makes the difference in the detector is that the two distinct areas inside it are no longer separated by a wire-level, compared to drift chambers, but by a micromesh. 
Typical MicroMegas detectors consist of a flat electrode (drift), a gas gap a few millimeters thick that acts as a conversion gap, and a thin wire mesh (micromesh), at a typical distance of 128μm from the readout anode, creating the amplification gap. A structure of cylindrical columns (pillars), made of insulating material and placed with a step of a few millimeters, defines the height of the amplification gap and supports the mesh.

\begin{figure}[hbt!]
 \centering 
 \includegraphics[width=0.8\textwidth]{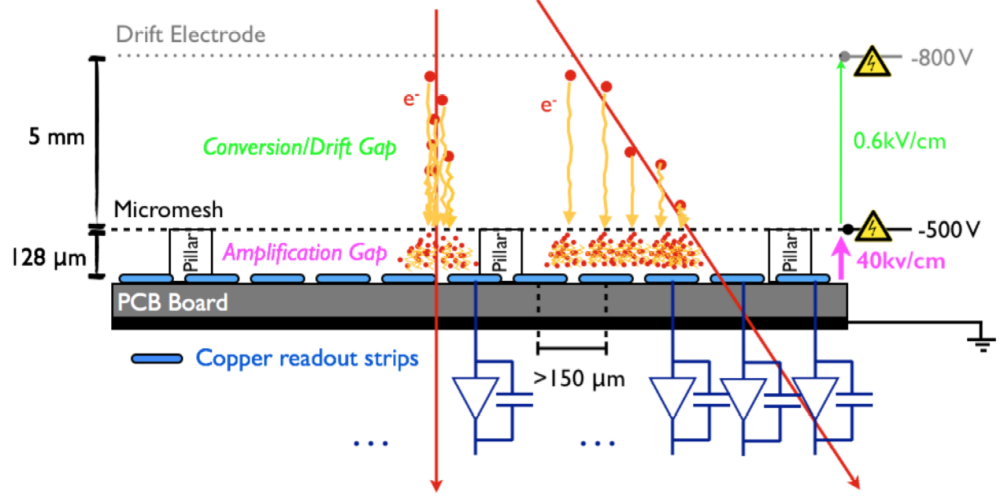}
\caption{Graphical representation of MicroMegas detector principle of operation \cite{Bouclier}.}\label{fig:micromegas}
\end{figure}

Initially, a particle crosses the drift electrode (cathode). Once there, the particle is in the conversion gap, which is a few millimeters away from the micromesh. During its passage, it interacts with the gaseous medium of the detector depositing part of its energy in the gas atoms through ionization and excitation. As a result, the ionization of the gas causes the formation of electrons and ions in the conversion gap, the secondary electrons under the influence of an electric field pass towards the mesh while the ions follow the opposite direction to the cathode electrode. Through the mesh, the electrons are led to the amplification region where due to the strong electric field that prevails (40kV / cm) they create avalanches. This electron avalanche continues in the amplification gap and eventually leads to the ascent where the electrons of the avalanche are collected by the strips. This avalanche takes place at a time scale of the order of 1ns, producing a rapid pulse in the readout strips. The ions produced in the avalanche process move towards the mesh lattice (in the opposite direction) at speeds about 200 times slower than the electrons (it takes more than 200ns to reach the mesh lattice). The majority of these ions reaching the mesh from the amplification region,  will be absorbed by this structure, in a process called \emph{neutralisation}. Some of them will pass through the mesh, consisting the ion back-flow. Those ions remaining in the amplification region, will induce a signal, proportional to the size of ion back-flow, resulting to the analogous space charge, with high efficiency and low speed. The typical structure and functional parameters of a MicroMegas are shown in the Figure \ref{fig:micromegas}.

The anode consists of copper strips with a typical width of 150μm and at a distance of 200μm between them, grounded through low noise preamplifiers, high performance in isolated layer (usually Kapton). The copper strips that structure the anode (strips) are made of copper (5μm) with a process based on the technique of photolithography, which allows us to print on them openings 25μm and at a distance of 50μm between them.

\begin{figure}[hbt!]
 \centering 
 \includegraphics[width=0.8\textwidth]{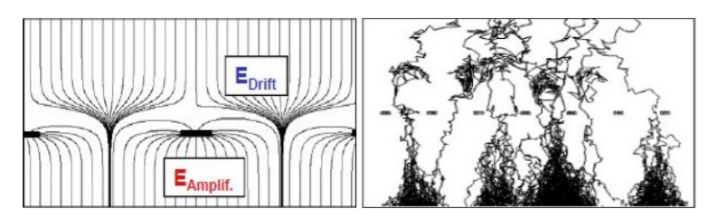}
\caption{Graphical representation of the electrical field of a MicroMegas detector \cite{Ntekas:2143887}.}\label{fig:micromegasfield}
\end{figure}

The role of the grid (micromesh) is multiple, and its usefulness is greater than just being the end of the conversion gap and the beginning of the amplification gap. By providing a smooth operation of the electronics, the micromesh prevents the ions produced by the avalanche from returning to the conversion gap. The potential difference applied to the conversion gap (above 500V) is such that the ratio of the intensity of the electric field in the amplification gap to the intensity in the conversion gap is very large. The higher this ratio, the higher the transmission of electrons in the amplification gap (in practice a ratio of 20 means full transmission-transparency). With an electric field in the amplification region 50 to 100 times stronger than the displacement field, the lattice is transparent to more than $95\%$ of the electrons, as shown in the Figure \ref{fig:micromegasfield}.
In addition, because the electric field in the conversion gap is created by applying high voltage to the micromesh and higher to the cathode electrode, this practically means that the micromesh is used as the middle electrode.
 By carefully selecting the high potential difference, we create an electric field of the order of 1kV/cm. The electric field in the amplification gap is achieved by keeping the anode strips grounded, creating a field 100 times stronger than that of the cathode electrode.
 
 Thus, the shape of the electric field is deformed near the mesh holes. Knowledge of the shape of field dynamics near the micromesh is a fundamental issue for the operation of the detector, especially for the percentage of electrons that penetrate the mesh, as well as how fast the area is emptied of positive ions. More specifically, a part of the dynamic lines of the field due to the load distribution in the conversion gap (above the mesh) creates a funnel-shaped shape, near the holes of the mesh. Having a thin mesh and since the amplification gap field is an order of magnitude larger than the conversion gap any dynamic line coming from the top of the mesh is not returned to the bottom. The consequence of this is that most of the electrons penetrate the mesh and are received completely from the plane of the anode (strips), as shown in Figure \ref{fig:micromegasfield}.

Despite the outstanding features of MicroMegas detector technology, and the promising industrial manufacturing process, the very thin amplification area along with the detailed readout structure make the detectors particularly vulnerable to sparks. The phenomenon of discharges occurs when the local concentration of electrons exceeds the $10^7$ particles. In this case, the detector due to the increase of the induced leads to a total evacuation of the mesh. The rate of these discharges is proportional to the rate of incoming particles. This phenomenon is not catastrophic for low particle flux, bat in a heavy particle high beam environment it significantly affects the operation of the detector, as it increases the dead time and contributes to the wear of the electronic readers. As a result, in order to deal with the phenomenon, it was decided to modify the detector in order to reduce its sensitivity to the specific phenomenon. 

\begin{figure}[hbt!]
 \centering 
 \includegraphics[width=0.8\textwidth]{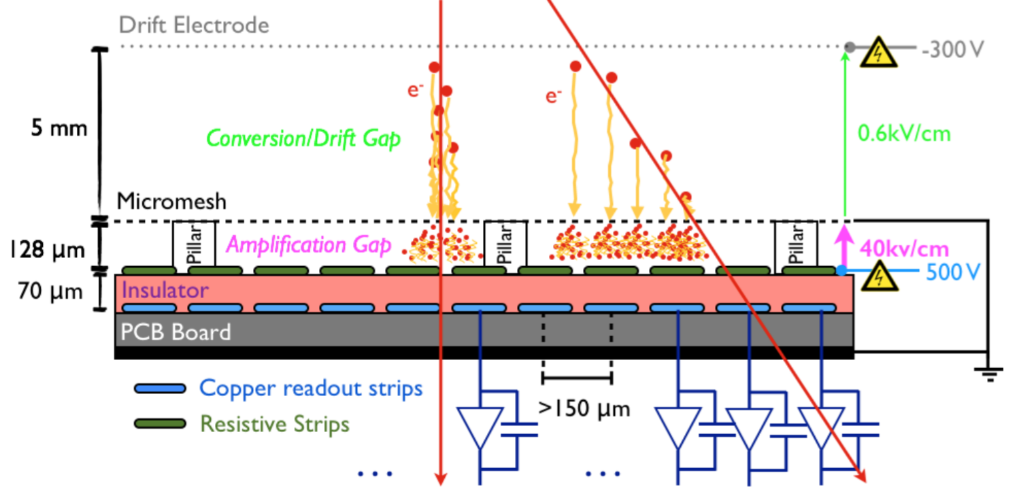}
\caption{Graphical representation of a resistive MicroMegas detector \cite{Alexopoulos:2011zz}.}\label{fig:resistive_MM}
\end{figure}

The aforementioned disadvantages have recently been overcome with the development of spark resistant MicroMegas technology. In this technique, the readout strips are covered with a durable high-strength insulating layer(equivalent to a 15-50M$\Omega$/cm resistor) and has the same geometry as the readout strips. This morphology is done to reduce the diffusion of the charged into many strips and to reduce the intensity of the discharge current. The idea of resistive MicroMegas is schematically described in Figure  \ref{fig:resistive_MM}.

\section{The PICOSEC MicroMegas Detector}\label{section:PICOSEC}
\paragraph{} The time resolution of the above described MicroMegas detector is limited because of the stochastic nature of ionization. This is explained because MIP after the first ionization and until the time it will create an avalanche will suffer from longitudinal diffusion effect, meaning scattering, that will result to a delay in timing. Moreover a study resulted to the fact that the electrons of the avalanche drift with higher drift velocities than its electron components, which means that the creation of avalanche in the early stages of the drift gap will influence the timing resolution of the detector. A second drawback of normal Micromegas is the random way the primary electrons are produced by the incident charged particle, this is why we need a way to produce synchronous electrons in the detector area. 

\begin{figure}[hbt!]
 \centering 
 \includegraphics[width=0.8\textwidth]{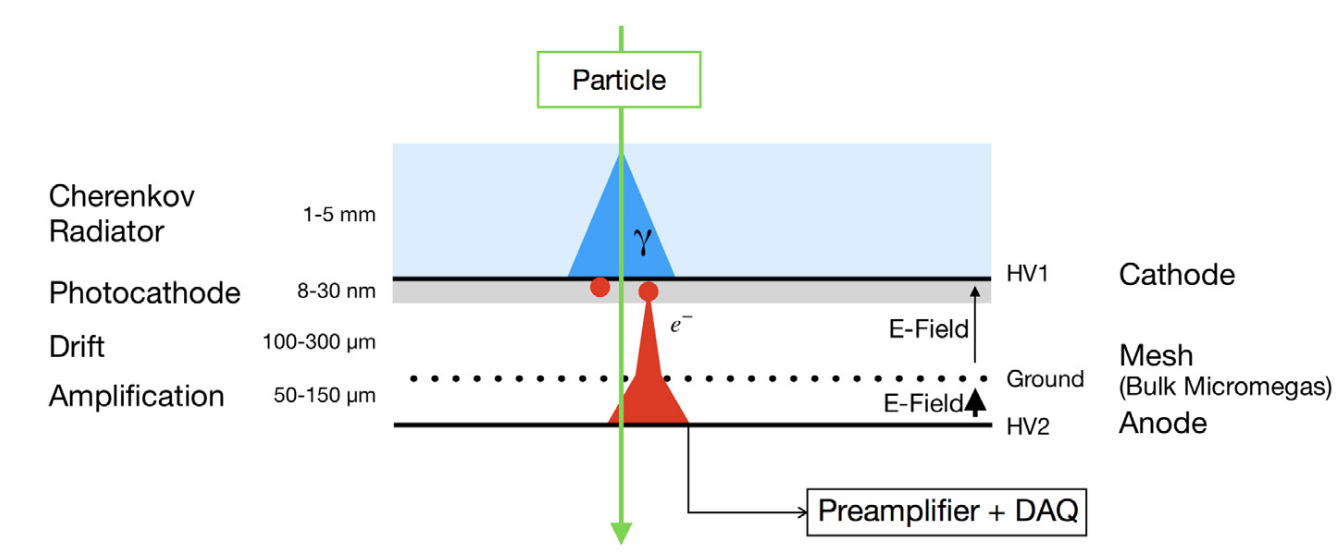}
\caption{Graphical representation of a PICOSEC MicroMegas detector \cite{BORTFELDT2018317}.}\label{fig:picosec}
\end{figure}

The PICOSEC detector is described in Figure \ref{fig:picosec}. It is a MicroMegas-based detector, and its aiming is to time charged particles with tens of picosecond precision. The main modification difference from a normal MicroMegas detector is the smaller conversion gap (from 3 mm to 200μm). This choice reduces the probability of a charged particle interacting with the gas to produce a primary electron, and then an avalanche, that will be detectable. This reduced conversion gap is a result of the need of creating avalanches as early as possible, which means we have to increase the probability of creating an avalanshe, e.g. the Townsend coefficient, which in its order means either an increase to the applied electric or a reduce to the drift gap. As a result this region from the cathode to the micromesh is so on called drift gap and the relevant voltage applied between them will be referred to as drift voltage and it will be high enough to cause the creation of avalanches in that gap. These avalanches from so on will be called as pre-amplification avalanches \cite{BORTFELDT2018317}.  

Reducing the drift gap region, is a choice that aims to increase the field and make pre-amplification avalanches possible. Additionally, the probability of the primary (charged) particle to interact with the gas of the drift region is significantly reduced . The solution comes by equipping the detector with a photocathode, instead of a simple cathode, acting as a photodetector with precise timing resolution. In that way, charged particles can become visible to the detector with a Cherenkov radiator on top of the photocathode. This happened with a 3mm thick $MgF_2$ crystal so that when relativistic charged particles pass through this radiator with greater speed than the speed of light in the medium, the radiator will emit Cherenkov photons. The advantage of this method is that these Cherenkov photons will interact with the photocathode, creating  synchronous photoelectrons and enter the gaseous volume. Thus, a great advantage is the prompt prompt production of photoelectrons. The fact that we apply strong electric field in this region, is allowing charge multiplication. This avalanche multiplication will continue to the micromesh, where a fraction of the electrons of the avalanche will transverse the mesh and will continue the multiplication in the amplification gap, which is held in higher field, resulting to an induced current on the anode. The readout electronics will read this signal and create a measurable pulse in our system. This procedure results to the minimization of the time jitter of a few \,ns that classic MicroMegas offer.

\begin{figure}[hbt!]
 \centering 
 \includegraphics[width=0.75\textwidth]{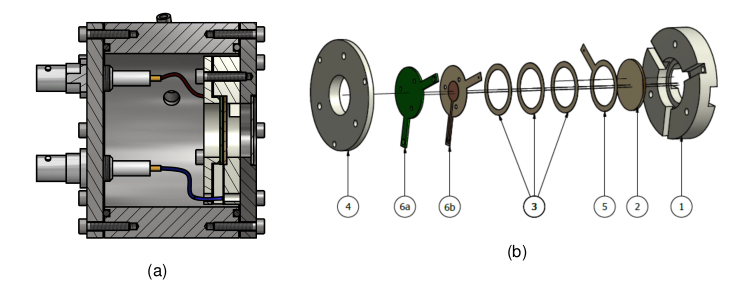}
\caption{(a) Transverse view of a MicroMegas mounted on the right place of this figure. (b) Expanded view of the MicroMegas and radiator mechanics. Part 1 and 4 are the supporting rings holding the whole apparatus straight. Part 6a and b is the MicroMegas. Part 3 and 5 consist the four rings defining the drift gap with a thickness of $50\mu m$, with the fourth one to having the additional copper ring that gives electrical contact to the cathode. The sketch is completed with the part 2, the Cherenkov radiator. Plot adopted from  \cite{sohl:tel-03167728}.}\label{fig:micromegas_2}
\end{figure}

\begin{figure}[hbt!]
 \centering 
 \includegraphics[width=0.55\textwidth]{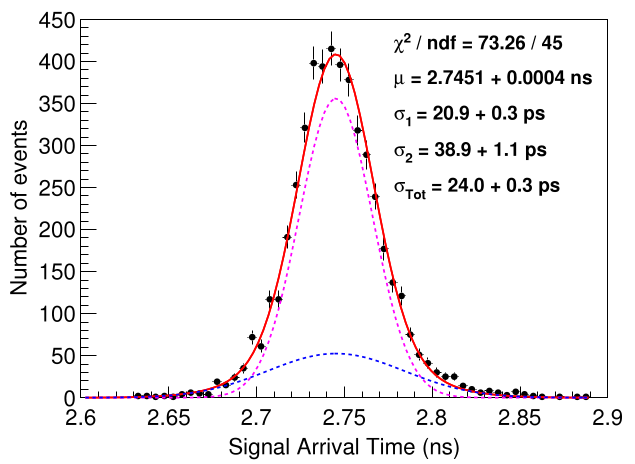}
\caption{Result of Signal Arrival Time distribution for 150\,GeV muon beam test. In the Figure is presented a superimposed fit of a Double Gaussian Distribution, with PICOSEC operating with anode voltage at 275V and drift voltage at 475V. Plot adopted from  \cite{Bortfeldt2019}.}\label{fig:24ps}
\end{figure}

\newpage
Previous studies focused on the performance of this PICOSEC-MicroMegas detector, resulted in the distribution of its Signal Arrival Time, as the one shown in Figure \ref{fig:24ps}, with black points. This histogram is fitted with a Double Gaussian distribution where the two distributions have the same mean values, the one shown with the solid red line. In the same plot, the dashed lines correspond to the relevant Gaussian components of the combined Double Gaussian. This Gaussian has a standard deviation of the order $\sigma$ = 24$\pm$ 0.3\, ps. This is the evidence that the PICOSEC-MicroMegas detector has a novel timing precision contrary to the rest of its kind gaseous detectors, and that with just a single measurement of a minimum ionizing particle.

\chapter{Precise Timing Techniques with PICOSEC-Micromegas Detector\label{cha:design}}
\vspace{-0.8cm}
\noindent\fbox{%
    \parbox{\textwidth}{%
        The PICOSEC Micromegas detector is a novel gaseous detector offering two orders of magnitude better timing resolution than any other gaseous detector. The objective of this chapter is to describe further the detector referring to its characteristics and operating conditions \cite{BORTFELDT2018317}. Thus, it is necessary to make the first steps in the sub-picosecond domain of timing. Later on in the chapter, we will provide detailed description of its structural details and the experimental setup of a laser test beam facility (here after called \emph{\say{Laser Test}}), that provided us the experimental data under study in the present thesis. Finally in the last section is described a summary of the timing techniques that have been used in the analysis in Chapter \ref{cha:implementation}.      
    }%
}

\subsubsection{The Picosec Prototype}\label{section:PicoPrototype}
\paragraph{}The PICOSEC Micromegas Prototype is a bulk Micromegas detector \cite{Giomataris:2004aa}, but in circular geometry, without micro-strips, and with a diameter of 1\,cm. As it happens to the standard Micromegas detector, and because this is a bulk design, the micromesh sits on top of the anode at a distance of 128\,$\mu$m . Micromesh is a wire mesh, with 63\,$\mu$m pitch size, witch defines the distance between the centers of each hole of the mesh. This property of the mesh implies another optical characteristic, its optical transparency, which is 50$\%$, and refers to the fraction of photons in visible region that can transverse the mesh. On the other hand, electron transparency refers to the fraction of electrons that can transmit the mesh, and this property is highly connected to the strength of the electric field. 

The mesh is held in a stable position by 6 cylindrical pillars with a diameter of 200$\mu$ \,m, and they are settled in a  hexagonal layout the center of the anode. 
\begin{figure}[hbt!]
 \centering 
 \includegraphics[width=0.8\textwidth]{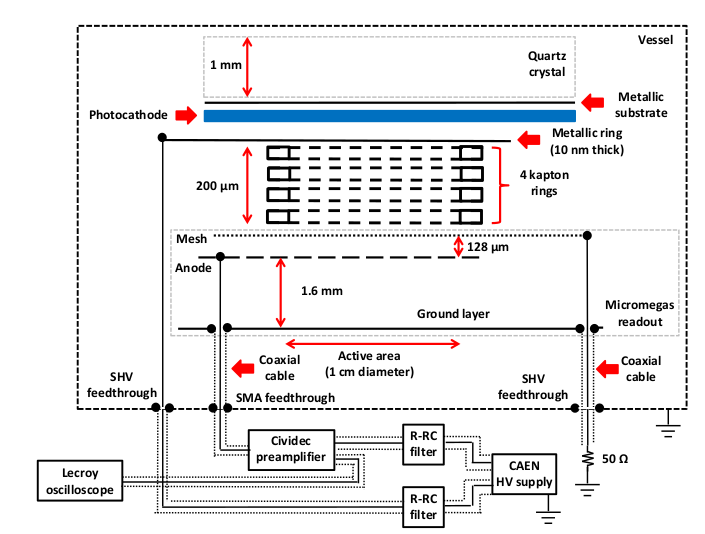}
\caption{Schematic representation of a PICOSEC Micromegas detector prototype and the Data Acquisition system \cite{BORTFELDT2018317}.}\label{fig:picoprototype}
\end{figure}

The drift gap is set to 194\,$\mu$m above the center of the mesh. As it can be seen from Figure \ref{fig:picoprototype} these 194\,$\mu$m are specified with four capton rings, each of 50\,$\mu$m thickness, placed one on top of the other, and which are placed on the insulating layers, during the lamination process of construction of the bulk technology. Later the photocathode and the Cherenkov radiator are placed one above the other, and the full assembly on top of the rings, there is also a 9\,mm Chromium layer below the photocathode, serving as the photocathode. The Cherenkov radiator is made of $MgF_2$ with 3\,mm thickness. The photocathode is placed on its center, which is made of $CsI$, with 18\,nm thickness and 1\,cm diameter. This configuration is settled in a steel vessel which is filled with gas mixture which is composed of $80\% Ne, 10\% CF_4, 10\% C_2H_6$. For prevention of reflections in mesh connection, it is grounded via a long coaxial cable with a BNC plug and a 50 Ohm termination resistor.  
\section{Experimental Evaluation with Laser Beams.}\label{section:LaserTest}
\begin{figure}[hbt!]
 \centering 
 \includegraphics[width=0.5\textwidth]{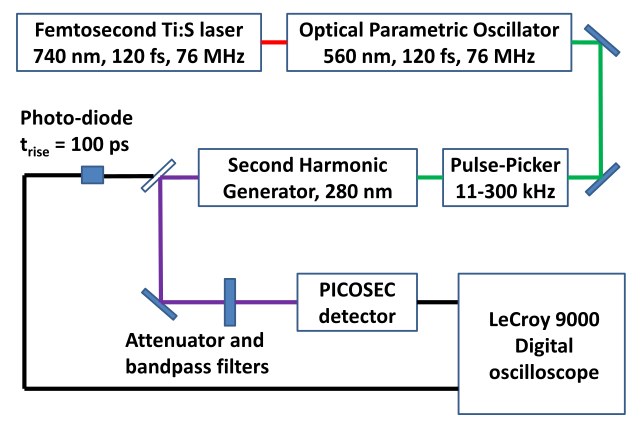}
\caption{Schematic representation of the experimental set up during the laser test \cite{BORTFELDT2018317}.}\label{fig:lasertest}
\end{figure}

\paragraph{} The first investigation of the detector's timing response will be with single photo-electrons. This measurement took place at the \emph{Saclay Laser-matter Interaction Center(IRAMIS/SLIC, CEA)}. Aiming to a picosecond order of timing resolution, the experimental setup, seen in Figure \ref{fig:lasertest}, is using a femtosecond laser (used to provide test beam to the PICOSEC-MicroMegas detector) with a pulse rate in a region of 9kHz to 4.7MHz, wavelengths from 267\,nm to 278\,nm and a focal length of 1\,mm. At the first point, the femtosecond laser beam is split into two equal beam parts, one is directed straight to a fast photodiode (PD0), which is used as a time reference, and the other one to the PICOSEC prototype. Both those signals are recorded simultaneously, giving their time difference as a result of the measurement. The high intensity of the laser and the fast rise time of the pulse of the PD0 device results in a reference time of 13\,ps. On the other hand, the intensity of the laser beam is reduced, before reaching the PICOSEC, by a number of light attenuators called $\text{electroformed fine nickel meshes}$ with $100-2000 LPI$(\emph{Lines per Inch}, which have values for the optical transmissions in the range $10\%-25\%$, yielding attenuation factor of 4. 

The attenuation procedure of the laser beam is adjusted such that only one photo-electron to be extracted from the photocathode. The PICOSEC signal is preamplified through a CIVIDEC device, before its digitization and the photodiode signal are both registered with a 2.5 GSamples/s oscilloscope and a digitization rate of $20GSamples/s$(i.e 50ps time step between each digitization). 
When it comes to the operation of PICOSEC, the anode voltage was scanned in a range of $450-525V$ with a voltage step of 25V, while the drift voltage was scanned in a range of $350-500V$ for the $CF_4$ and in the range of $200-425V$ for the COMPASS gas with the same voltage step. The combinations of anode-drift voltages were chosen with the prerequisites of having a clean signal, distinguishable from noise, and a stable operation of the PICOSEC \cite{BORTFELDT2018317}. 

\begin{figure}[hbt!]
 \centering 
 \includegraphics[width=0.8\textwidth]{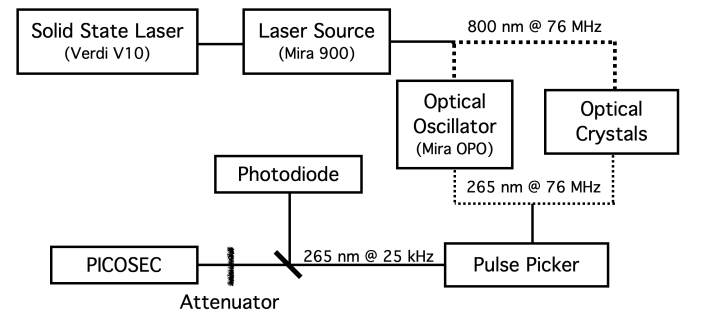}
\caption{An alternative of the experimental set up during the laser test \cite{sohl:tel-03167728}.}\label{fig:lasertestshol}
\end{figure}

In our analysis, data were collected from Dr. Lukas Sohl during his Ph.D. Thesis \cite{sohl:tel-03167728} at the same laboratory with the LYDIL laser of CEA-IRAMIS. The motivation of this test was to lead to a better understanding of the performance of the PICOSEC Micromegas detector prototypes through changes in detector parameters and as a result on its timing resolution. For this reason, an alternative form of the setup was used, like the one presented in Figure \ref{fig:lasertestshol}.  Our data collection comes from RUN 291 with PICOSEC-MicroMegas drift gap reduced to 119\,$\mu m$ \cite{BORTFELDT2021165049}, gas mixture of Neon-Ehtane-C$F_4$ : $80\% Ne, 10\% CF_4, 10\% C_2H_6$, anode voltage of 275V, and cathode voltage of 525V. This prototype was constructed and tested in Saclay, and with these operation characteristics led to the best timing resolution from single photoelectron measurement. 

The difference in this testing apparatus is the use of UV laser light, which can create single photoelectrons in the photocathode. The laser used in this setup is a tunable laser (MIRA 900 \cite{MIRA} with the name of \emph{Ti:sapphire laser}($Ti:Al_2O_3 - titanium-sapphire$). Laser usually needs another laser to be pumped from with a wavelength range between [514,532]\, nm. In our case, the choice was a solid-state laser operating at 532nm (VERDI V10\cite{VERDI}). Another characteristic of the Ti:sapphire laser is that it belongs to the mode-locked laser type. That means it can generate ultra-short pulses with a duration of a few picoseconds to tens of femtoseconds, and that they operate efficiently at 800mm. This laser source results in exiting pulses of maximum wavelength at 800nm and repetitive every 76MHz. In the next step, the laser beam needs to be adjusted to 265nm, but this can be achieved using nonlinear optical processes \cite{sohl:tel-03167728}. Two methods are indicated for this procedure. The first method demands an optical parametric oscillator (MIRA OPO \cite{MIRAOPO}), after the Ti:Sapphire laser. This optical oscillator contains a second-harmonic generation material, which after the excitation process these crystals emit non-linearly polarised light with doubled oscillation frequency than the pumped light \cite{sohl:tel-03167728}. Moreover, it needs a ring cavity for visible light. The oscillator provides a pulse with a tune less than 200\, fs at 580\, nm and a range of [530,660]\,nm. Up next there is a 5mm frequency doubling crystal which reduces the wavelength to the desired value of 265\, nm. An alternative method is a direct mix of the 800\,nm pulse coming out of the Ti:Sapphire laser with an array of non linear optical crystals to the desired wavelength \cite{sohl:tel-03167728}. This final method leads to a more stable pulse, without any need of fine-tuning. 

The fact that our detector has an ion collection time in the photocathode, less than 76MHz, it means that we need a lower repetition rate before the mixing process with the crystals. This reduction process can be realized with a pulse picker which adjusts the repetition rate in a range of [4.76MHz, 25kHz]. For that reason an acousto-optic modulator is required, and the process is described in detail in \cite{sohl:tel-03167728}. The exiting pulses have intensity that has been reduced to their 40$\%$ and an energy in the range of $[40,18]pJ/pulse$ at 265nm. After all this, the procedure goes as previously mentioned, this beam is split, with its one part is sent to the photodiode(here a DET10 A/M \cite{DET} is used) and the other part is sent to the PICOSEC-Micromegas detector. The photodiode's signal triggers the DAQ system and it is used as a time reference. In order to have a control of the number of photoelectrons exiting the photocathode of the detector, a combination of attenuators is used as well as in the setup of the Laser Test Beam described previously. 

The last thing that has been changed in this data taken procedure was the photocathode of the PICOSEC-Micromegas prototype, its material is a 10nm Aluminium layer because of its work function of 4.25$\pm$0.05\,eV that is lower that the photon energy that gives the laser $E_{\lambda} = 4.68$\,eV, and its lower radiation damage in high intensity environments. That property can help us to achieve better quantum efficiency for the setup. 

\begin{figure}[hbt!]
 \centering 
 \includegraphics[width=0.7\textwidth]{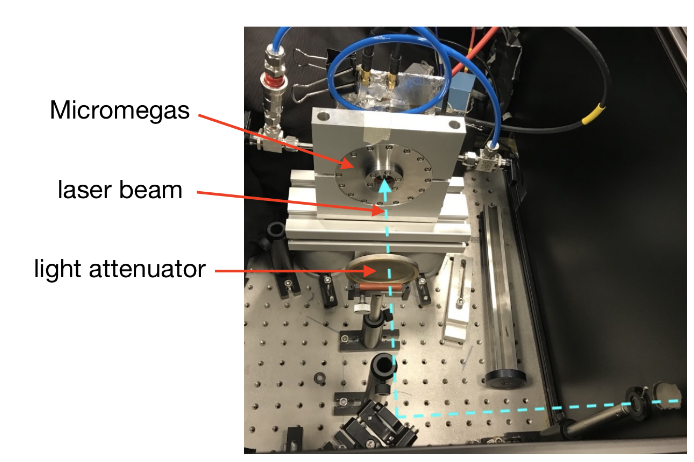}
\caption{Representation of the Laser Test Beam set-up, with the proper position of PICOSEC-MicroMegas detector according to the laser beam and attenuator filters). Image adopted by \cite{Sohl2020}.}\label{fig:laserbeam}
\end{figure}

\section{Signal Processing}\label{section:signalprocessing}
\begin{figure}[hbt!]
 \centering 
 \includegraphics[width=0.7\textwidth]{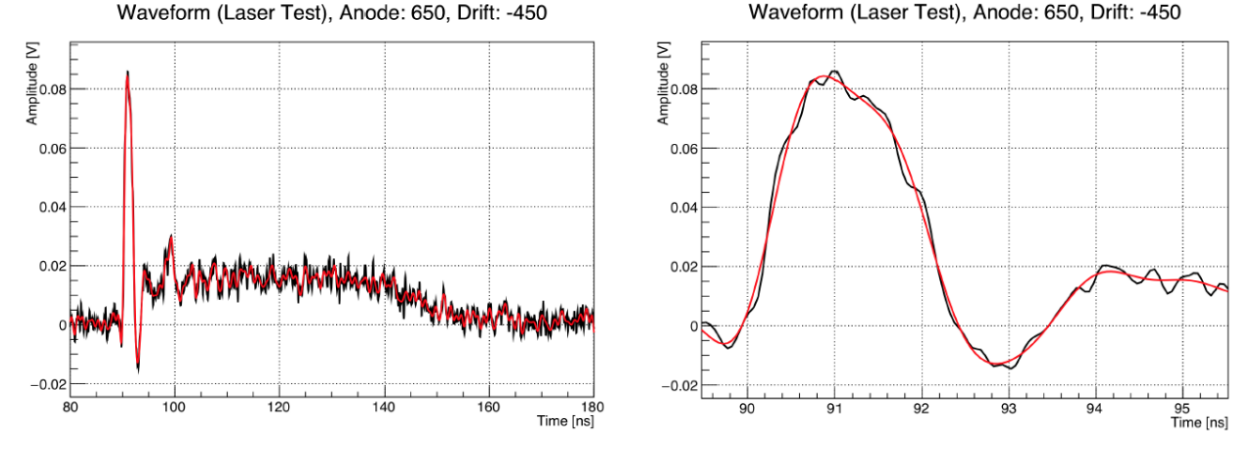}
\caption{A typical pulse of PICOSEC detector, giving its amplitude with respect to time. On the left is given an example of a full pulse while on the right is given a zoom representation of the electron peak, after Fourier filtering at 2GHz(red line). \cite{Bortfeldt2019,Paraschou:2020max}.}\label{fig:waveform}
\end{figure}

\paragraph{} Moving charges in the active volume of the detector, give rise to electrical signal on the readout electrodes. Thus, in our set up its activation causes an electrical signal, which is described by a \emph{waveform}. A typical pulse of PICOSEC detector is shown in Figure \ref{fig:waveform}. 

The waveform consists of two easily recognizable components, the sharp peak of the electron contribution (hereafter called \emph{e-peak}) and the extended \emph{ion tail}. The electron peak is a fast signal, a shape expected  due to the well-known movement of electrons in the gas \cite{BORTFELDT2021165049}. The ion tail is a slower component and is induced from the positive ions created in the amplification gap and drift to the micro-mesh, and after transmitting the mesh continues to move through the cathode. The red line in the same Figure, is an example of the waveform that comes out of a Fourier-based filter. Usually the data acquisition introduce a non zero baseline in the waveforms, which has to be measured and corrected on an event-by-event basis. The correction procedure in the case of specific data set was performed according to the following procedure. The oscilloscope was adjusted with a 75\,ns buffer, in order this area of the waveform not to contain any signal. In this signal, we can calculate on an event-by-event basis, the average amplitude, for all the pulses of the data-set, according to a Gaussian fit with a mean value which determines the \emph{baseline offset} and standard deviation(RMS) value that determines the \emph{baseline noise}. Then, the baseline offset is subtracted on a point-by-point basis, moving the pulse to the real zero amplitude in the region without signal, like the one mentioned in the next Figure \ref{fig:noise}.    
\begin{figure}[hbt!]
 \centering 
 \includegraphics[width=0.5\textwidth]{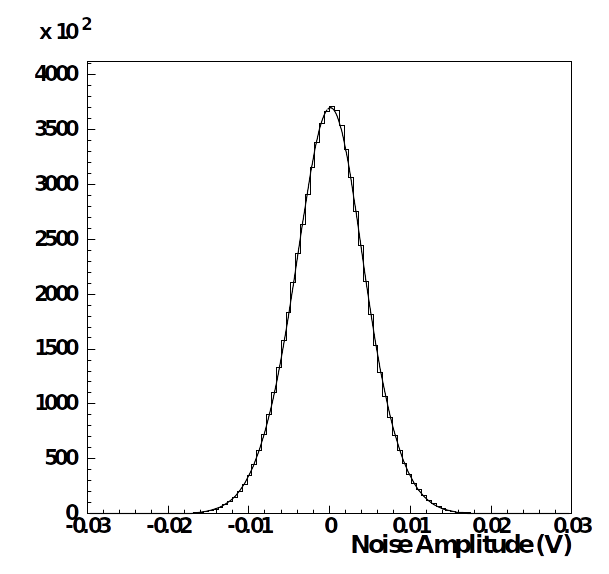}
\caption{Example of the random noise distribution, from which the baseline offset is determined from the mean value, and the baseline noise is determined from the standard deviation of the Gaussian fit.}\label{fig:noise}
\end{figure}

\subsection{Electron Peak Size}\label{subsection:ElectronPeakSize}

The pulse size can be fully described by two quantities, the electron-peak amplitude and the electron-peak charge. The electron-peak amplitude is the amplitude at the highest amplitude point of the waveform, and it is related to the multitude of the secondary electrons created in the detector, or the gain of the detector. The electron peak charge, is the integral of the electron peak, from its starting point to its end. The measurement unit of this value can be transformed to (pico)Coulombs using the termination resistance value of $50\Omega$. This variable is more reliable value than the e-peak amplitude, because the noise is averaged out in the integral.      
\begin{figure}[hbt!]
 \centering 
 \includegraphics[width=0.5\textwidth]{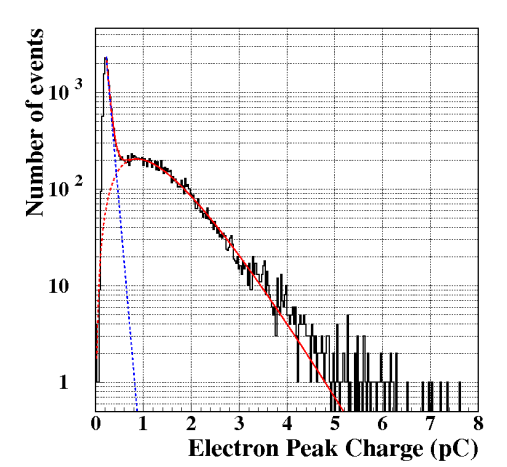}
\caption{Electron Peak Charge distribution of single photo-electron events, fitted with Polya distribution. Plot adopted from \cite{Bortfeldt2019}, mentioned in Section \ref{section:LaserTest}.}\label{fig:single}
\end{figure}
Both the distributions of electron peak and amplitude are parameterized with the Polya distribution of Equation \ref{eq:N-polya}. The typical distribution of single electron peak charges is shown above in the Figure \ref{fig:single}, where the black points are the experimental data of the laser test, the red line corresponds to the Polya fit and the blue line gives the noise contribution, fitted with an exponential that discribes the data.  
\section{Timing Techniques}\label{section:TimingTechniques}
\paragraph{} In addition to electron peak charge and electron peak amplitude, timing is also an important characteristic of the waveform. Unfortunately, there are some statistical errors affecting the resolution of timing. Accurate measurements in small time intervals that range from a few picoseconds to a few microseconds, require special timing techniques to be taken into account.  In this section we will discuss timing techniques, which are usually applied either on a real time or offline emphasizing the statistical and systematical errors implied.
\subsection{Constant Threshold Discriminator}\label{subsection:CoT}
\subsubsection{The Time Jitter Effect}\label{subsubsection:slewing}
\begin{figure}[hbt!]
 \centering 
 \includegraphics[width=0.4\textwidth]{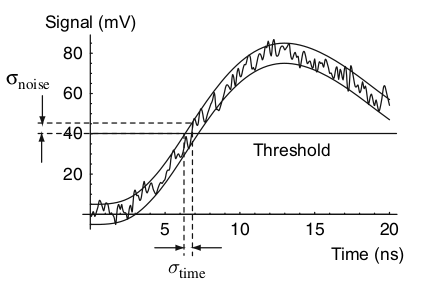}
\caption{Time jitter geometrical representation. Plot adopted from \cite{BLUM}.}\label{fig:jitter}
\end{figure}

\paragraph{} The first and most important statistical error, common to most timing techniques, is the one known as \emph{"time jitter"}. The aid of electronic devices in order to translate the signal to a readable form also involves electronic noise, which is difficult to avoid in the first place. Due to those fluctuations, electronic noise in amplitude will give rise to a timing noise, called \emph{jitter}, presented in the Figure \ref{fig:jitter}. This is the reason why, seemingly identical waveforms won't trigger at the same point. 

As it can be seen clearly from this representation, the amplitude noise is "translated" in timing noise according to the Equation \ref{eq:noise}.
\begin{equation}\label{eq:noise}
    \sigma_{time} = \bigg|\frac{dV}{dt}\bigg|^{-1}\cdot \sigma_{Vnoise}
\end{equation}
As a result, the spread of timing resolution (RMS) will depend on the signal slope, giving the fast pusles the advantage of having greater slope, meaning faster rise-time and finally better timing resolution. This technique obliges us to assume uniform signals when it comes to their amplitude and shape, and that the discrimination procedure of two identical signals, will happen with the observation of them crossing a fixed threshold. Thus, the time resolution will be identical with the time jitter $\sigma_{time}$, using this timing technique. Although this is an error that we can not eliminate with the offline analysis, as it can happen with the correction of time walk effect. 

\begin{figure}[hbt!]
 \centering 
 \includegraphics[width=0.4\textwidth]{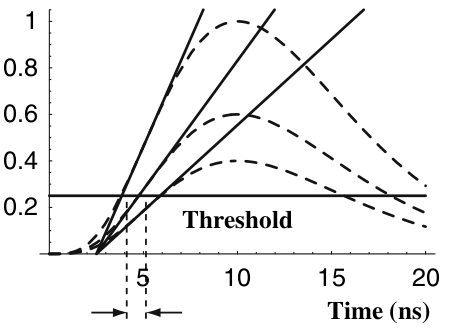}
\caption{Time walk geometrical representation, crossing a fixed threshold. Plot adopted from \cite{BLUM}.}\label{fig:timewalk}
\end{figure}

In a more realistic case, where pulses are not in a fixed amplitude, the pulses with higher amplitude, will cross the applied fixed threshold earlier in time, than smaller pulses (in amplitude). The so called \emph{time walk effect}, expresses exactly this dependence between timing and amplitude size (Figure \ref{fig:timewalk}). The contribution of this effect to the final timing resolution of the waveform, is much more than the one of the time jitter effect, thus, ignoring this, would lead to less precise timing. Fortunately, the time walk effect on pulses, can be corrected later, on the off-line analysis. 
In fact, techniques there have been developed in order to avoid this off-line correction, such as the \emph{constant fraction discrimination}.
\subsection{Constant Fraction Discrimination}\label{subsection:CDF}
\paragraph{} The constant fraction discriminator technique can help us to avoid the pulse-height dependence that constant threshold technique includes. In this method, the logic signal is produced at a specific fraction of the peak height, for example the time at the $20\%$ of the peak height. The only condition that continues to hold on this technique, is the identical shapes of the signals that we need. If all the above can be applied to our problem, then constant fraction discriminator technique can lead to the determination of timing resolution eliminating time-walk effect. However, the time walk effect still exists, as we will see later, and it is caused due to the charge of the pulse shape and underlying stochastic physical processes occurring in the drift region of the detector \cite{BORTFELDT2021165049}.

Supplementary to the constant fraction discrimination timing technique, there are some additional techniques, however, they may be suffer from time walk effect, but also can be used in order to time accurate the pulse, using proper corrections. Thus, there is a number of methods that can be used to time a signal, and it is convenient to use multiple of them in order to make sure that we have consistent results. 

Fist attempt can be a so called \emph{naive} version of constant fraction discrimination. In this case, someone has to find two successive points, on the left and on the right of the relevant point of the waveform crossing the threshold we have set, e.g. at $20\%$ of the peak amplitude. Thus, applying a linear interpolation between the points, we can find the timing at this fraction of the amplitude, overcoming constraints due to the digitisation rate.

As an alternative of the first attempt, assuming again the naive version of constant fraction discrimination, but now using more than two points, in total, close to the crossing threshold point of the waveform. An example can be the cubic interpolation that needs four points inserted, i.e two on the left and two on the right of the, e.g. $20\%$, of the peak amplitude. Again this is the the right timing of the signal. 

Last but not least, the technique of \emph{parameterization}. In this technique, the effort is to adjust a curve to the experimental data, i.e the digits of the signal. For this reason someone can use a parametric function to fit the leading edge of the waveform and then apply the the constant fraction discrimination technique to the parametric function. A special function called the \emph{sigmoid function}(logistic)  seems to fit our leading edge of the waveform. Its functional equation is:
\begin{equation}\label{eq:logistic}
    f(x;p_0,p_1,p_2,p_3) = V(t) = p_3 + \frac{p_0}{1+e^{-(x-p_1)p_2}} 
\end{equation}
where $p_0$ and $p_3$ are the maximum and minimum values, $p_1$ is the point where the derivative of the slope changes sign, and $p_2$ is the rise-time of the waveform. As an example the time at the $20\%$ of the peak amplitude for the constant fraction discrimination technique, replacing on the Eq.\ref{eq:logistic}, $V(t) = 0.2V_{max}$, $x \rightarrow t$ and solving for the time, is \cite{BORTFELDT2018317}:
\begin{equation}\label{eq:CDF20}
    t_z = p_1 - \frac{1}{p_2}\ln{\Bigg(\frac{p_0}{0.2\times V_{max}-P_3}-1}\Bigg)
\end{equation}

\begin{figure}[hbt!]
 \centering 
 \includegraphics[width=0.6\textwidth]{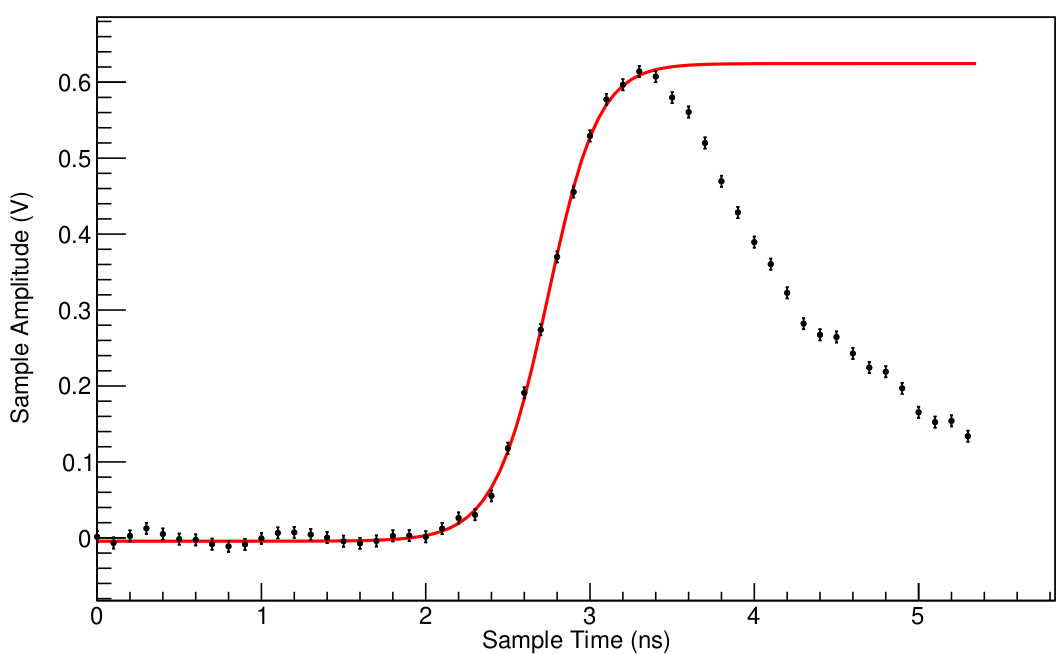}
\caption{The sigmoid function fitted to data points on the leading edge of a PICOSEC-Micromegas signal. Plot adopted from \cite{sohl:tel-03167728}.}\label{fig:sigmoid}
\end{figure}

Another similar approach is the generalized logistic function in order to fit the leading edge of the pulse, with the functional form of Eq.\ref{eq:genlogistic}. 
\begin{equation}\label{eq:genlogistic}
    f(x;p_0,p_1,p_2,p_3) = V(t) =  \frac{p_0}{(1+e^{-(x-p_1)p_2})^{p_3}}
\end{equation}
giving timing of :
\begin{equation}\label{eq:CDF20-2}
    t_z = p_1 - \frac{1}{p_2}\ln{\Bigg(\Big( \frac{p_0}{0.2\times V_{max}}\Big)^{1/p_3} - 1\Bigg)}
\end{equation}
As it is clear from both the Eq.\ref{eq:CDF20} and Eq.\ref{eq:CDF20-2} we need a very precise measurement of the peak amplitude, in order to achieve precise timing. But it is very difficult to determine the maximum amplitude, due to the electronic noise, we have mentioned before, and as a result, this variant adds a bias in our estimation. An interesting improvement would be, to find a parametric function to fit in the whole electron peak, for the estimation of the maximum amplitude. This parametric function is seems to be the difference of two generalized logistic functions as it can be seen in the Eq.\ref{eq:logisticdifference}
\begin{equation}\label{eq:logisticdifference}
f(x;p_0,p_1,p_2,p_3,p_4,p_5,p_6) = \frac{p_0}{(1 + e^{-(x-p_1)p_2})^{p_3}} - \frac{p_0}{(1 + e^{-(x-p_4)p_5})^{p_6}}  
\end{equation}
The only drawback of this expression is that it has no analytical solution, but we can apply numerical techniques to find a solution, to find estimators for the timing of the signal, the peak maximum and the peak charge through the integration of the function of Eq.\ref{eq:logisticdifference}. More information on the exact process analysis and how to apply these techniques can be found in the next Chapter. 

\chapter{Analysis of the PICOSEC Micromegas Signal\label{cha:implementation}}
\vspace{-0.8cm}
\noindent\fbox{%
    \parbox{\textwidth}{ This chapter gives an analytical description of the analysis procedure, that has been followed in the offline analysis of the Laser Test data, estimating timing resolution with different timing techniques, that have been mentioned in Chapter \ref{cha:design}, as a proposal of alternative timing techniques.
    }%
}
\section{The Standard Constant Fraction Discrimination Timing Technique}\label{subsubsectionCDF}
It is proved that PICOSEC Micromegas Detector can provide precise timing measurements. In this section, the data from a Laser test run, mentioned in the previous chapter have been analyzed under an offline analysis, and as a part of the Research and Development of the PICOSEC AUTH-Group, to rise alternative techniques on timing procedure.

To determine the precision of the calculation of the arrival time of a signal that came from PICOSEC detector, taken from the experimental apparatus of the Section \ref{section:LaserTest}, we need a more precise device as a reference time. Hence, the time difference of the detector's signal  and the reference's, for many measurements, will follow a normal distribution. The standard deviation (RMS) of this normal distribution determines the timing resolution of the PICOSEC-MicroMegas detector. 

\begin{figure}[hbt!]
 \centering 
 \includegraphics[width=0.5\textwidth]{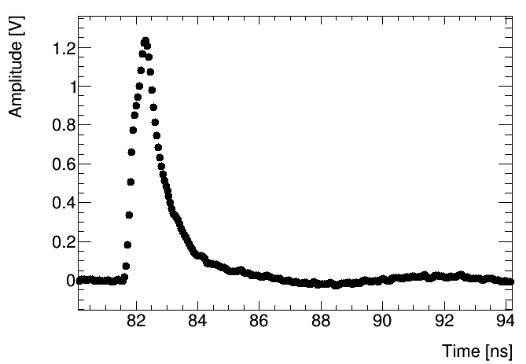}
\caption{Photodiode typical waveform shape, on an event-by-event basis. }\label{fig:photodiodewaveform}
\end{figure}

The reference time in our experimental set-up is given from a fast photodiode. Its signal has a time resolution of about 3\,ps, so it has the necessary precision to study the time resolution of our detector in a level of picosecond. In Figure \ref{fig:photodiodewaveform}, a digitized with 20GSample/s, photodiode's signal is presented, with the Constant Fraction Discrimination technique being used to time this at $20\%$ of peak amplitude using the logistic function to fit the leading edge or a linear interpolation to find the timing of the waveform. 

\begin{figure}[hbt!]
 \centering 
 \includegraphics[width=0.4\textwidth]{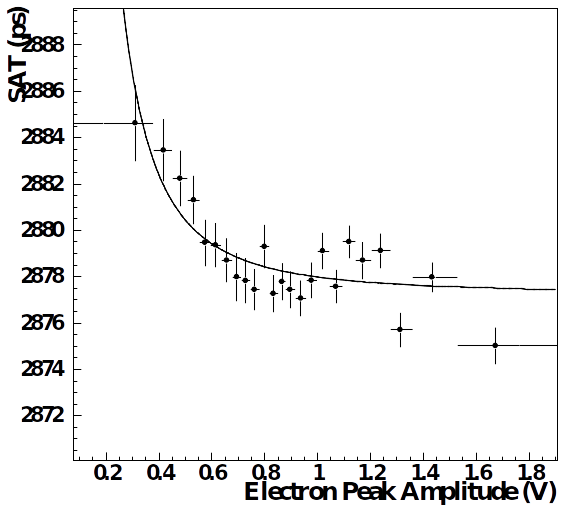}
\caption{Graphical representation of mean of Signal Arrival Time with respect to the electron peak amplitude, without any time walk corrections.}\label{fig:slewinggpeak}
\end{figure}

The same technique, as it is described in section \ref{subsection:CDF}, i.e. the standard Constant Fraction Discrimination at 20$\%$ of peak's amplitude, is being used to the PICOSEC's signal. After applying this procedure, separately to both PICOSEC and photodiode's signals, and subtracting on an event per event basis, the arrival time of the signal to the reference photodiode from the arrival time of the signal to the PICOSEC detector, we evaluate the \emph{Signal Arrival Time-SAT}). It can be seen that this distribution is asymmetric. This asymmetry is an indication of a systematic effect of the signal, the well-known to us as \emph{time walk effect}. This can be represented by the Figure \ref{fig:slewinggpeak}, expressing the dependence between the mean signal arrival time and the size of the electron peak. 
\begin{figure}[hbt!]
 \centering 
 \includegraphics[width=0.5\textwidth]{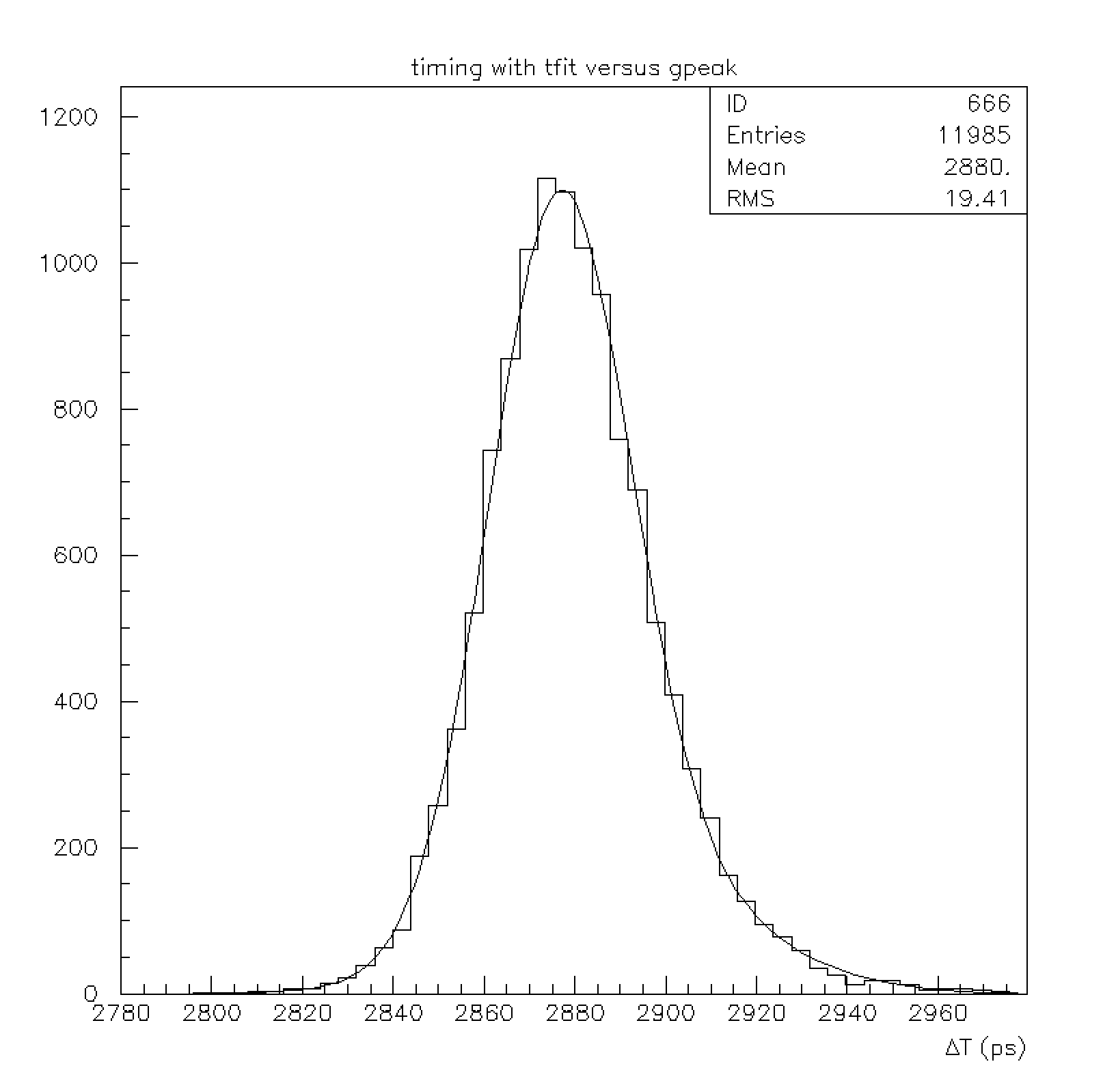}
\caption{Time difference between timing of the PICOSEC's signal and photodiode's signal, fitted with a double Gaussian Distribution.}\label{fig:resolutionwithout}
\end{figure}

To be able to quantify the dependence of the mean SAT on the electron peak size, we divide the total data-set collection in bins of electron peak amplitude, e.g. we select events in peak of charge or voltage. After that, we apply Gaussian fit to the time distribution of events contained in each bin, we keep the interesting parameters of the fit, the mean value and standard deviation of each Gaussian fit with their corresponding errors, and we store, the mean SAT and the rms, in two different data sets of the corresponding mean amplitude and mean charge of each bin. 

From evidence, the fact that all the events not having the same resolution, the corresponding distribution (in Figure \ref{fig:resolutionwithout}) will deviate from Gaussian shape, thus it will not fit accurate the experimental data. For this reason, it is used a Double Gaussian with five parameters, as a sum of two Gaussian with the same mean($\mu$), sigma($\sigma_1,\sigma_2$) and normalization constants ($c_1, c_2$) like in the Eq.\ref{eq:doubleGauss}: 
\begin{equation}\label{eq:doubleGauss}
    G(t;c_1, \mu_1, \mu_2, \sigma_1, c_2, \sigma_2) = \frac{c_1}{\sigma_1}e^{\frac{(t-\mu_1)^2}{2\sigma_1^2}} + \frac{c_2}{\sigma_2}e^{\frac{(t-\mu_2)^2}{2\sigma_2^2}}
\end{equation}
giving a combined total resolution to the double Gaussian fit :
\begin{equation}\label{eq:total_resolution}
    \sigma_{tot} = \Big[\sum_{i=1}^{N}c_i^2\sigma_i^2 + \sum_{i=1}^{N} \sum_{j=i+1}^{N}c_i c_j\Big(\sigma_i^2 + \sigma_j^2 + (\mu_i - \mu_j)^2\Big) \Big]^{1/2}
\end{equation}
which in case of the same mean takes the form: 
\begin{equation}
    \sigma_{tot} = \sum_{i=1}^N c_i \sigma_{i}^2
\end{equation}
and an error given :
\begin{equation}
    \delta_{V[\Delta T]} = \Big[\sum_{i=1}^N c_i \cdot \delta_i^2 + 4 \sum_{i=1}^N c_i^2\cdot\Big(\mu_i - \sum_{j=1}^N c_j\cdot \mu_j \Big)^2 \cdot w_i^2 \Big]^{1/2}
\end{equation}
where $c_i$ are the normalization factors of the Gaussians, $\delta_i$ is the error in mean value, and $w_i$ is the error in variance. 
In the way of using two Gaussians with different mean values we can have an simple estimation of the resolution, e.g. in Figure \ref{fig:resolutionwithout} is about 19.4$\pm$ 0.5\, ps, but this way will not give us a fully successful fit, before the time walk correction. 

Starting with the representation of SAT as a function of electron peak size (either its peak amplitude, or its charge), we fit with the following curve, defined as a power law plus a constant term: 
\begin{equation}\label{eq:powerlaw}
    g(x;a,b,w) = a + \frac{b}{x^w}
\end{equation}
where as x we can set both the quantities of the electron peak size, the electron peak amplitude or the electron peak charge, $a$ is a constant term and $a,b,w$ are the parameters of the fit. 

\begin{figure}[hbt!]
 \centering 
 \includegraphics[width=0.4\textwidth]{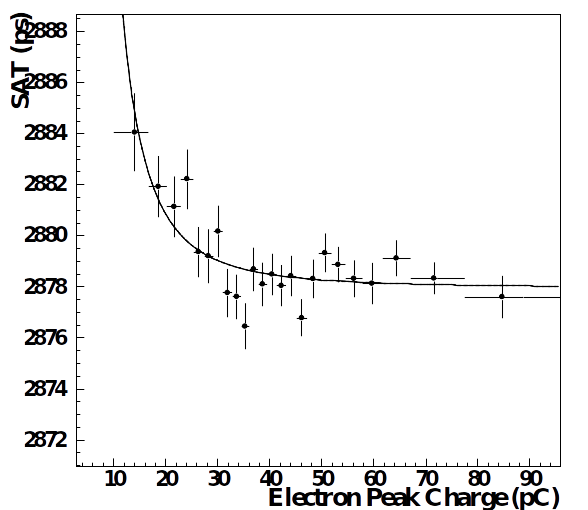}
\caption{Graphical representation of the mean of Signal Arrival Time with respect to the electron peak charge, without any time walk corrections.}\label{fig:slewingqe}
\end{figure}

As an analogous of representation of the Signal Arrival Time as a function of electron peak amplitude, in Figure \ref{fig:slewingqe} we can see the SAT as a function of electron peak charge. Where the electron peak charge is found from the integration of the electron peak region. Defining this region we need a starting and an ending point. As a starting point of the pulse we define the first point, of the leading edge, that has an amplitude greater than three times the baseline noise. As ending point we assume a local minimum of before the ion tail. By this way, we can separate our experimental data into bins of charge distribution, creating the relevant \say{groups} as before.    

Our study now is oriented to correct our data points that suffer from the time-walk effect. This has to be applied on an event-by-event basis, where the points of each Figure \ref{fig:slewinggpeak}, and \ref{fig:slewingqe},  are fitted with the power law of the Eq.\ref{eq:powerlaw}, and then subtracted with the mean SAT, through the expression :
\begin{equation}\label{eq:slewing_corre}
    \text{Corrected SAT} = \text{SAT} - \frac{a}{(\text{Pulse Amplitude})^b} +c
\end{equation}

\begin{figure}[hbt!]
 \centering 
 \includegraphics[width=0.5\textwidth]{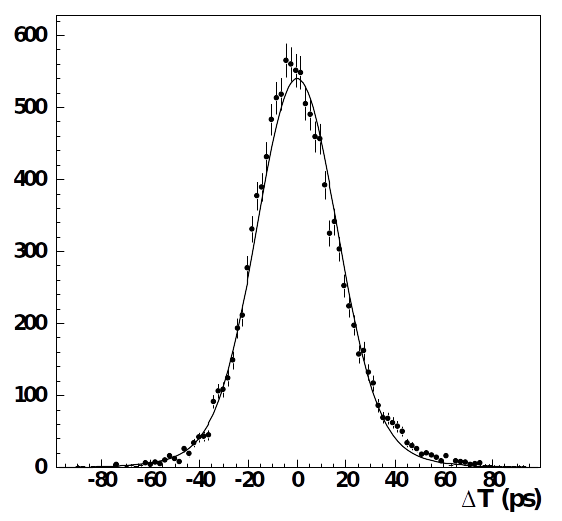}
\caption{Signal Arrival Time after time walk corrections, on an event-by-event basis.}\label{fig:datacorr}
\end{figure}

\begin{figure}[hbt!]
 \centering 
 \includegraphics[width=0.5\textwidth]{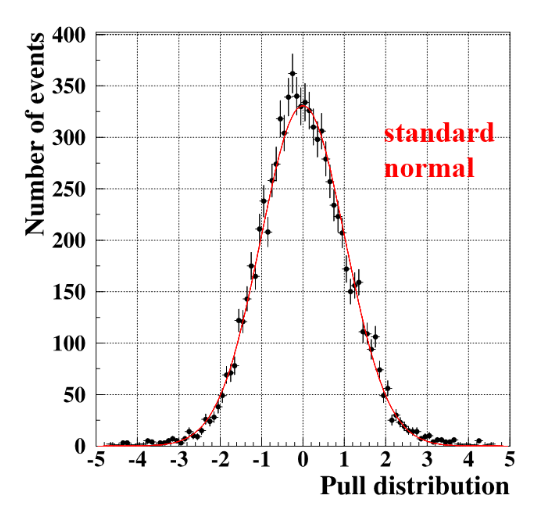}
\caption{The Pull Distribution with the respective Gaussian fit, resulting to a mean value of 0.0006 and a standard deviation of 0.996, as a support that our SAT results are accurate.}\label{fig:pull}
\end{figure}

The result of this procedure is a corrected SAT distribution with a mean value around zero, as it can clearly be seen at the Figure \ref{fig:datacorr}. Notice, that after time-walk corrections the distribution of signal arrival time is almost symmetric. If the Pull Distribution been checked, as in Figure \ref{fig:pull}, the fact that it is in a good approximation with a normal Gaussian, gives a strong support of the evidence of the RMS dependence on electron peak charge. This distribution help us consider our offline analysis, as accurate as the standard deviation of the Pull distribution, where this distribution can be created after the SAT distribution where from each data point we subtract the mean value of SAT distribution and divide by each $\sigma$ : $Pull = \frac{SAT - \langle SAT \rangle}{\sigma}$.   

After the time walk correction, the error of the calculated timing resolution uncertainty, which is reflected in the variance and is given by :  
\begin{equation}\label{eq:total_error}
\delta_{V(\Delta T)} = \Bigg[\alpha_1\cdot \delta_1^2 + \alpha_2 \cdot \delta_2^2 + 4 \alpha_1^2\cdot\Bigg(\mu_1 - \alpha_1\cdot \mu_1 - \alpha_2\cdot \mu_2\Bigg)^2 \cdot w_1^2 + 4 \alpha_2^2\cdot\Bigg(\mu_2 - \alpha_1\cdot \mu_1 - \alpha_2\cdot \mu_2\Bigg)^2 \cdot w_2^2\Bigg]^{1/2} 
\end{equation}
where $\delta_i$ is the error of the mean value and $w_i$ is the error of the variance, giving in what precision we can predict the time walk effect and the timing resolution. This procedure gives a timing resolution of 18.3 $\pm$ 0.2 \,ps for the PICOSEC Micromegas Detector, operating on multi-photoelectron Laser Test Beam (RUN 291), with 275V anode and 525V drift voltage.   

\begin{figure}[hbt!]
 \centering 
 \includegraphics[width=1.0\textwidth]{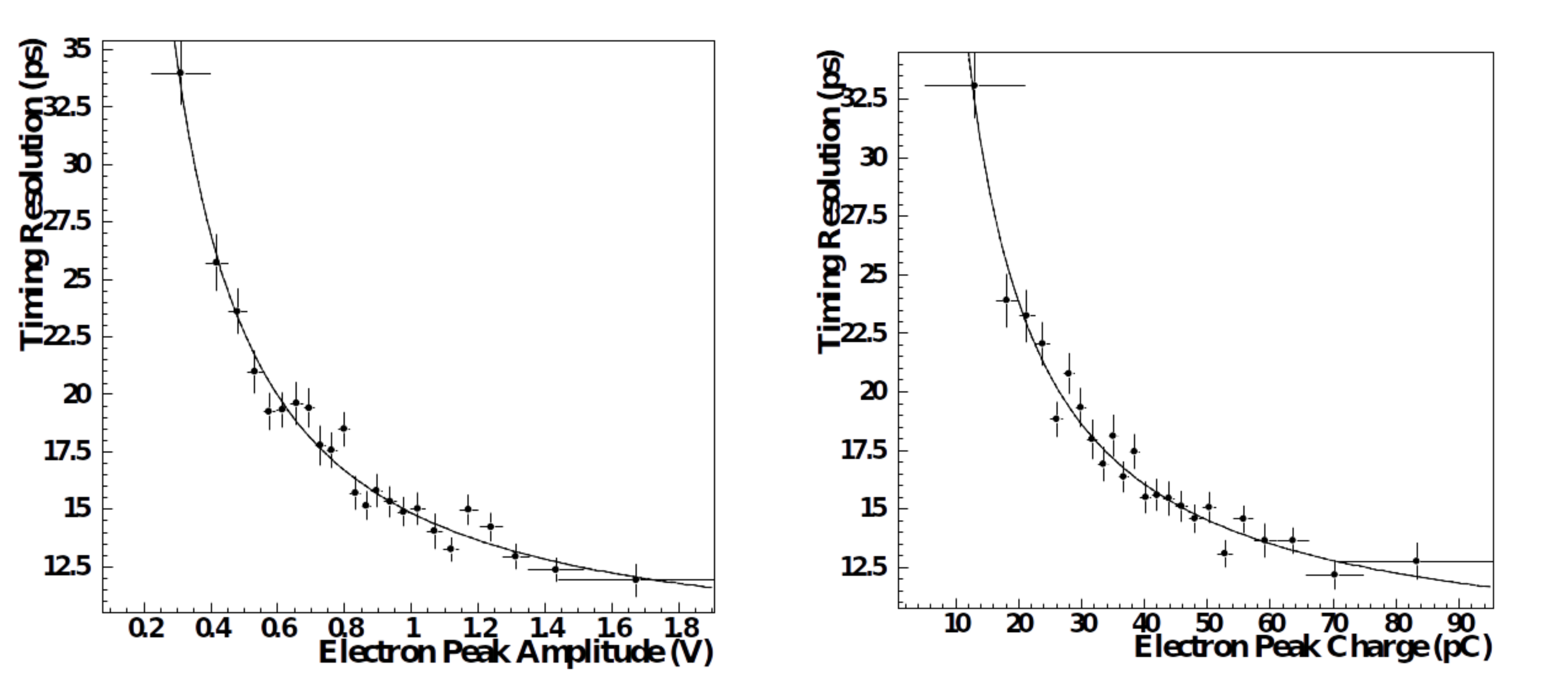}
\caption{Graphical representation of Resolution of Signal Arrival Time with respect to the electron peak amplitude(left), and electron peak charge(right), after time walk corrections.}\label{fig:datacorre}
\end{figure}

In conclusion, as it can be observed in Figure \ref{fig:datacorre}, the higher the pulse amplitude the earlier its signal arrival. This behavior reveals, that the time-walk effect is affecting the pulse's arrival time, causing a shift of the whole electron peak, for reasons that are not related to the experimental technique or procedure of data taken, but it has its source in pulse formation dynamics.  For this reason, this behavior was studied, and it was found that there is a  microscopic parameter that explains the observed SAT. This is the average time of the fraction of the preamplification electrons that passing the mesh, giving the same mean and RMS values as SAT. Its statistical properties are influenced by the passing time of the photoelectron's first ionization point and the avalanche's drift time. Both these times appear a linear dependence on the avalanche length, and their sum, which corresponds to the movement in the drift region before passing the mesh, also has a linear dependence on the avalanche length. The corresponding transmission time through the mesh has the same behavior, which differs only by a constant number. That means that the photoelectron and the avalanche are moving with different drift velocities, while the rms of the transmission time of the photoelectron increases with the length of the drift path, the respective time of the avalanche is saturated. Thus, to achieve better timing resolution, we need to start the ionization as soon as possible. More information about this work can be found in \cite{BORTFELDT2021165049,Paraschou:2020max}. Even though we have reduced the drift gap of our detector to reduce this phenomenon it still exists and gives a power law dependence to the SAT with peak size. For now on this analysis with CFD technique will be our reference, in order to cross check our new techniques that are going to be proposed.

\section{A new Constant Threshold Discrimination Timing Technique}

In our effort on exploring alternative timing techniques, using the existing electronic devices to reduce the cost of experimental set-up, we studied the possibility of using a constant threshold for timing the signals. Although, it is well-known that such a timing suffers from time walk, we prove that there is experimental information that can be exploited to correct for this effect, achieving very high resolution. We improved the idea of the multi-Threshold timing \cite{article}, using instead of time over threshold measurements, the integrals of the parts of the electron peak which exceed certain thresholds. Hereafter, such integrals will be referred as \emph{Qup}. 

\begin{figure}[hbt!]
 \centering 
 \includegraphics[width=0.7\textwidth]{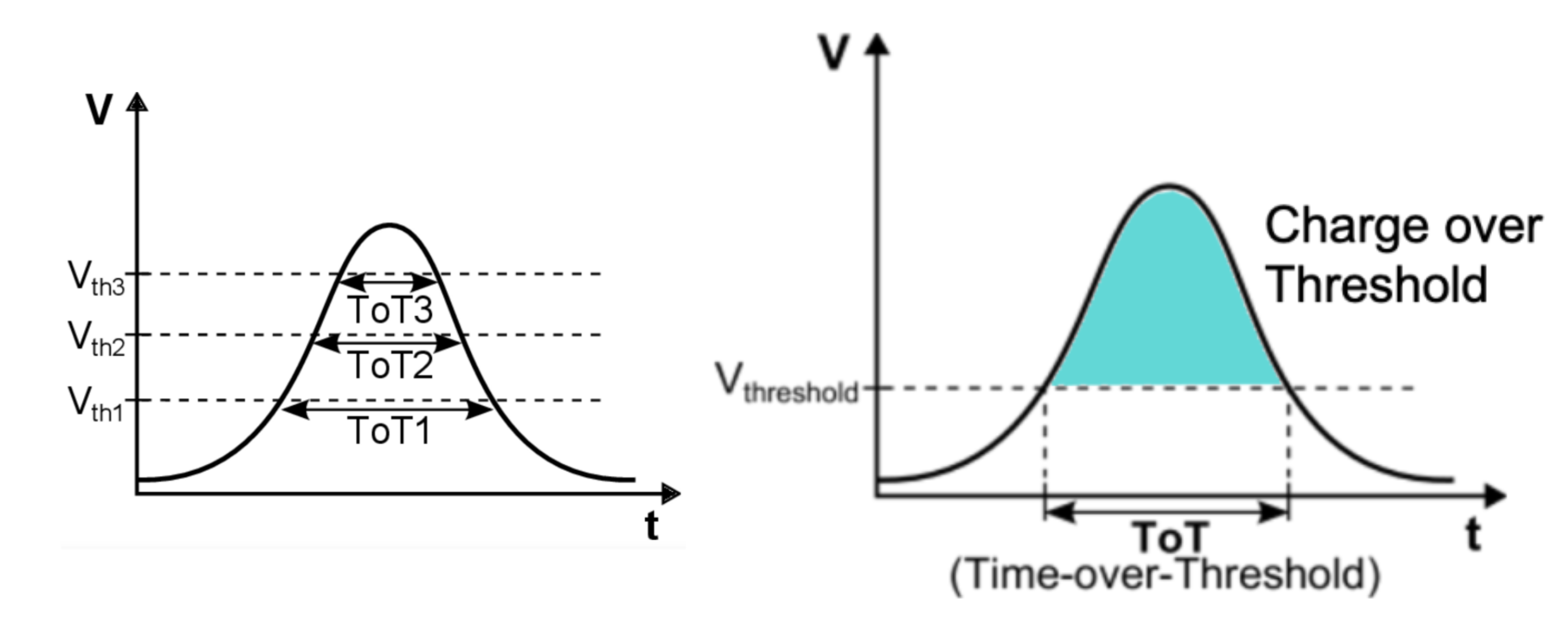}
\caption{Graphical demonstration of multi-thresholds crossing a waveform (on the left) and the definition of the charge above relevant threshold(on the right).}\label{fig:multiToT}
\end{figure}

Specifically, as presented on the Figure \ref{fig:multiToT}, we define four amplitude thresholds 100\,mV, 200\,mV, 400\,mV and 600\,mV. Those will cross the waveform at two points, on both sides of the electron peak (one on the leading edge and the other on the falling edge), and the charge above those thresholds will give an estimation of the electron peak amplitude, alternative of using the time over threshold measurements. 

\begin{figure}[hbt!]
 \centering 
 \includegraphics[width=0.5\textwidth]{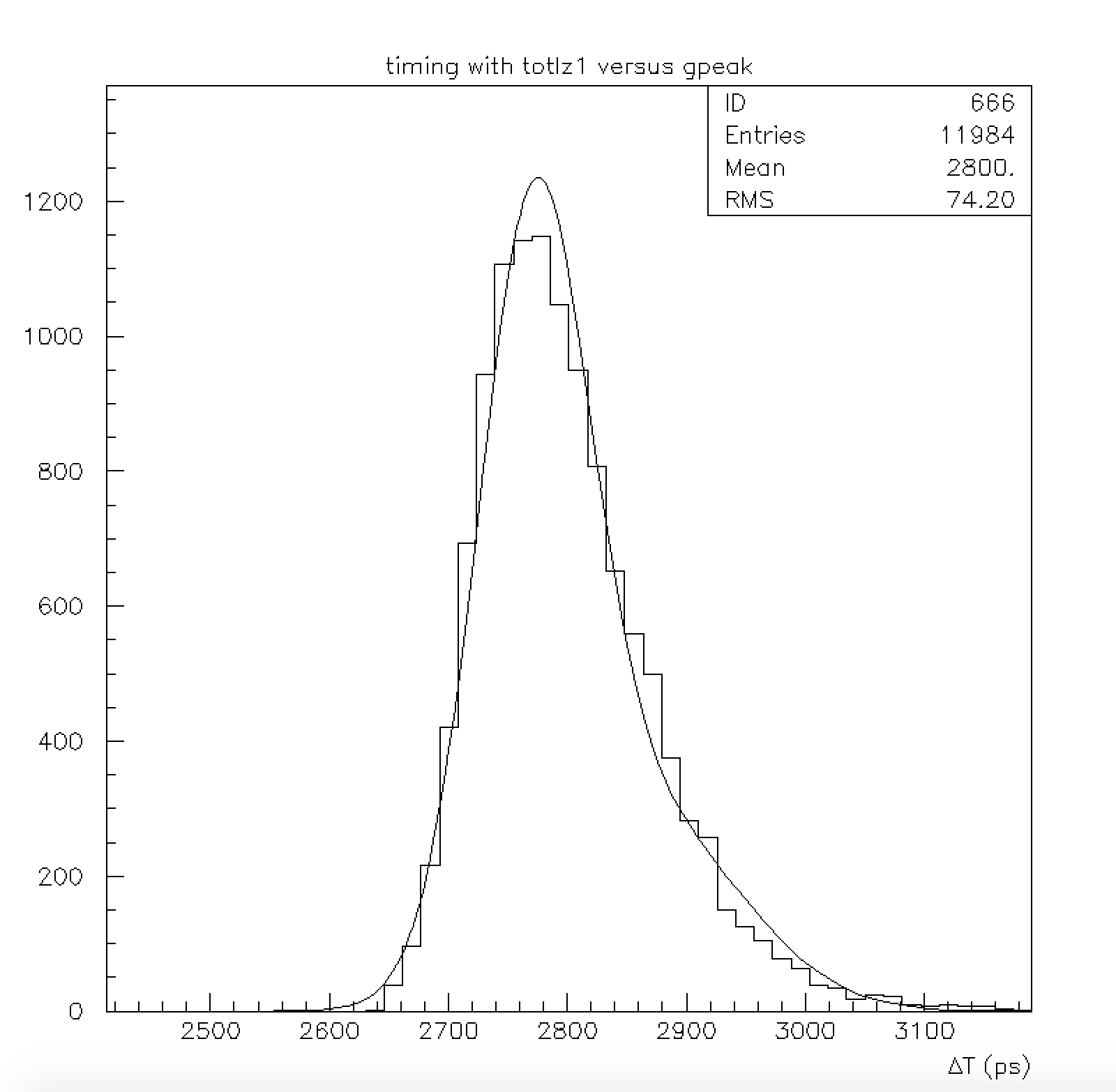}
\caption{Signal Arrival Time distribution, using constant threshold at 100\,mV, before time walk corrections. Notice that the Gaussian fit on the data, results to a timing resolution of the order 74.2$\pm$0.6\,ps. }\label{fig:consThreshold}
\end{figure}

In the following we define as Signal Arrival Time the time that the leading edge crosses the lowest applied threshold(e.g. 100\,mV), with respect to the photodiode's reference time. As it is shown in \ref{fig:consThreshold}, such a timing is influenced by time walk effects and the Signal Arrival Time distribution, is asymmetric, with an RMS of 74.2$\pm$0.6\,ps, corresponding to a bad timing resolution. The idea is to correct this timing information for the systematic errors (time walk) in such a way that the corrected timing will reach the same accuracy as the CFD technique. In the case that the electron peak amplitude is well measured, we could use this information to correct for time walk. 

\begin{figure}[hbt!]
 \centering 
 \includegraphics[width=1.0\textwidth]{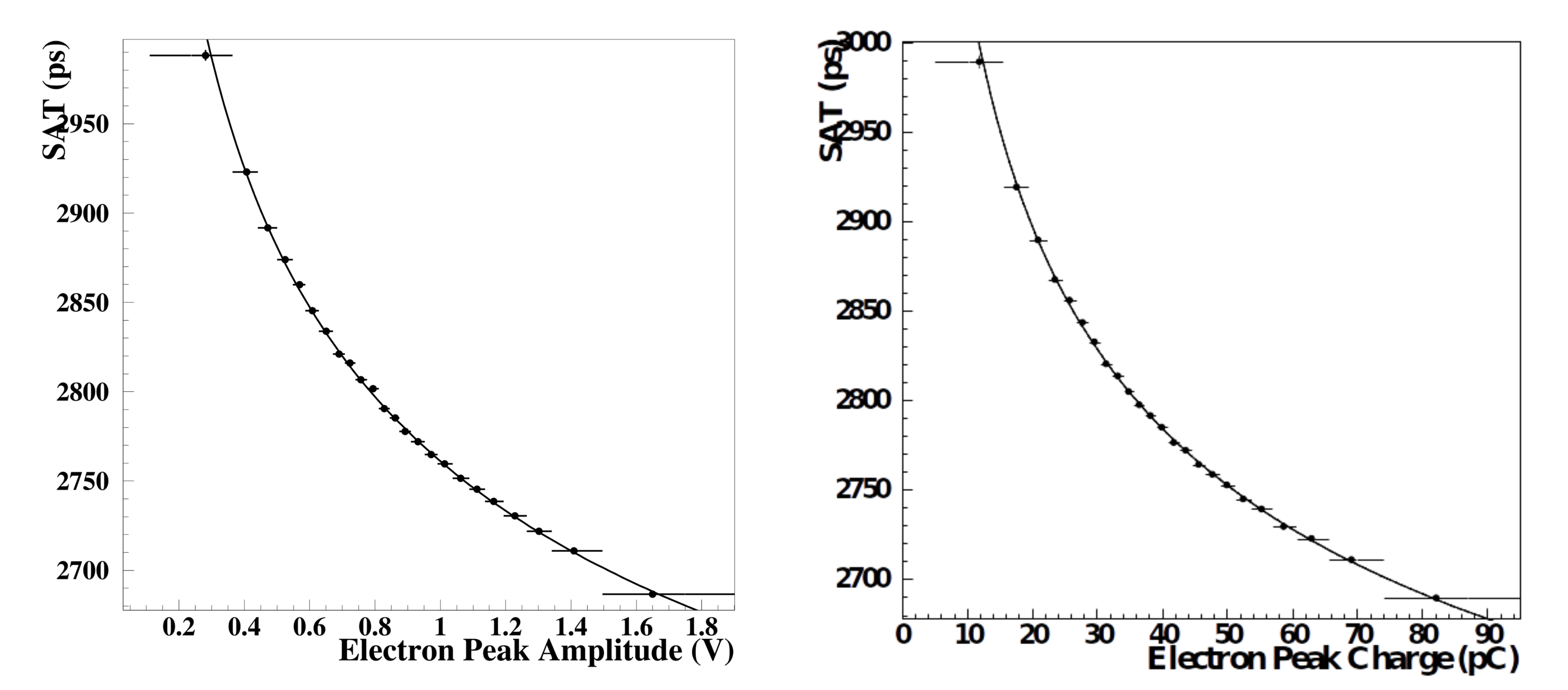}
\caption{Graphical representation of the mean Signal Arrival Time with respect to the electron peak amplitude(left), and electron peak charge(right), before time walk correction at 100mV .}\label{fig:slewingToT}
\end{figure}

Indeed, by dividing the data-set in bins of peak amplitude (or the charge of the electron peak), as in the CDF analysis, and estimated the mean value of SAT, as a function of the peak amplitude, we can evaluate correction curves as shown in Figure \ref{fig:slewingToT}. 

\begin{figure}[hbt!]
 \centering 
 \includegraphics[width=0.5\textwidth]{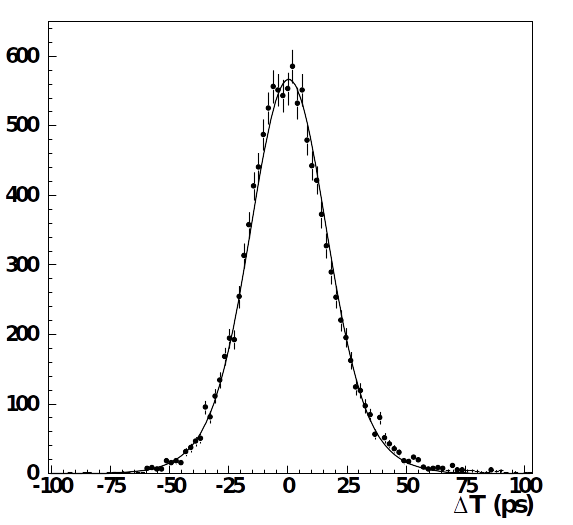}
\caption{Signal Arrival Time after time walk correction, using one Constant threshold timing technique.}\label{fig:pull_corre_ToT}
\end{figure}

We parameterize the dependence of the mean SAT on the size of the electron peak (peak amplitude or electron peak charge), as a power law plus a constant term, according to the Eq. \ref{eq:slewing_corre}, and we use this to correct the systematical shift due to the time walk effect. After correction, the SAT distribution  can be seen in the Figure \ref{fig:pull_corre_ToT}, where it is apparent that the corrected timing results to a symmetric distribution with an RMS (resolution of the detector) of 18$\pm$ 0.3 \,ps, which is the same resolution as the one reached with the Constant Fraction Discrimination Technique.   

\begin{figure}[hbt!]
 \centering 
 \includegraphics[width=0.5\textwidth]{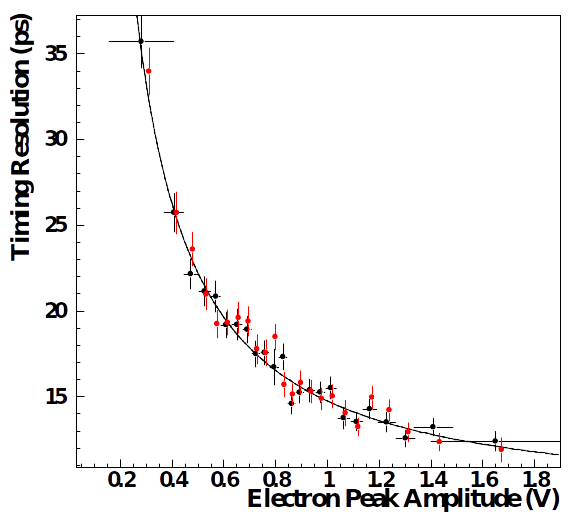}
\caption{Resolution as a function of electron peak amplitude after global time walk correction, using both Constant Fraction Discrimination (red points) and  multi-ToT timing techniques (black points).}\label{fig:resol_corre_ToT}
\end{figure}

As a further demonstration, we compare here the resolution as a function of electron peak amplitude, as achieved by the two methods, the Constant Fraction Discrimination and the Constant Threshold at 100\,mV.  From Figure \ref{fig:resol_corre_ToT}, we can see that both techniques resulting to the same behavior of the timing resolution as a function of electron peak amplitude. 

In conclusion, we have proved that by using a Constant Threshold to time the signals, can result to precise timing, when the measurement is corrected for time walk effects, using the peak amplitude as the correction parameter.

However, the determination of the peak amplitude (e.g. highest point of the electron peak) is very difficult to be available by the hardware. Usually, the time over threshold (ToT) is used as an indirect measure of the pulse size. There are existing electronic devices like NINO and NINO-2 chips \cite{NINO}, that can give us the ToT information, but it has been studied explicity that ToT technique results to less precise timing than the CFD technique. However, at a low threshold level the electron peak of the PICOSEC-MicroMegas waveform, is not always separated by the ion peak generating thus systematic uncertainties. We have introduced the Qup, as an alternative way estimate of the electron peak size. In the following we will demonstrate that by using multiple thresholds and calculate the charge above each threshold, we can achieve very accurate timing measurements. 

\subsubsection{Evaluation of Data with multi-Charge Threshold Technique} \label{subsubsectionqup}
This timing technique is an alternative method of Constant Threshold. In this case, we will use the multiple amplitude thresholds, defined above, for their two crossing points on the waveform (on the leading and the falling edge) to integrate this region to extract the information of the charge of each waveform, called as \emph{Charge over Threshold}, seeing in right image of the Figure \ref{fig:multiToT}.

To produce the information of electron peak amplitude, we need readout electronic devices to digitize the waveform with specific time intervals. On a large-scale device, this will imply the need for a big amount of different channels and as a result, increases the cost of detector and experiment development. On the other hand, if a pulse integrator can be applied (like the NINO-chip \cite{NINO}, or other new generation frond end electronics providing the pulse integral), and if the timing technique can achieve the resolution of the Constant Fraction Discrimination Technique, then the cost will remain as low as possible. 

\begin{figure}[hbt!]
 \centering 
 \includegraphics[width=1.0\textwidth]{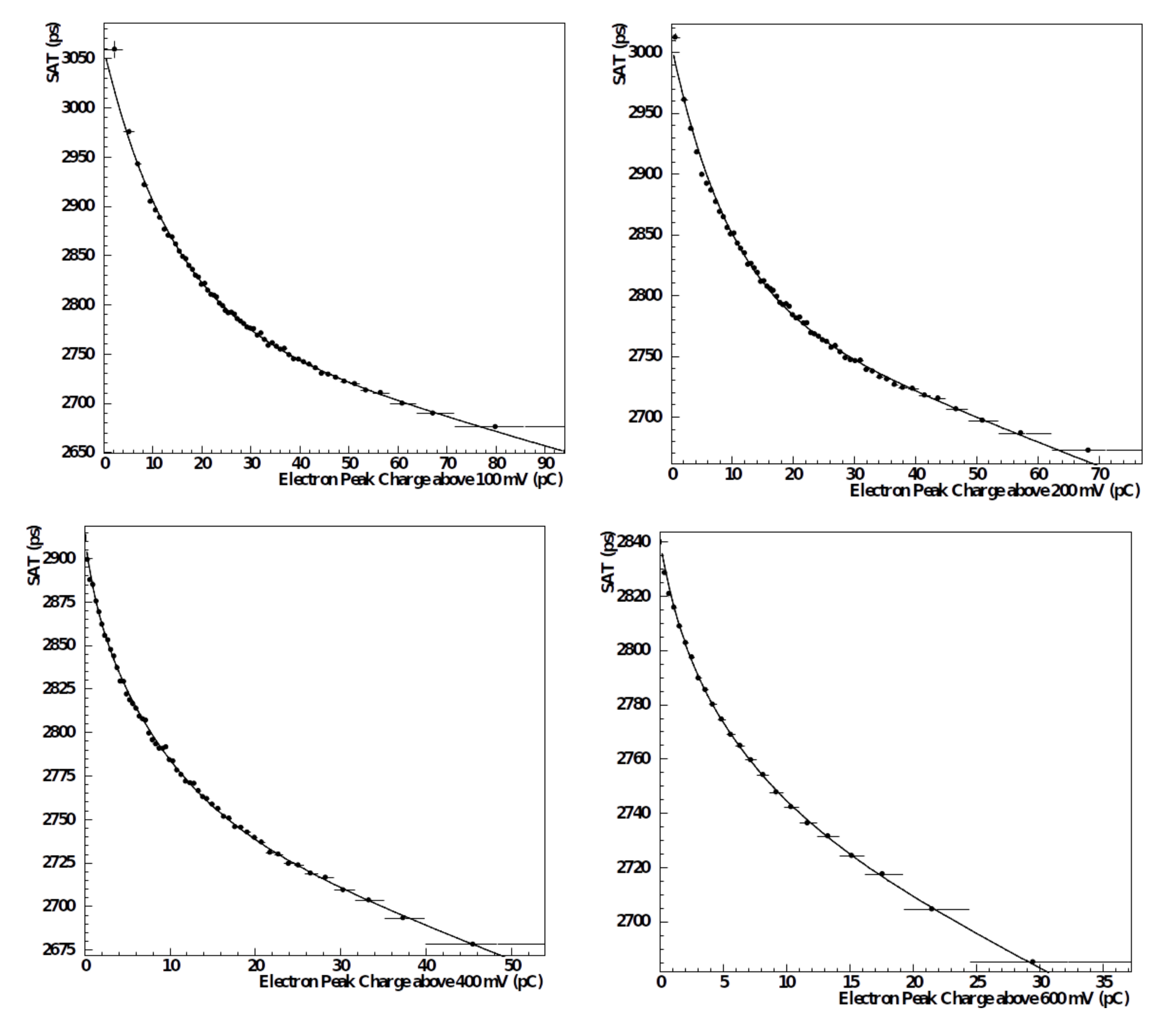}
\caption{Signal Arrival Time as a function of Charge above Threshold for the four different thresholds of the analysis, fitted with the corresponding power law curve.}\label{fig:qup_slewing}
\end{figure}

The procedure demands again the division of the data-set in bins of electron peak size, fitting each bin with the corresponding Gaussian and storing the mean value of Signal Arrival Time once crossing the lowest threshold and once crossing the highest available threshold, in analogous to the electron peak amplitude, defined by its charge. In this way, we can create four different Signal Arrival Time distributions as a function of electron peak charge, each one for every threshold. Thus, we can fit them with the corresponding power law curve, keeping the fitting parameters, as we also did in the Constant Threshold technique. The behavior of Signal Arrival Time, as a function of the electron peak charge above multiple thresholds can be seen in Figure \ref{fig:qup_slewing}. From this Figure we keep the parameters of each fit in order to correct the Signal Arrival Time of each pulse on each highest threshold. This Figure once again, reveals the time walk effects which are present in this timing technique.

\begin{figure}[hbt!]
 \centering 
 \includegraphics[width=0.75\textwidth]{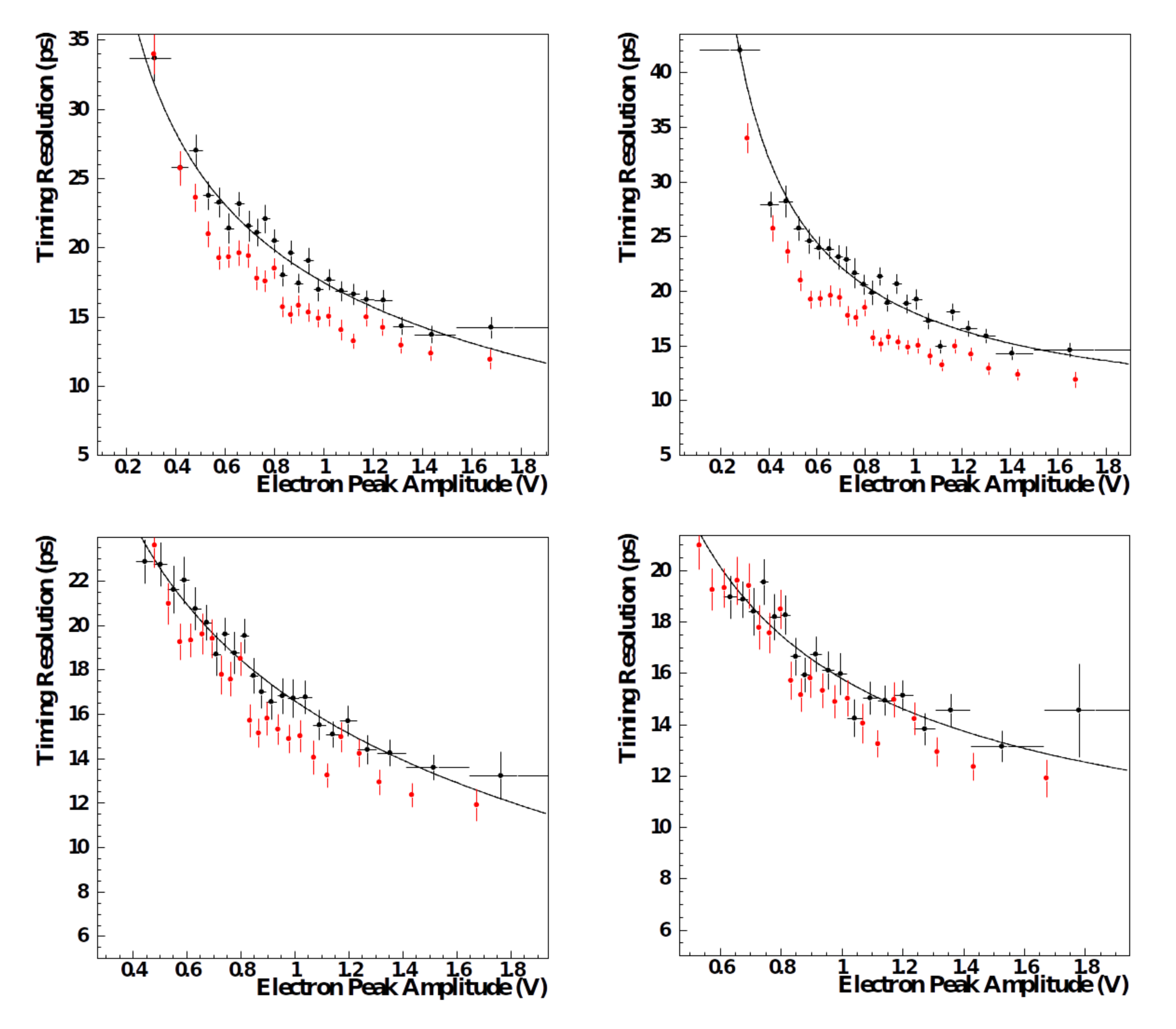}
\caption{Resolution as a function of electron peak amplitude after time walk correction, using both Constant Fraction Discrimination (red points) and  multi-Charge over threshold timing techniques (black points), for the four different amplitude thresholds.}\label{fig:corre_qup}
\end{figure}

Correcting the timing resolution, due to the different Signal Arrival Time, parameterised with charge above threshold for the four different applied thresholds, results to the black points of the Figure \ref{fig:corre_qup}. The idea is to use every threshold (100\,mV, 200\,mV, 400\,mV, 600\,mV) as both for timing and the correction threshold in order to conclude for the final choice of using the smaller threshold for timing while using the higher for the Qup corrections. The result on each peak amplitude region in comparison of the CDF Technique and the multi-Charge over threshold, can be seen in this Figure. 

The distribution of time walk effect in the detector, will be corrected, globally, in every amplitude range set by the threshold choice, simultaneously, using the corrections of Eq.\ref{eq:slewing_corre}. For every waveform, using the Signal Arrival Time at the lowest threshold, we will correct for time walk effects according to the fitted power law curve of the highest threshold crossed by this one.

\begin{figure}[hbt!]
 \centering 
 \includegraphics[width=0.5\textwidth]{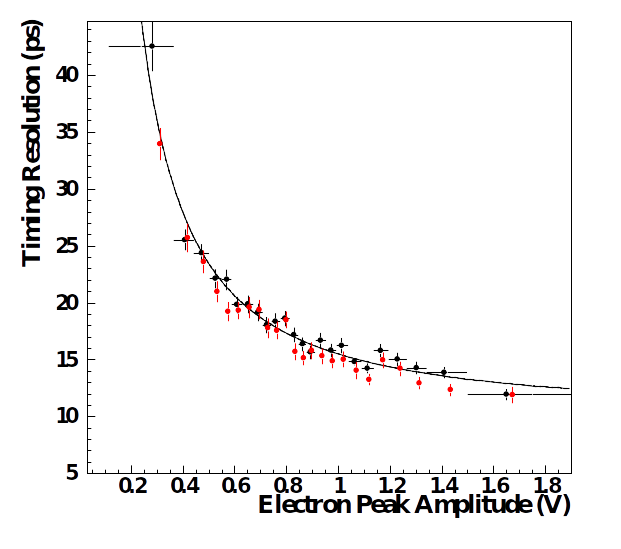}
\caption{Resolution as a function of electron peak amplitude after time walk correction, using both Constant Fraction Discrimination (red points) and  multi-CoT timing techniques (black points).}\label{fig:corre_global}
\end{figure}

Summing up all the information of SAT corrections in the different peak amplitude bins, we can conclude to the Figure \ref{fig:corre_global}, representing the timing resolution behavior as a function of electron peak amplitude for the Constant Fraction Discrimination and the multi-Charge over Threshold method. As it can be seen, both methods achieve similar timing resolution behavior. 

\chapter{Timing with Artificial Neural Networks\label{cha:simulation}}
\begin{figure}[hbt!]
 \centering 
 \includegraphics[width=0.5\textwidth]{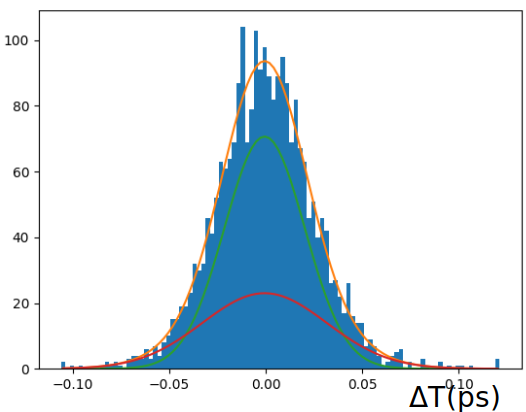}
\caption{First results of our Artificial Neural Network, resulted on a resolution of the order 24.1$\pm$0.6\,ps. First indication that our evaluation is consistent with the full offline analysis, reported on \cite{Bortfeldt2019}.}\label{fig:nn_24ps}
\end{figure}

\vspace{-0.8cm}
\noindent\fbox{%
    \parbox{\textwidth}{On our way to New Physics discoveries, our experiments demand increasingly accurate measurements, meaning that our well-known technologies need to be modified. Aiming to precise timing measurements, our research concludes to Artificial Intelligence as a strong candidate. Firstly in this chapter, a simulation model is being represented, as a prerequisite for developing an Artificial Neural Network timing technique. Basic information and a brief overview of ANN performance is described in Appendix \ref{appendix:A}.
    }%
}
\paragraph{}In order to achieve results of timing resolution of gaseous detectors online, or just to minimize the information that needs to be saved during data acquisition, the development of an Artificial Neural Network(ANN) needs to be implemented. As a first step, to have an indication if our objective can be achieved, we tried to reproduce the experimental results of timing resolution, using data of the Muon Test Beam taken in 2017, mentioned in \cite{Bortfeldt2019}. Because of the limited amount of experimental data, the best choice to train and test the Neural Network was to use the k-fold cross validation technique, (the basic information about ANN is given in Appendix \ref{appendix:A}). In this method, the data-set is divided into ten folds, using nine for training and one for testing (validation) sets respectively. This procedure was performed cyclically, resulting in ten different models, and by extension, ten different prediction samples, which were combined, to test the timing resolution of the algorithm. This technique resulted in a resolution of the order of 24.1$\pm$0.6\,ps, shown in Figure \ref{fig:nn_24ps}. 

Reproducing the same results, as with the full offline timing analysis, but with our Neural Network, it is considered to be a good indication of the development of a novel timing technique. For a complete test without folding technique, we need the means to produce large samples of training sets, to increase our statistics. For this purpose, a simulation model was developed, which uses fully digitized single photoelectron pulses, acculturated during spacial laser calibration runs, to generate more multi-photoelectron ones, in order for our Network to be trained properly. 

\section{The Simulation Model}
The ANN requires a great amount of data to be trained on, before testing unknown (to the ANN) data, in order to reduce biases. During the UV Laser Test that represented in section \ref{section:LaserTest}, two sets of data(e.g. waveforms) were collected, one with single photoelectrons exiting the photocathode and one with multi-photoelectrons entering the active volume of the PICOSEC-Micromegas detector. The multi-photoelectron data-set consisted of  12000 events only, a number that was considered really low for both the training and the testing process of the ANN. This is why, there was a rising need for more data, to be used for training. But it is obvious that, by advancing a simulation model we offer a general method of producing training sets to be used.  
\subsection{Creation of multi-photoelectron pulses}
The first data set, hereafter called \emph{SPE-set}, contains waveforms from single photoelectrons, hereafter called \emph{single-pes}, in this run 8000 events of single-pes were collected. These measurements were achieved using multiple attenuation filters and observing the oscilloscope's monitor, with the additional information of electron peak amplitude distribution. As we keep adding filters we observe the distribution on the oscilloscope with a reducing mean value. When we reach the baseline of one photoelectron exiting the photocathode of our detector, the mean value of the distribution will remain constant, while its maximum amplitude will be reduced. 

The target of the first steps of this simulation model is to create multi-photoelectron pulses given only the single photoelectron pulses. During the same run, another event collection was made, hereafter called EXP-set, which contains multi-photoelectron pulses. Our simulated pulses should be similar to the pulses of the EXP-set of data and should give the same timing resolution, those data were analyzed earlier with multiple timing techniques on the Chapter \ref{cha:implementation}. 

\begin{figure}[hbt!]
 \centering 
 \includegraphics[width=0.55\textwidth]{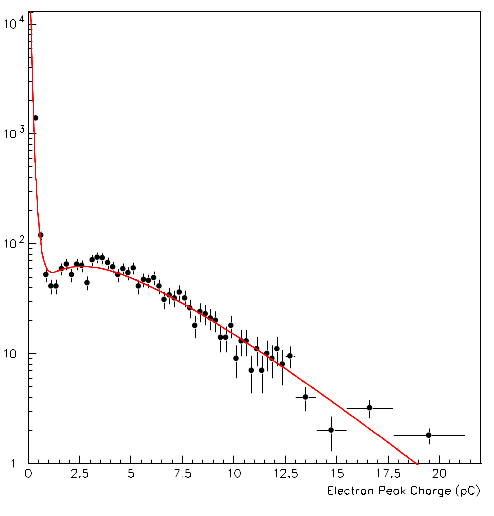}
\caption{Single Photoelectron Distribution fitted with Polya .}\label{fig:singlepe}
\end{figure}

Starting with the development of multi-photoelectron pulses, using the single ones, it is convenient to mention the statistical properties that affect our study. In Figure \ref{fig:singlepe} the distribution of charge of the SPE-set can be seen, where the red line corresponds to the fitting curve, which contains an exponential fit for the falling edge of the peak, which corresponds to the noise, and the well-known Polya distribution for the rest of the data, Eq. \ref{eq:N-polya}.

Assuming that single photoelectrons charge distribution is given by a probability density function as:
\begin{equation}\label{eq:singlecharge}
    \mathcal{P}_{e}(Q;\Bar{Q_e},V_e) = \begin{cases} \text{a positive function}, &Q\geq0 \\
  0, & Q<0    
\end{cases}
\end{equation}
where Q is the accumulated charge with mean value and variance : 
\begin{align}
    \Bar{Q}_e &= \int_{-\infty}^{\infty}Q \mathcal{P}_e(Q;\Bar{Q}_e,V_e)dQ \\
    V_e &= <Q^2>-\Bar{Q^2_e} = \int_{-\infty}^{\infty}(Q-\Bar{Q_e}^2) \mathcal{P}_e(Q;\Bar{Q}_e,V_e)dQ 
\end{align}

The probability density function of the charge of n photoelectrons will be a convolution of the charge distribution of Eq. \ref{eq:singlecharge}, for n times:
\begin{align}\label{eq:multipedistrib}
\mathcal{P}_{ne}(Q_n;\Bar{Q}_e,V_e) &= \underbrace{\mathcal{P}_{e}(Q;\Bar{Q}_e,V_e) \otimes  \mathcal{P}_{e}(Q;\Bar{Q}_e,V_e) \dots \otimes \mathcal{P}_{e}(Q;\Bar{Q}_e,V_e)}_{n\textup{-times}} \\
\int_{-\infty}^{\infty} \mathcal{P}_{ne}(Q_n;\Bar{Q}_e,V_e)dQ_n &= 1
\end{align}

Where the expected value and variance of $Q_n$ can be calculated from Eq.\ref{eq:multipedistrib}, but it is easier to consider charge as a sum of multiple charges (with variable name $x_i$) created by equal number of photons ($Q_n = x_1+x_2+\dots+x_n$) and distributed according to Eq.\ref{eq:singlecharge}. Then:
\begin{equation}\label{eq:mean}
    \Bar{Q}_{ne} = \int_{-\infty}^{\infty}Q_n\mathcal{P}_{ne}(Q_n;\Bar{Q}_e,V_e)dQ_n = n\cdot \Bar{Q}_e
\end{equation}
\begin{equation}\label{eq:variance}
V_{ne} = \int_{-\infty}^{\infty}(Q_n - \Bar{Q}_{ne})^2 \mathcal{P}_{ne}(Q_n;\Bar{Q}_e,V_e)dQ_n = n\cdot V_e 
\end{equation}

Assuming also a noise contribution as a function of charge for zero photoelectrons with mean zero and variance $s^2$ as $\mathcal{P}_{0e}(Q;\Bar{Q}_e,V_e) = f(Q)$, and a probability density function of the form $g(n;\mu,\vec{M})$, which could be Poissonian, for the n photoelectrons to have: 
\begin{equation}\label{eq:nmean}
    \mu = <n> = \sum_{n=0}^{\infty}n\cdot g(n;\mu,\vec{M})
\end{equation}
\begin{equation}\label{nvariance}
V_g = <n^2> - <n>^2 = \sum_{n=0}^{\infty}(\mu - n)^2\cdot g(n;\mu,\vec{M})
\end{equation}

Hence, the total accumulated charge, e.g. an amount proportional to the integral of the corresponding electron peak in the waveform, is distributed according to :
\begin{equation}
\mathcal{G}(Q;\mu,\Bar{Q}_e,V_e,M) = \sum_{n=0}^{\infty} g(n;\mu,\vec{M})\cdot\mathcal{P}_{ne}(Q;\Bar{Q}_e,V_e)  
\end{equation}
\begin{equation}
    \int_{-\infty}^{\infty} \mathcal{G}(Q;\mu, \Bar{Q}_e,V_e,\vec{M})dQ = 1 
\end{equation}
Thus, the total charge will have an expectation value given by:
\begin{equation}\label{eq:meantotalcharge}
\begin{split}
E[Q] &= \int_{-\infty}^{\infty}Q\cdot \mathcal{G}(Q;\mu, \Bar{Q}_e,V_e,\vec{M})dQ = \sum_{n=0}^{\infty}g(n;\mu,\vec{M})\cdot \int_{-\infty}^{\infty}Q\cdot\mathcal{P}_{ne}(Q_n;\Bar{Q}_e,V_e)dQ \\&= \sum_{n=0}^{\infty}g(n;\mu,\vec{M})\cdot n\cdot \Bar{Q}_e \\&= \mu\cdot \Bar{Q}_e
\end{split}
\end{equation}
And the variance of this will be: 
\begin{equation}\label{eq:vartotalcharge}
\begin{split}
V[Q] &= \int_{-\infty}^{\infty} (Q - \mu\cdot \Bar{Q}_e)^2 \cdot \mathcal{G}(Q;\mu, \Bar{Q}_e,V_e,\vec{M})dQ \\&= \sum_{n=0}^{\infty} g(n;\mu,\vec{M})\cdot \int_{-\infty}^{\infty}(Q - \mu\cdot \Bar{Q}_e)^2 \cdot \mathcal{P}_{ne}(Q_n;\Bar{Q}_e,V_e)dQ \\&= \sum_{n=0}^{\infty} g(n;\mu,\vec{M})\cdot \int_{-\infty}^{\infty}\{ Q^2 + (\mu\cdot \Bar{Q}_e )^2 - 2Q\mu \cdot Q\}\cdot  \mathcal{P}_{ne}(Q_n;\Bar{Q}_e,V_e)dQ \\&= \sum_{n=0}^{\infty} g(n;\mu,\vec{M})\{nV_e +n^2 \Bar{Q}_e^2 +(\mu\cdot \Bar{Q}_e)^2 - 2\mu\cdot n \cdot \Bar{Q}_e^2\} \\&= \sum_{n=0}^{\infty}g(n;\mu,\vec{M})\{nV_e + \Bar{Q}_e^2(\mu - n)^2\} \\&= \mu V_e + \Bar{Q}_e^2 V_g 
\end{split}
\end{equation}

The Eq. \ref{eq:vartotalcharge} describes the variance of the charge of a multi photoelectron electron waveform, which is equal to the mean number of photoelectrons times the variance of the single photoelectron charge distribution plus the product of the square of the single photoelectron mean charge with the variance of the number of photoelectrons.  Assuming that the number of photoelectrons follows a Poisson Distribution then $V_g = \mu$ and then the Eq.\ref{eq:vartotalcharge} becomes: 
\begin{equation}\label{eq:vartotalchargenew}
    V[Q] = \mu V_e + \Bar{Q}_e^2 V_g = \mu V_e + \mu \Bar{Q}_e^2 = \mu(V_e + \Bar{Q}_e^2) 
\end{equation}

That allows us to estimate the mean number of photoelectrons corresponding to an experimental spectrum of multi-photoelectrons. Using the technique of the Eq.\ref{eq:vartotalchargenew}, we only need the information of two moments (mean value and variance) of the charge distribution of single and multi-photoelectron distributions, to give us an indication of the mean number of photoelectrons in the laser beam producing the whole data-set. Subsequently, a more naive approach, as an quick check of the mean number of photoelectrons, is to divide the mean value of the charge distribution of the multi-photoelectrons with respect to the mean value of the charge distribution of the single photoelectrons, or the relevant RMS values. These calculations should be consistent to those resulted from the Eq. \ref{eq:vartotalchargenew}.

However, there is a more accurate technique, the one that uses the whole distribution that describes our data. Knowing the information of the single photoelectron distribution, we can simulate pulses corresponding to the PICOSEC response to many photoelectrons. The deconvolution algorithm, described in \cite{Manthos2020}, is used as a way of estimating the number of photoelectrons producing the experimental data-set waveforms. This algorithm uses the electron peak charge distribution of the two sets, the single photoelectron, and the multi-photoelectron, to estimate the number of photoelectrons that created the EXP-set waveforms. 

\begin{figure}[hbt!]
 \centering 
 \includegraphics[width=0.5\textwidth]{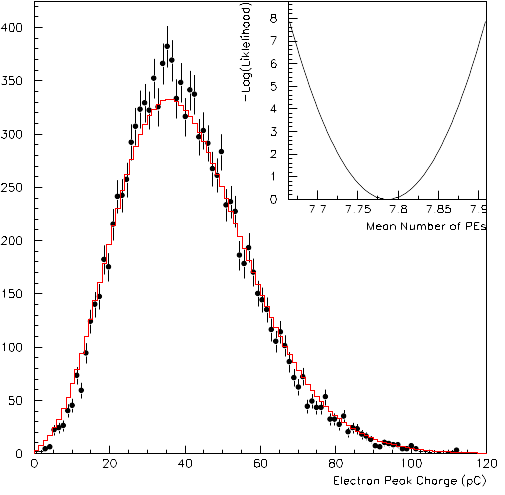}
\caption{The fit of the described convolution process. Red line is the result of this fit, and the black points are the experimental data.}\label{fig:deconvolution}
\end{figure}

Assuming that the number of photons (n) produced in the active volume of the Cherenkov radiator, due to the passage of a charged particle, follows a Poisson Distribution :
\begin{equation}
    \text{Poisson}(n;\mu) = \frac{\mu^n e^{-\mu}}{n!}
\end{equation}
and that each of these photons has either the probability to interact in photocathode or just to pass and escape. This probability can be described by a Binomial distribution:
\begin{equation}
    \text{Binomial}(k;n,\epsilon) = \frac{n!}{k!(n-k)!}(\epsilon^k (1-\epsilon)^k)
\end{equation}
Thus, the probability to observe k photoelectrons is given as the infinite sum of the Binomial distribution of the n photons to interact producing photoelectrons and the Poissonian distribution of those photons hit the photocathode.
\begin{equation}
    f(k;\mu,\epsilon) = \sum_{n=0}^{\infty} \text{Binomial}(k;n,\epsilon)\cdot \text{Poisson}(n;\mu)
\end{equation}
which after a few easy calculations finally lead to be a new Poissonian Distribution:
\begin{equation}
    f(k;\mu,\epsilon) = \text{Poisson}(k;\mu\cdot \epsilon)
\end{equation}
Hence, the k photoelectrons, starting simultaneously, will produce a signal, whose charge is distributed according to the k-Polya distribution. But on the case where k=0, there is a noise charge distribution. The parameters of the k-Polya distribution, is the mean value of the single photoelectron charge distribution and its RMS.
As  a result, the probability to detect an electron peak with charge q, is given as the infinite sum of Poisson and single-Polya distributions. In Figure \ref{fig:deconvolution}, the fit of the described convolution of the Poissonian and Polya distributions to the charge distribution of the experimental multi-photoelectron pulses is presented. The free parameter of the fit, is just the mean value $m = \epsilon \cdot \mu$ of the produced photoelectrons. We use a Log-Likelihood estimation and we found that, the mean number of photoelectrons  produced is $7.8\pm 0.1$.

In the next Table \ref{tab:numpes}, the above described approaches are being presented, creating the same charge distributions with different bin size, in order to find any systematic effects existing in the data set. 
\begin{table}[hbt!]
\begin{center}
\resizebox{16cm}{!}{
\begin{tabular}{|c|c|c|c|c|c|c|}
 \hline
\textbf{\begin{tabular}{@{}c@{}}Mean Value \\ Single Pes\end{tabular}}&\textbf{\begin{tabular}{@{}c@{}}RMS \\ Single Pes \end{tabular}}&\textbf{\begin{tabular}{@{}c@{}}Mean Value \\ Multi-Pes\end{tabular}}&\textbf{\begin{tabular}{@{}c@{}}RMS \\ Multi-Pes \end{tabular}}&\textbf{\begin{tabular}{@{}c@{}} Number of \\ Photoelectrons\end{tabular}}&\textbf{\begin{tabular}{@{}c@{}c@{}}
Number of\\ Photoelectrons\\(Eq.\ref{eq:vartotalchargenew})\end{tabular}}&\textbf{\begin{tabular}{@{}c@{}c@{}}Number of \\Photoelectrons\\(Deconvolution)\end{tabular}}\\
 \hline
 5.3 & 3.56 & 40.76 & 26.84 & 7.69 & 7.54 & 7.73\\
 \hline
 5.2 & 3.65 & 40.71 & 27.78 & 7.84 & 7.62 & 7.88\\
 \hline
5.17 & 3.66 & 40.74 & 28.04 & 7.88 & 7.66 & 7.93\\
 \hline
5.14 & 3.75 & 40.76 & 26.94 & 7.93 & 7.59 & 7.98 \\
 \hline
 5.26 & 3.55 &  40.71 & 27.09 & 7.75 & 7.64 & 7.79\\
 \hline
 5.29 & 3.53 & 40.733 & 26.83 & 7.71 & 7.6 & 7.74\\
 \hline
 5.3 & 3.5 & 40.76 & 26.67 & 7.69 & 7.62 & 7.73\\
 \hline
\end{tabular}}
\caption{\label{tab:numpes} Mean Value of the Number of Photoelectrons, using different approaches, indicating the consistency of calculations.}
\end{center}
\end{table}

\begin{figure}[hbt!]
 \centering 
 \includegraphics[width=1.0\textwidth]{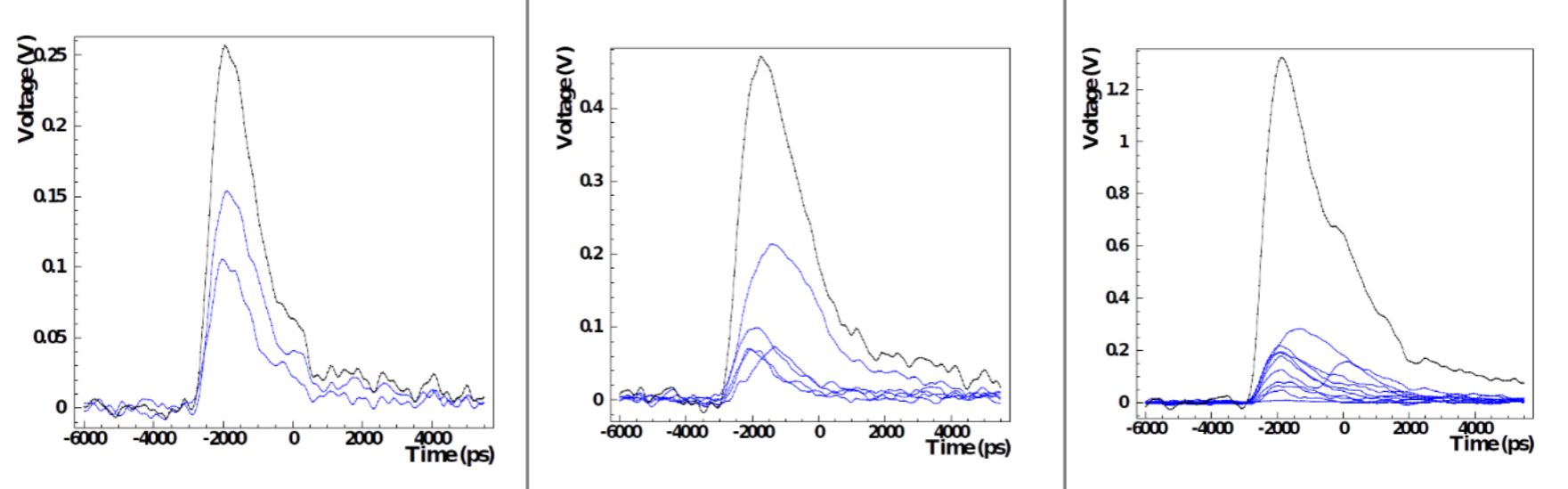}
\caption{Simulated pulses of the PICOSEC response to 2(left), 5(center) and 10(right) photoelectrons, by summing waveforms of the SPE-set. }\label{fig:spe-sum}
\end{figure}

In the way of producing multi-photoelectron pulses, assuming that PICOSEC-MicroMegas is a linear device, not having the opposite evidence, there is a need of summing up the corresponding number of single photoelectron pulses, assuming that the complete set of information needed is saved in one data-file. The prerequisite of this sum is to make the single pulses synchronous, this can be achieved with synchronizing their reference time, which in our case is set to be zero. After this is done, a $3^{rd}$ degree polynomial interpolation between digitization was made and the so on synchronous pulses can be added together to create a new multi-photoelectron pulse. Some examples of summing procedures with a different number of single photoelectron pulses can be seen in Figure \ref{fig:spe-sum}. 

\begin{figure}[hbt!]
 \centering 
 \includegraphics[width=0.6\textwidth]{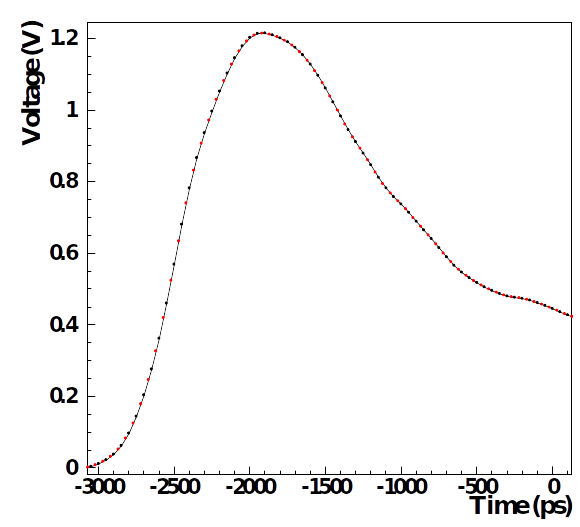}
\caption{Example of a simulated multi-photoelectron pulse shifted within $\pm$50\,ps. Black points are the original digits of the waveform, while the red ones are the shifted digits.}\label{fig:wave_shift50}
\end{figure}

Setting the reference time to zero, has a negative effect on the created multi- photoelectron pulses, having exact the same digitization for all the cumulative pulses that are being made(i.e. first digit on $50\,ps$, second on $100\,ps$ and so on). In fact, this means that the signal of the trigger and from digitization device are recorded simultaneously, without any time jitter. To simulate the reality, on a digit-by-digit basis, and on every sum-up event created, it is necessary to shift its digits randomly within $\pm$50\, ps. An example of this approach is represented in Figure \ref{fig:wave_shift50}, for the two cases before and after this shift.   

Because of the fact that the number of simple photoelectrons that can add up and give pulses to many photoelectrons is limited, the number of possible combinations during these additions is also quite specific, with the result that the final pulses are highly correlated, due to the fact that there will be some multi-photoelectron pules that will share the same single-photoelectron pulses. We used the bootstrap technique \cite{ramachandran2009mathematical}, which uses a metric for random sampling with replacement, in our case we followed the Poissonian distribution, for the number of single photoelectron pulses to be added up each time.

In order for the simulation model to have the same response as the EXP-set, e.g. with multi-photoelectron pulses, it has to be developed with the same mean value of photoelectrons. 

Thus, pulses are generated with such number of photoelectrons (e.g. number of pulses that are going to be summed) as a Poissonian distribution with a mean value of 7.8 photoelectrons. This random generator tells us how many single photoelectron pulses we will need, but after that the choice of who these pulses will be, is performed with just a simple random generator. That means that our algorithm follows a selection criteria for the number of single-photoelectron pulses, from that Poissonian distribution, in order to make the cumulative waveforms.   

\subsubsection{Noise contribution}
\begin{figure}[hbt!]
 \centering 
 \includegraphics[width=1.0\textwidth]{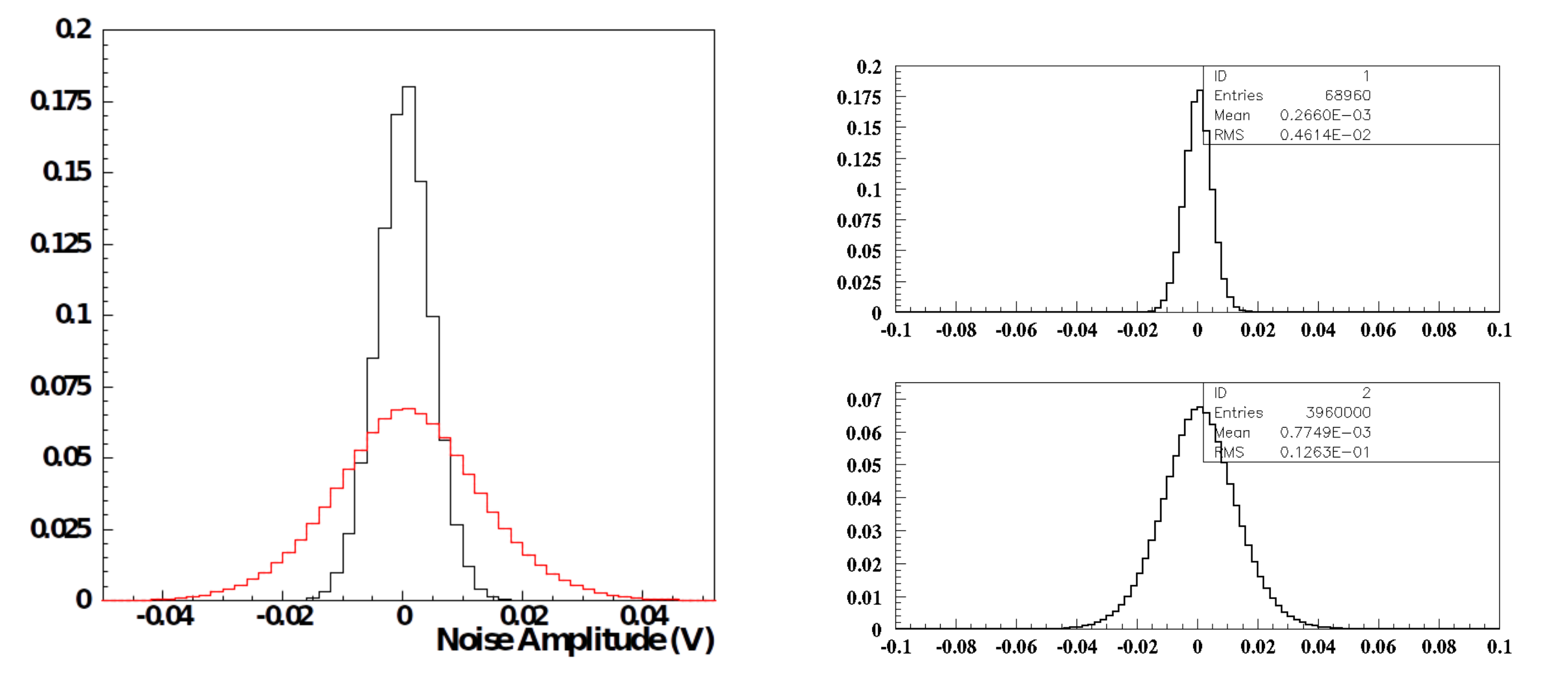}
\caption{Combine histogram of average noise contribution on single photoelectrons and simulated multi-photoelectrons(left). On the up-right histogram the noise on the single photoelectrons is being presented, while the bottom-right histogram presents the noise contribution on the case of simulated multi-photoelectrons.}\label{fig:spe-noise}
\end{figure}

\paragraph{} Despite the fact, that from single photoelectron pulses the electronic noise has been subtracted before the summing procedure, it can be seen, from Figure \ref{fig:spe-sum}, that the multi-photoelectron pulses suffer from the remaining random noise more than the single ones. Due to the fact that we add up together multiple waveforms that suffer the random noise, the resulting waveform will have a corresponding noise with the same cumulative property. For this purpose, the first 2\,ns of the resulting pulses, are chosen for the noise contribution estimation. Putting all the noise amplitudes of those digits in a histogram, will result to a distribution like the upper right of the Figure \ref{fig:spe-noise}. As it can be seen, the mean value of both these distributions is close to zero, nevertheless the rms is rising, confirming the higher noise contribution to the simulated multi-photoelectron pulses.   

\begin{figure}[hbt!]
 \centering 
 \includegraphics[width=0.6\textwidth]{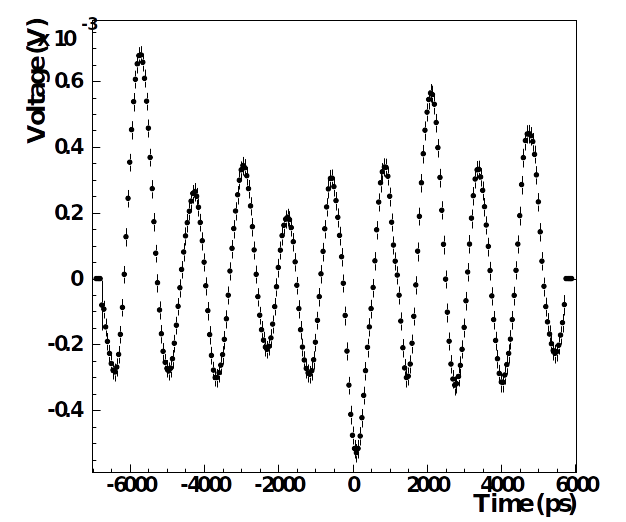}
\caption{Correlated noise of the laser.}\label{fig:noise_aver_shape}
\end{figure}

On the other hand, laser is a mechanical device which gives high intensity signals, e.g. electromagnetic waves, which in our case are being attenuated, but it also has its own electronic noise. With the attenuation process, we cut-off the light but on the oscillator can still reach signal, but as noise, because of the quantum efficiency of the device, that limits our capability to detect every photoelectron created in the active volume, which is hidden in the electronic noise of the laser. 
This is a correlated noise, containing on the empty events of the Figure \ref{fig:singlepe}, corresponding to events with charge lower than 1\,pC, or 30\,mV amplitude. Representing those event amplitudes on a histogram, as a function of time, results to a distribution like the one seen on Figure \ref{fig:noise_aver_shape}. As the maximum value of this correlated noise it is considered to be 0.7\,mV, which is not a significant noise to spoil the measurement of resolution.     

In conclusion, the simulated pulses, generated with the above method, contain information of the SPE-set of waveforms with electron peak charge greater than 1\,pC, or amplitude 30\,mV , in order to eliminate the noise contribution. But this software cut, eliminates number of events, hence the missing part of this spectrum needs to be replaced. This replacement is realised by scaling down larger pulses (1-2\,pC), according to Polya distribution. 

\begin{figure}[hbt!]
 \centering 
 \includegraphics[width=1.0\textwidth]{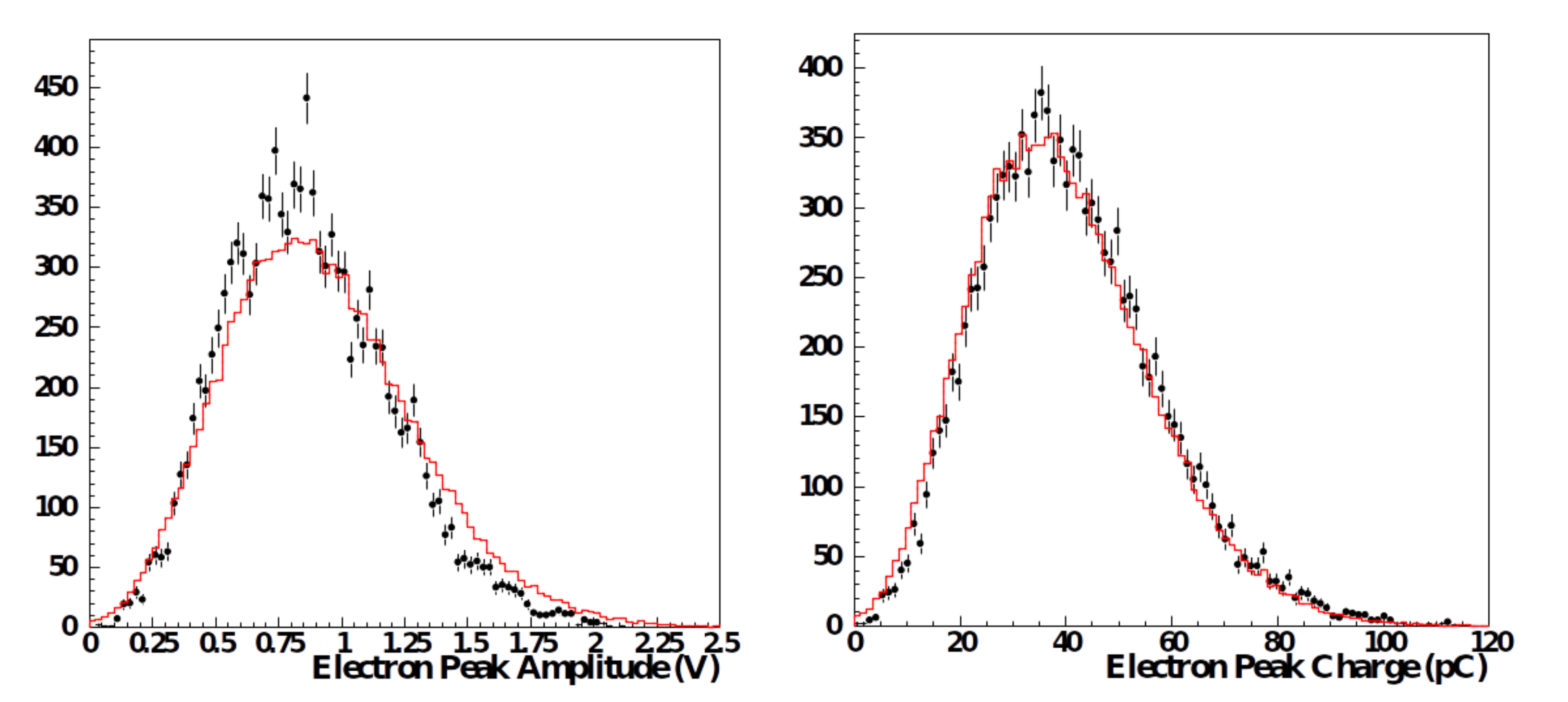}
\caption{Electron Peak Amplitude(left) and Electron Peak Charge(right) distributions for both the simulated-set and the EXP-set of waveforms. With black, the simulated points are represented, and with red the simulated ones.}\label{fig:charge_gpeak_simul}
\end{figure}

As a cross reference, in order for us to compare the simulated results, with the results of the EXP-set, we generate the distributions of the electron peak charge, and peak amplitude. Results can be seen in Figure \ref{fig:charge_gpeak_simul}.  

Someone on Figure \ref{fig:charge_gpeak_simul} can spot the difference between the EXP-set waveforms and the simulated waveforms, when it comes to the distribution of electron peak amplitude. Defining the maximum amplitude of the pulse, requires the localization, of the point with the greatest amplitude in the waveform. It is obvious that this measurement is biased from noise contribution. The greater the noise (it is known that we suffer from greater noise in the simulated pulses) on the waveform, the greater the peak amplitude we determine. No matter that, we have seen in Figure \ref{fig:spe-noise} has a mean value close to zero, while its sigma is rising. That will affect the total distribution, which will be forced to greater amplitudes, as the sigma of the noise contribution is rising with the square root of the number of photoelectron. This is a limitation of our simulation model, that we have to take into account, in our later on analysis. On the other hand, the electron peak charge distribution is a more reliable parameter for the simulation, that reproduces the same distribution as the real data.

The second point at which the data will be checked, before to be assumed suitable for training set for the Neural Network, is the timing properties they carry. For this reason a full timing analysis will take place in the next pages. 
\subsection{Timing with Constant Fraction Discriminator technique at \texorpdfstring{20$\%$}{Lg}}

\begin{figure}[hbt!]
 \centering 
 \includegraphics[width=1.0\textwidth]{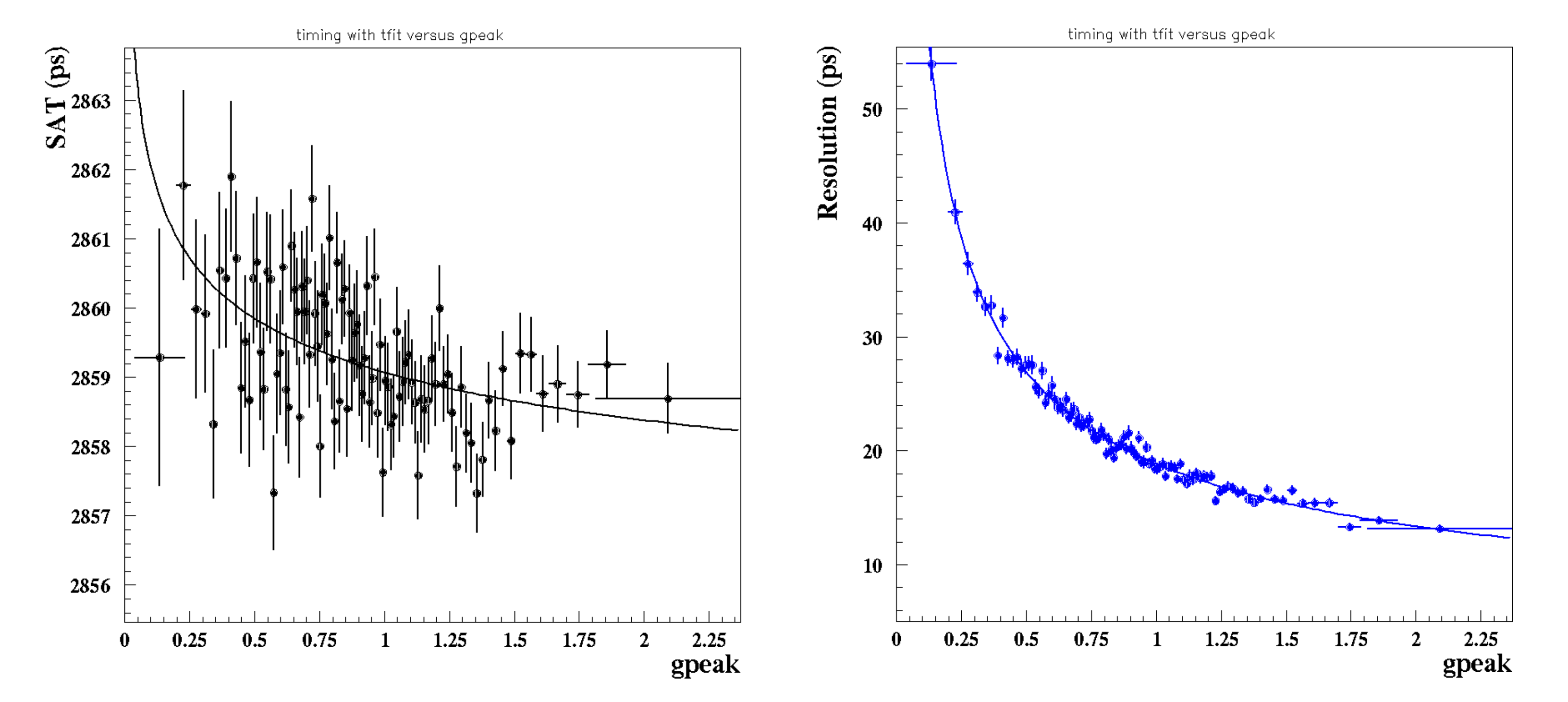}
\caption{Signal Arrival Time (left) and Resolution (right) of the simulation model, as a function of electron peak amplitude, without time walk corrections with CDF timing technique.}\label{fig:slewing_gpeak_simul}
\end{figure}

Following the same analysis procedure as with the real data, but with the difference of picking bins of amplitude every 1000-2000 events, described in the Section \ref{subsubsectionCDF}, the  representation of Signal Arrival Time as a function of electron peak amplitude, parameterised in bins of peak amplitude for the whole set of the simulated data, can be seen in Figure \ref{fig:slewing_gpeak_simul}(left), where simulated data points have been fitted with the corresponding power law, revealing the existence of time walk effect, as it was expected. On the right plot of the same Figure, the timing resolution is being presented.   

\begin{figure}[hbt!]
 \centering 
 \includegraphics[width=1.0\textwidth]{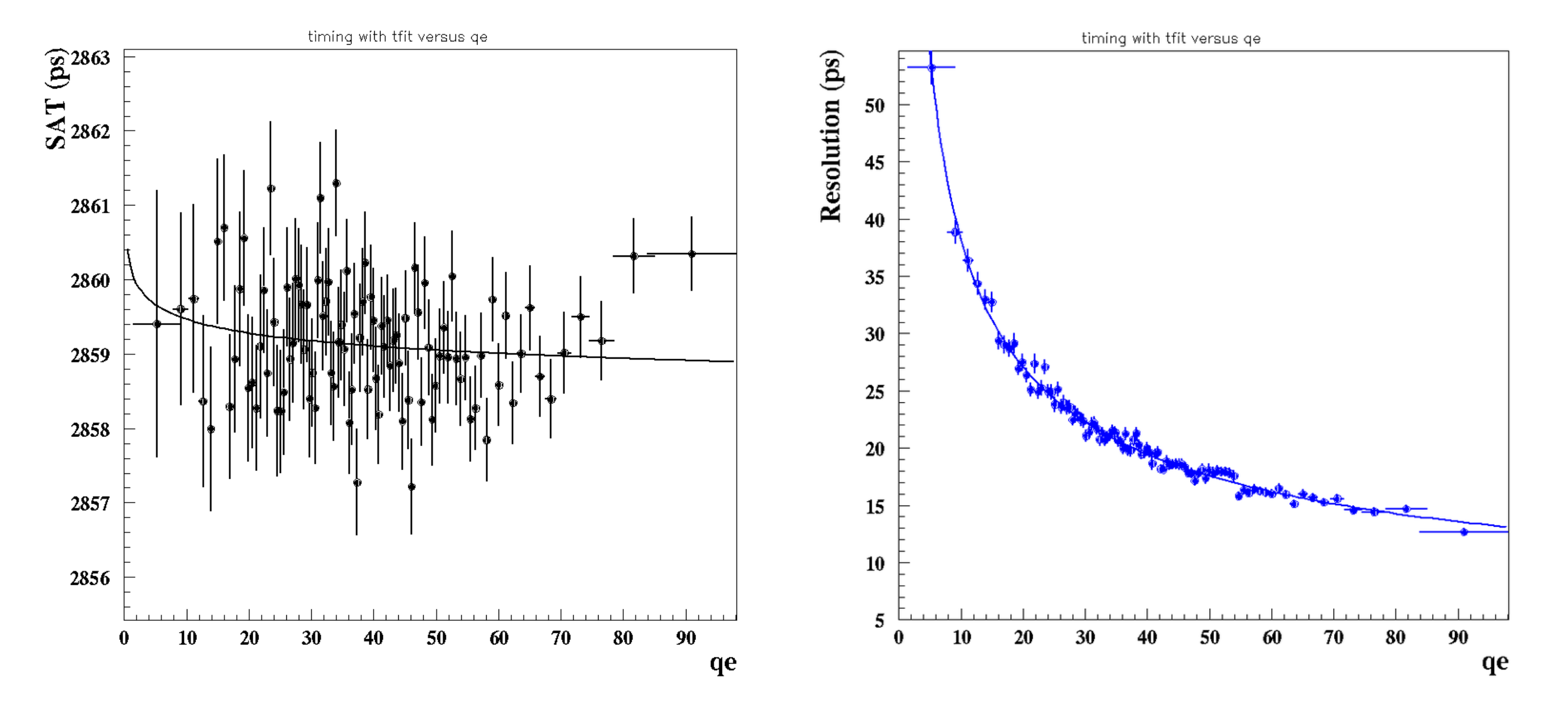}
\caption{Signal Arrival Time (left) and resolution (right) of the simulation model, as a function of electron peak charge, without time walk corrections with CDF timing technique.}\label{fig:slewing_charge_simul}
\end{figure}

Figure \ref{fig:slewing_charge_simul}, is a result of the same procedure, but this time dividing the events in bins of charge, representing the same behavior for the Signal Arrival Time and the timing resolution. 

\begin{figure}[hbt!]
 \centering 
 \includegraphics[width=0.5\textwidth]{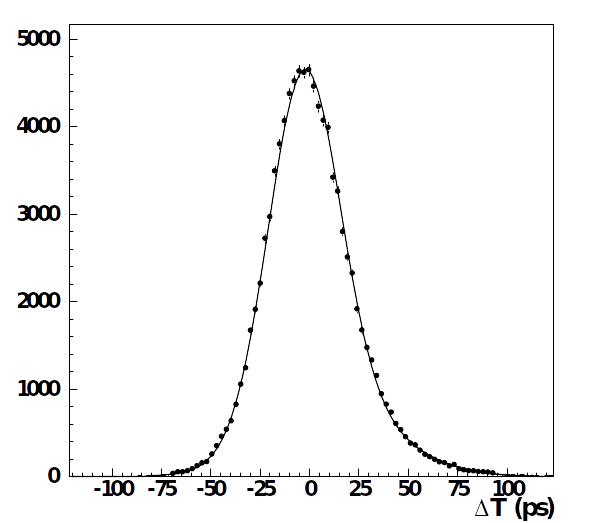}
\caption{Signal Arrival Time of the simulation model, after time walk corrections, resulting to a timing resolution of 21.3$\pm$0.6\,ps.}\label{fig:simul_dt_corr}
\end{figure}

A full timing analysis of the simulated pulses, after time walk corrections in the determination of Signal Arrival Time, results to timing resolution of 21.3$\pm$0.6\,ps, seeing in the Figure \ref{fig:simul_dt_corr}, which is worse than the results of the EXP-set data by 3\,ps. This small difference is caused from the fact that the addition of N single photoelectrons, accumulates extra noise on the simulated waveforms, and that the photodiode timing jitter is added quadratically as a function of the number of photoelectrons. 

\begin{figure}[hbt!]
 \centering 
 \includegraphics[width=1.0\textwidth]{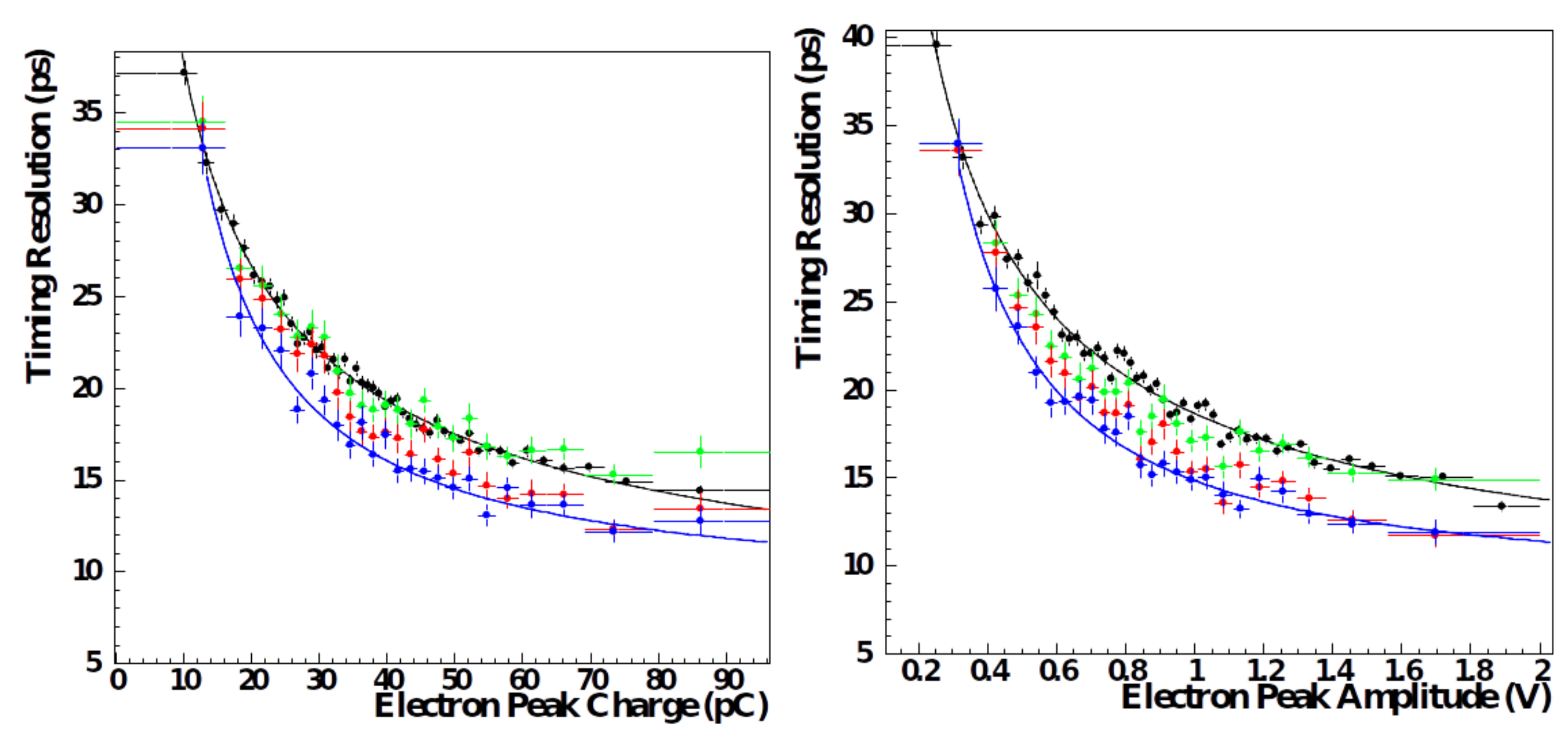}
\caption{Resolution of the simulation model, after time walk corrections as a function of electron peak charge (left) and electron peak amplitude(right).With blue color the real data are represented, with red color the data after the added noise corresponding to number of photoelectrons, with green color the red data with the extra added due to the resolution of photodiode and with black color the simulated data, for CDF timing technique.}\label{fig:resolution_simul_tot_global}
\end{figure}

The corresponding timing resolution curve, is the black colored one on the Figure \ref{fig:resolution_simul_tot_global}, after time walk correction, where the deterioration in the timing resolution is also presented. The simulated data suffer of the noise contribution, as it is underlined earlier. Having produced this way the simulation data, and knowing the limitations of our method, we will try to move the real data adding the relevant contribution of noise as error, due to the number of photoelectrons and the resolution of photodiode timing reference.  

\begin{figure}[hbt!]
 \centering 
 \includegraphics[width=1.0\textwidth]{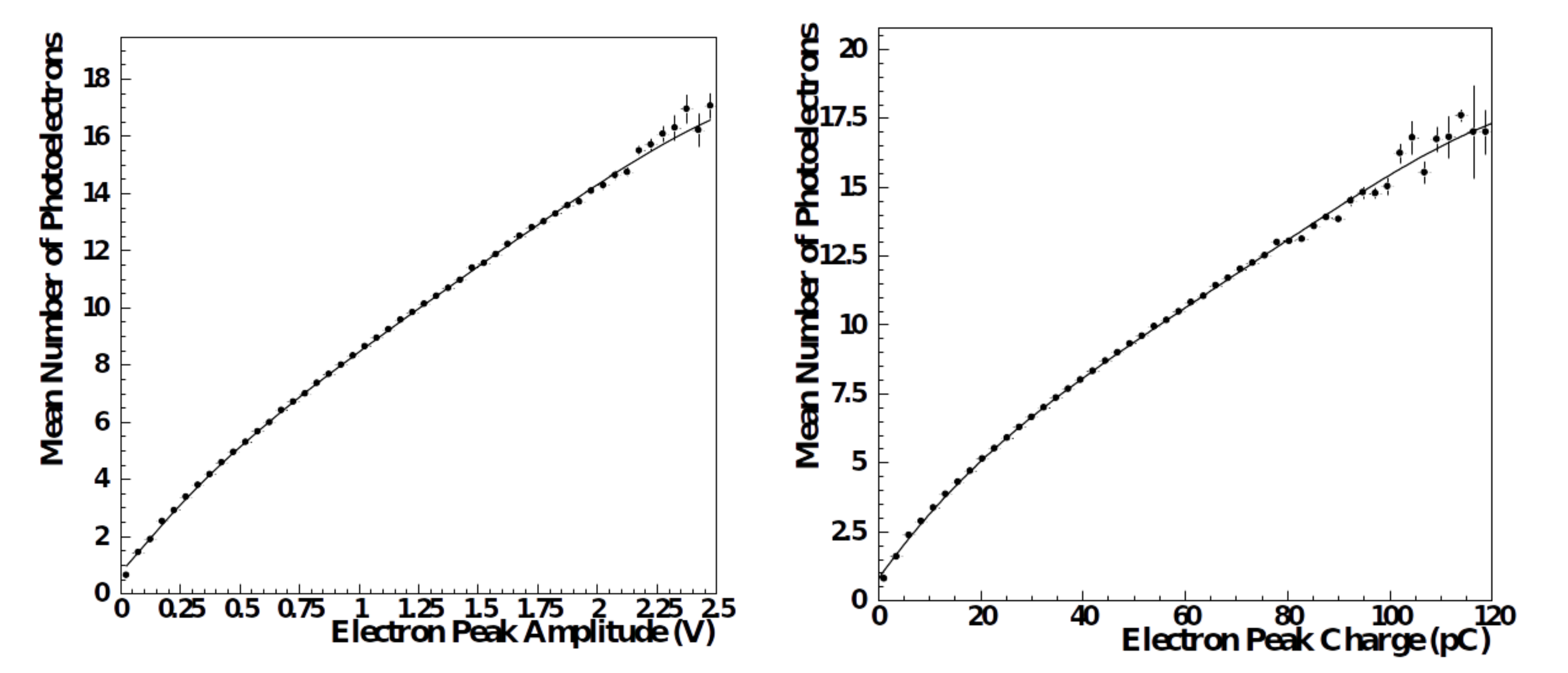}
\caption{Determination of mean number of photoelectrons as a function of electron peak amplitude(left) and electron peak charge(right).}\label{fig:mean_num_pes}
\end{figure}

Firstly, it is convenient to describe the behavior of mean number of photoelectron as a function of electron peak size. For this purpose, the electron peak charge distribution it is used, and also a charge distribution weighted with the number of photoelectrons. Divided those two distributions, results to the right plot of the Figure \ref{fig:mean_num_pes}, in those data the error bars are adopted from the weighted charge distribution over the square number of photoelectrons because of the fact that $rms = \langle N^2 \rangle - \langle\Bar{N}\rangle^2$. Fitting this curve with a $4^{rth}$ degree polynomial. 

Subsequently, on an event-by-event, of the real data, finding the corresponding charge of the pulse, going back to the right plot of Figure \ref{fig:mean_num_pes}, results to a specific number of photoelectrons in the waveform, and hence every digit of the waveform can be moved randomly through a Gaussian with mean value to zero and sigma: $\sigma = \sqrt{N_{pe}}\cdot\sigma_{1pe}$, where the $\sigma$ of one photoelectron is taken from the upper right plot of the Figure \ref{fig:spe-noise}, and it was found to be $\sigma_{1pe} = 0.0046$. This procedure results to the red colored data in the Figure \ref{fig:resolution_simul_tot_global}. 
 
Although the data move towards the direction of the simulated one, they still can't reach the same resolution. For this purpose, we have to include the photodiode's error, correlated with the number of photoelectrons. As it is mentioned before, data of the right plot of \ref{fig:mean_num_pes} are fitted with a $4^{rth}$ degree polynomial. Assuming a variable describing all these effects on data to be: $x = x_{data} + x_{\text{data with extra noise}} + x_{\text{effect of photodiode}} = y + x_{\text{effect of photodiode}}$, the effect of photodiode is related with the number of photoelectrons through the expression $x_{\text{effect of photodiode}} = n\cdot z$, where z is the photodiode's effect at single photoelectron. Thus, the variance of the total effect on the data will correspond to the value : $V[\text{Data}] = V[\text{Data with noise}] + n^2\cdot\sigma_{phd}^2$. Where the number of photoelectrons in this expression will be taken of the fit of the corresponding plot of Figure \ref{fig:mean_num_pes}, depending to the relevant resolution we need to correct.   

Following this procedure, with the mean value of photoelectrons as a function of electron peak charge, fitted with the equation:
\begin{equation}
    n = 0.7973+0.215\cdot q_e + 0.29437\cdot e^{-2}\cdot q_e^2  + 0.2796\cdot e^{-4}\cdot q_e^3 - 0.10017e^{-6}q_e^4 
\end{equation}
then, we move the red colored data through the error $n^2 \cdot \sigma_{phd}^2 = n^2\cdot(2.6)^2$, resulting to the green colored data on the left Figure \ref{fig:resolution_simul_tot_global}.

Same steps should have been followed in the case of representation of resolution, with respect to the electron peak amplitude, using this time the fitted equation of the mean number of photoelectrons as a function of electron peak amplitude of the left plot of the Figure \ref{fig:mean_num_pes} to be: 
\begin{equation}
      n = 0.66506+10.907 \cdot V_{peak} - 4.9459\cdot V_{peak}^2 + 2.2305 \cdot V_{peak}^3 - 0.38889\cdot V_{peak}^4
\end{equation}
resulting to the green colored data of the right plot of the same Figure. 
\subsection{Timing at \texorpdfstring{$100\,mv$}{Lg} with Charge over Threshold corrections}

\begin{figure}[hbt!]
 \centering 
 \includegraphics[width=1.0\textwidth]{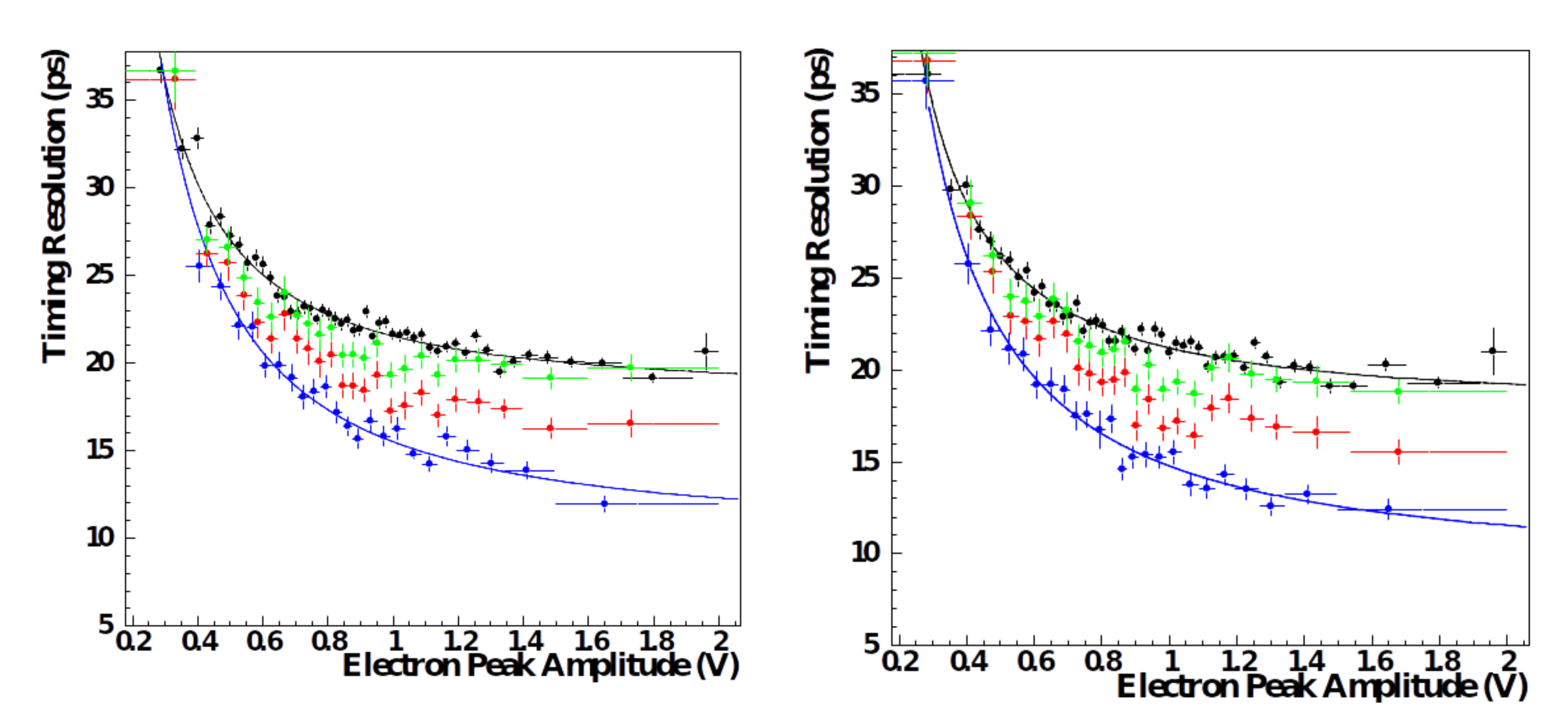}
\caption{Resolution of the simulation model, after time walk corrections as a function of electron peak amplitude, with timing at 100\,mV and parameterization with electron peak amplitude parameter(left) and with parameterization with maximum available charge over threshold(right). With blue color the real data are represented, with red color the data after the added noise corresponding to number of photoelectrons, with green color the red data with the extra added due to the resolution of photodiode and with black color the simulated data. With resolution of 23.24$\pm$0.6\,ps.}\label{fig:resolution_simul_tot_global_qup}
\end{figure}

Full analysis process, in this case, has been described in Section \ref{subsubsectionqup}, using as timing threshold the lowest one, where it is used 100\,mV, and correct for time walk effect in the highest available threshold with the integral of the pulse, e.g. its charge. This results in the Figure \ref{fig:resolution_simul_tot_global_qup}, where the same corrections to the real data have been implemented, in order to move towards the direction of the simulated data. The difference between the two plots is that on the left, the parameter of the time walk correction was chosen to be the electron peak amplitude, of the highest threshold crossed, while on the right plot the charge of the electron peak was used instead. 

These results, present another difference between the simulated data and the real ones. Resolution reaches a plateau faster for the simulated data, and this is an indication of the noise contribution to the greater pulses. The greater the pulse the more dominant becomes from noise. The Charge over Threshold method, uses as the only information needed the charge of the pulse(single measurement), which contains all the errors of the included digitization points. On the case of Constant fraction discrimination technique, we have used the logistic function to fit the leading edge of every pulse. This procedure gives an average estimation of the noise in every fitted function, and which corresponds to the minimum noise effect on the pulse. This is the reason why in CDF resolution plots, as a function of electron peak size, follows a falling behavior, instead of reaching a plateau. The proof is also the fact that by adding the appropriate noise we approach exactly the behavior of the simulated events.

\section{Using Simulation pulses for ANN training and timing results} 
\begin{figure}[hbt!]
 \centering 
 \includegraphics[width=0.5\textwidth]{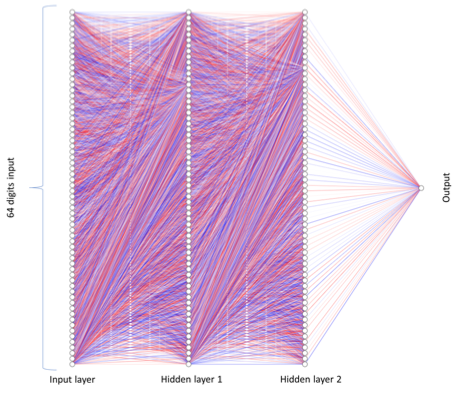}
\caption{An Artificial Neural Network architecture, consisting of one input and two hidden layers, and comprising one output node. The number of input and hidden nodes is defined by number of selected Electron Peak digitisation.}\label{fig:neural_sketch}
\end{figure}

\paragraph{} Our Neural Network, was implemented as a Feed Forward neural network, following the design described in Appendix \ref{appendix:A}. The choice of Network architecture depends on the specific problem, and it is chosen to be one, by trial and error. Our architecture consists of an input layer, two hidden layers with 64 neurons each, and an output layer. This was found to give satisfactory results, after trial and error, as it can be seen in the Figure \ref{fig:neural_sketch}. In order for the network to obtain the optimal architecture for the specific problem, one can turn to brute force approaches, such grid search, or more sophisticated ones, such Bayesian optimization of the hyperparameters\cite{snoek2012practical}, or hyperparameter optimization with Generic Algorithms \cite{net}. 

For all nodes, instead of sigmoid activation function, \emph{ReLU} activation function was used apart from the output layer\cite{agarap2019deep}, something essential for regression tasks. The loss function was used as the mean squared error described in the Appendix \ref{appendix:A} too. 

The training set that was used in each case was split in training validation test, with ratio 0.2. The purpose of this split was to validate the performance of the network in the training phase, with a data-set that was not previously seen by the network. All models were trained for 1000 epochs and the model with the best performance regarding the validation set is returned as the optimal model of the training. 
Subsequently, two training scenarios were performed in our study. 
\begin{itemize}
    \item Firstly, we studied the predictive capabilities of the Feed Forward Neural Network for the experimental data (EXP-set), where a k-fold cross validation, with five folds, was used, to see if the model can generalize independently for the given data-set. That needs the whole data-set o be split in five folds, where the four folds will be used for training and one fold for testing. This procedure was performed cyclically, resulting in five different models, and by extension, to five different prediction samples, which were combined, to test the timing resolution of the algorithm. This was the first technique used before the development of the simulation model, were the experimental data, were used both for the training and the testing steps. 
    \item After the implementation of the simulation model, producing larger data-set of pulses, the total amount of these pulses was used as a training set. The training procedure is analogous to the previous case, using k-fold validation, while the test of the neural network was based on the experimental data. 
\end{itemize}

\begin{figure}[hbt!]
 \centering 
 \includegraphics[width=0.5\textwidth]{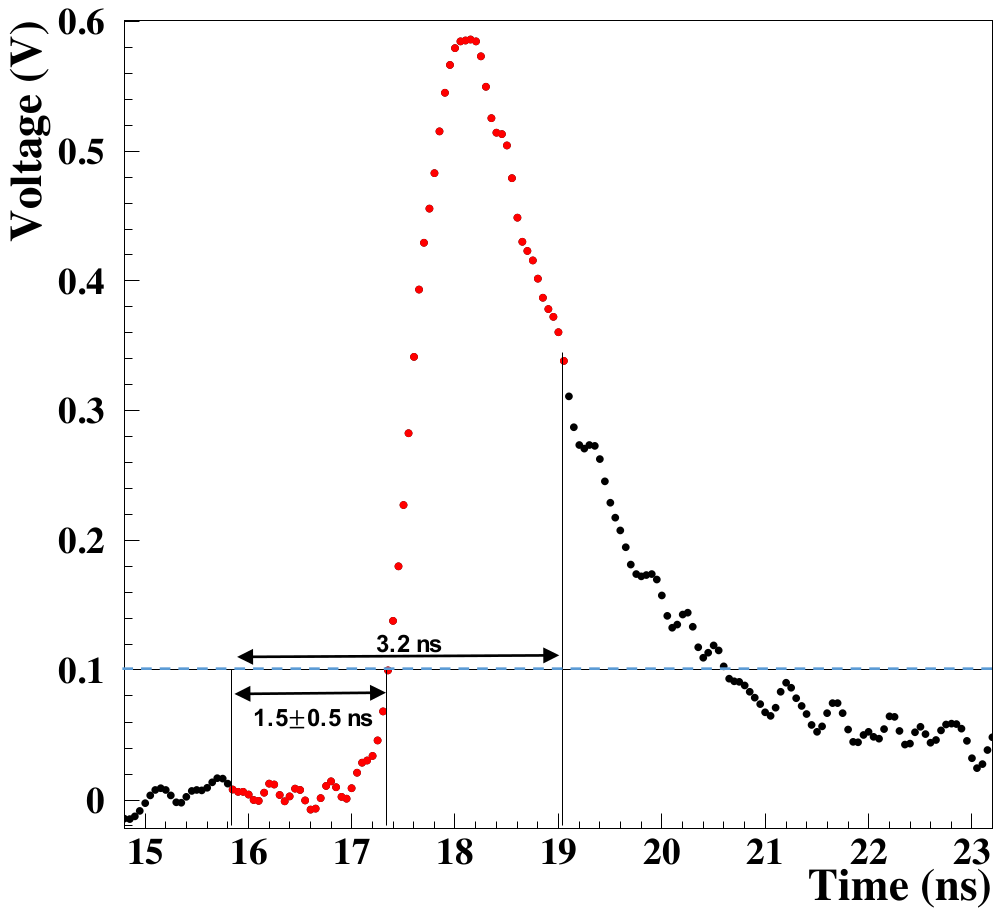}
\caption{Typical PICOSEC-Micromegas digitised waveform. Red points denote the digitised information that is presented to the ANN, in this case the first 64 digits, corresponding to 3.2\,ns timing interval.}\label{fig:nn_wave}
\end{figure}

Moving on to the running part of the Neural Network, this was fed with pulses like the one seen in Figure \ref{fig:nn_wave}. In the Figure, with red points are signaled the digits of the pulse that represent the input layer of the neural network, corresponding to 3.2\,ns timing interval, and with a starting point of the input that corresponds to  1.5\,ns back of the crossing point of a threshold trigger at 100\,mV.  This corresponds to 30 digits, since the time spacing of the digits is 50\,ps.
As arrival time, is defined the time of the first digit, measured by the reference photodiode detector. Various input lengths were tested, and it was found that the best performance was achieved when the input include the leading edge and the least possible part of the pulse after the peak, since this is the part that contains the useful timing information. After finding the first digit, in order to the network to be unbiased of the starting point timing position, this point was redefined by randomly choosing a digit around it, using a uniform distribution with a standard deviation of $\pm 500$\,ps. This point simulates the behavior of the trigger response, giving the sign that a pulse has been recorded, and by default, it inputs an extra delay of the signal of the order of 0.5\,ns.  

\begin{figure}[hbt!]
 \centering 
 \includegraphics[width=0.5\textwidth]{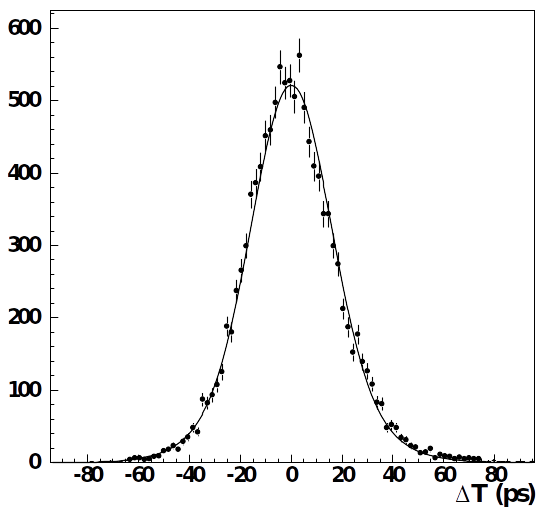}
\caption{The PICOSEC Signal Arrival Time, evaluated in the Neural Network, resulting to a Gaussian distribution with rms that determines the global timing resolution at 18.5 $\pm$ 0.6 \,ps.}\label{fig:gauss_nn}
\end{figure}

Using the first training technique with k-fold cross validation, using only the experimental data, also for the testing period, the Neural Network results to a resolution of 18.5$\pm$0.6\,ps, as it can be seen in Figure \ref{fig:gauss_nn}, same as the full signal processing analysis with Constant Fraction Discrimination Technique, which is a very promising result, that implies that the network works well. 

\begin{figure}[hbt!]
 \centering 
 \includegraphics[width=1.0\textwidth]{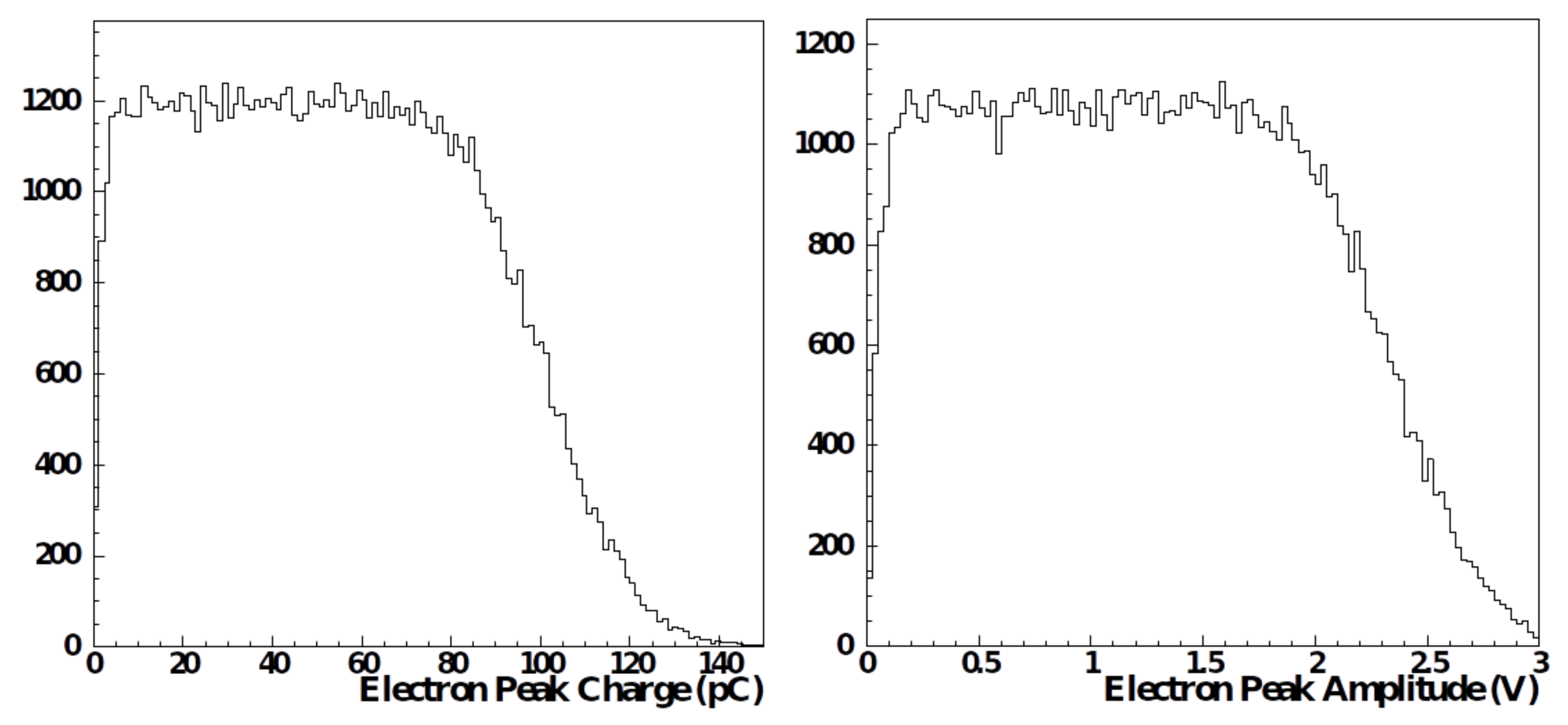}
\caption{Training of the ANN with uniform distribution of electron peak amplitude(right), and electron peak charge(left), in order to reduce the possibility of biases in the network, because of the electron peak size of the pulses.}\label{fig:flat_nn}
\end{figure}

This technique engages the danger of biases in the resulting timing resolution, because of using the same data-set both for the train and test procedures. This is why, we developed the previous described simulation model, in order to build independent set of data to train and test our network.  In order to reduce our fear of bias in the training process, we generate our simulated pulses for the training set with uniform distribution of charge and amplitude, as it can be seen in Figure \ref{fig:flat_nn}. This will eliminate network tendency to recognise specific bin charges due to their greater value, e.g. learn to recognise constant shape of the waveform.

\begin{figure}[hbt!]
 \centering 
 \includegraphics[width=1.0\textwidth]{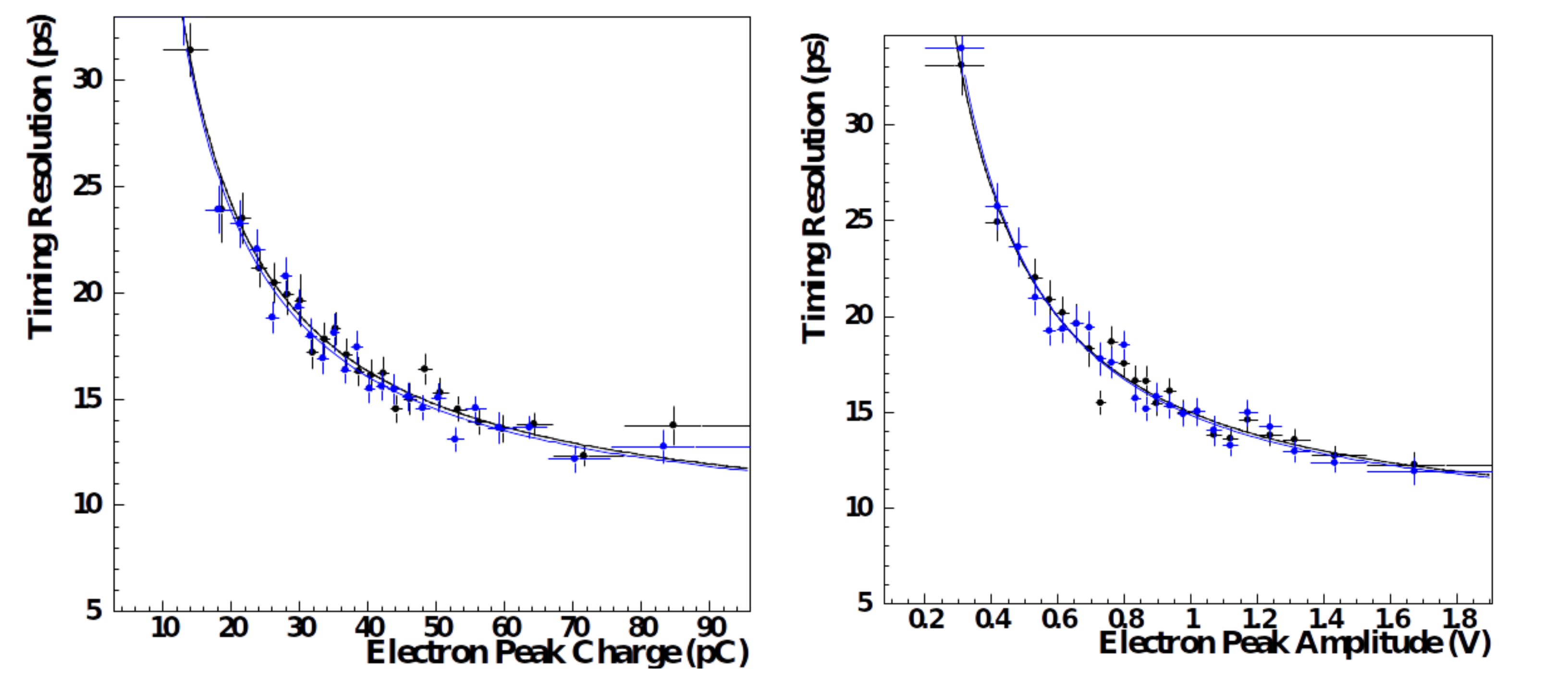}
\caption{Left :Resolution as a function of electron peak charge, in which the results of the ANN are compared with the results of the Constant Fraction Discrimination Technique. Right : Resolution as a function of electron peak amplitude, coming out of network, for the same timing method. Blue points correspond to the results of the Constant Fraction Discriminator technique while black points to the neural network.}\label{fig:resolution_nn}
\end{figure}

Following the full analysis procedure, results to a timing resolution corrected for the time walk effect, as the one seen in Figure \ref{fig:resolution_nn}. In this Figure, the resolution is a result of training with simulated data, in a uniform distribution either electron peak charge or amplitude, and tested in real experimental data, giving the desired resolution of 18.5 $\pm$ 0.6 \,ps.

Since we have strong evidence that our network works properly with the simulated data it was trained on, we still need to clarify its unbiased performance. It has been shown that injecting small noise to input training data, can prevent neural network from overfitting and may result in neural networks that are insensitive to noise in the data. Adding a noise of the order of 15\,mV ti input digits during the training phase, resulted in better performance with respect to timing resolution and absence of time walk effect.

\begin{figure}[hbt!]
 \centering 
 \includegraphics[width=1.0\textwidth]{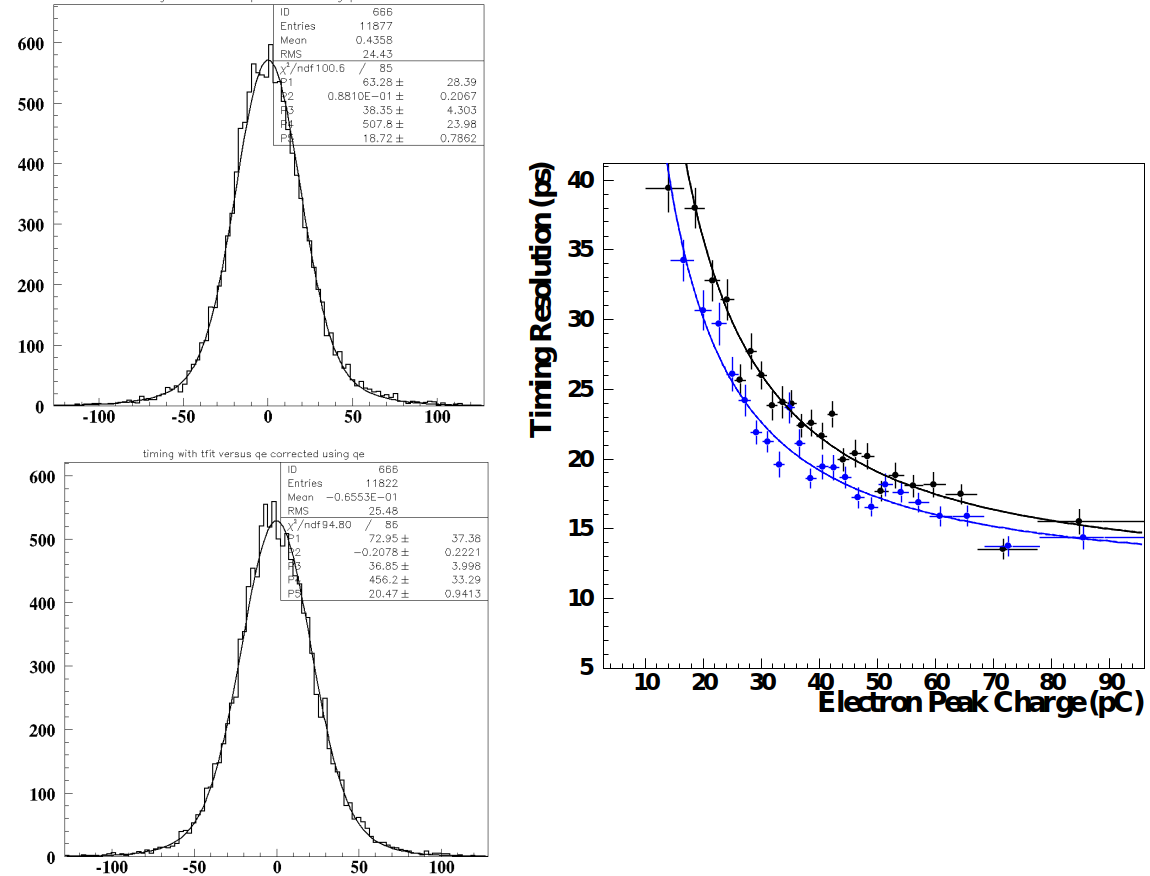}
\caption{Left (up) : Gaussian distribution of CDF timing technique on experimental data with an additional noise of 30\,mV, which RMS is 23.5 $\pm$ 0.5\,ps. Left (down) : Gaussian distribution of timing with the neural network, using the same data as the input layer, which RMS is 24.7$\pm$0.6\,ps. Right : Resolution as a function of electron peak charge both for the CDF data, colored in blue and for the data of the neural network, colored in black. }\label{fig:nn_30mV_noise}
\end{figure}

Our desire though, is to add significant noise to the testing data-set, to see if the resolution will be affected, which will mean that our neural network really understands those changes and is unbiased. The process corresponds to \say{destroy} a bit the experimental data, adding on the digits of the experimental waveforms, noise created randomly within a Gaussian distribution with mean zero and $\sigma$ 30\,mV. Using those noisy pulses in our full analysis algorithm for the Constant Fraction Discrimination timing technique, results to a resolution 23.6$\pm$0.5\,ps. For the same data given to neural network as testing set, results to a resolution 24.7$\pm$0.6\,ps. In the Figure \ref{fig:nn_30mV_noise}, on the left side Gaussian distributions of both the CDF technique(up) and the neural network timing (down) are represented, while on the right side, the relevant timing resolution can be seen, with blue the data points of the CDF and with black the points of neural network.

\begin{figure}[hbt!]
 \centering 
 \includegraphics[width=0.9\textwidth]{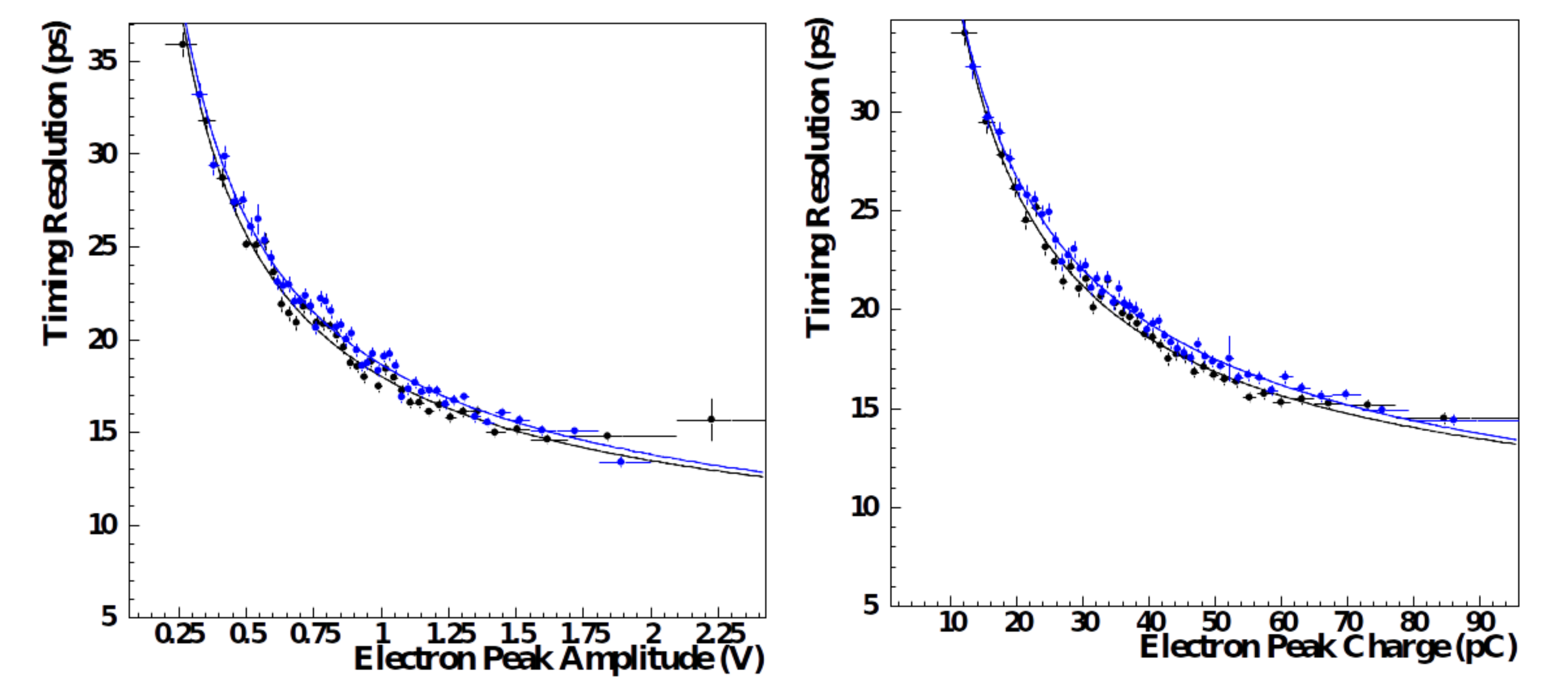}
\caption{Evaluation of performance of the Neural Network, in the simulated data both for train and test processes, resulting to a resolution 21.3 $\pm$ 0.6\,ps, comparing with the relevant results of the Constant Fraction Discrimination Technique(blue points) of the same data.}\label{fig:nn_evaluation}
\end{figure}

Next step in our way to check the consistency of our neural network, is to test it in the simulation data as well. For this purpose we use as training set the simulated data with the uniform distribution in charge and amplitude and then check it in the real simulated data with their proper charge and amplitude distribution. This is called evaluation test, which should lead to a resolution similar to the one came from the full analysis of the simulation pulses with the Constant Fraction Discrimination Technique. Indeed, this behavior was reveled by the neural network, resulting to a resolution of 21.3$\pm$0.6\,ps, seen in Figure \ref{fig:nn_evaluation}. 

\begin{figure}[hbt!]
\centering 
\includegraphics[width=0.9\textwidth]{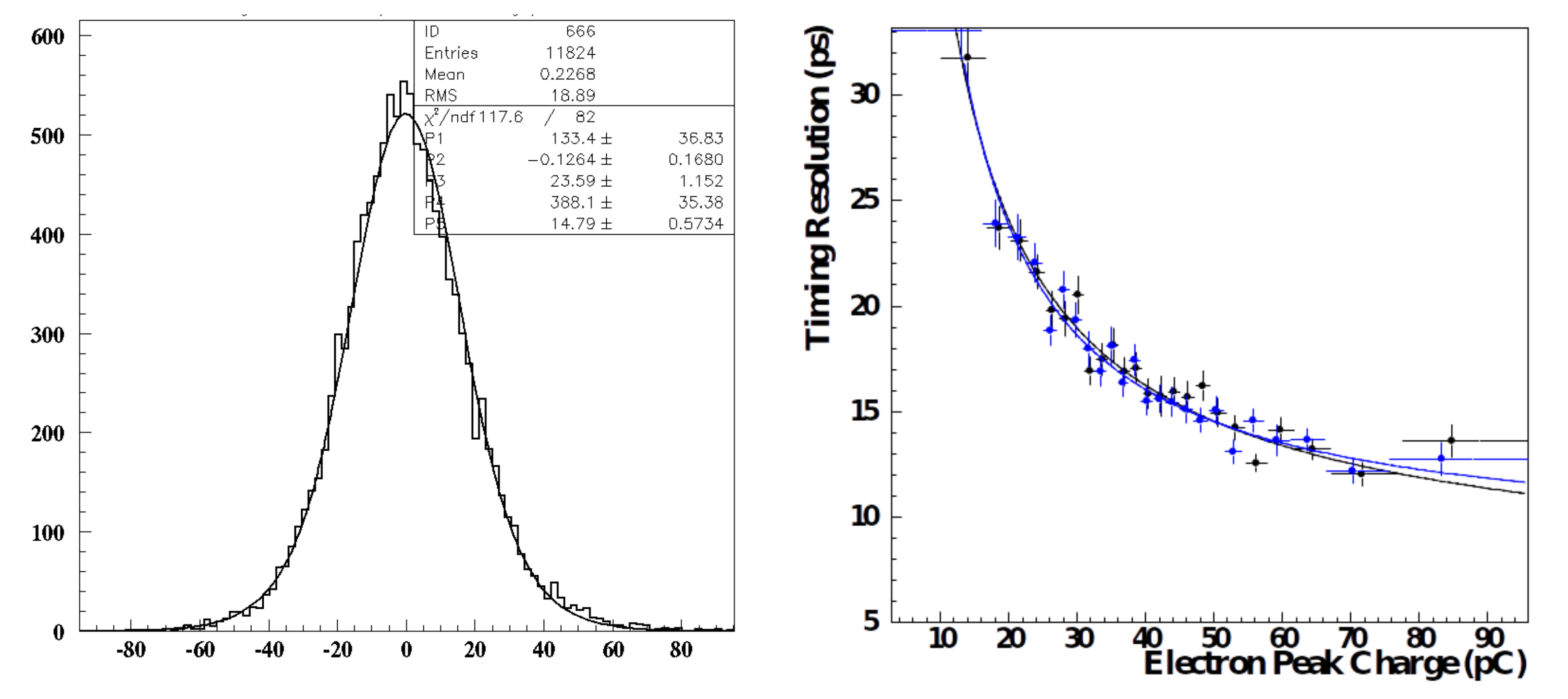}
\caption{Test of the Neural Network giving as a starting input point the one in 1.25\,ns before the threshold crossing point at 100\,mV resulting to a timing resolution of 18.4 $\pm$ 0.6\,ps. On the left the Signal Arrival Time of the Neural Network timing difference of the reference time. On the right the timing resolution curve as a function of electron peak charge both for the Constant Fraction Discrimination Technique (blue colored points) and for the neural network results(black colored points).}\label{fig:nn_new_start}
\end{figure}

\begin{figure}[hbt!]
\centering 
\includegraphics[width=0.9\textwidth]{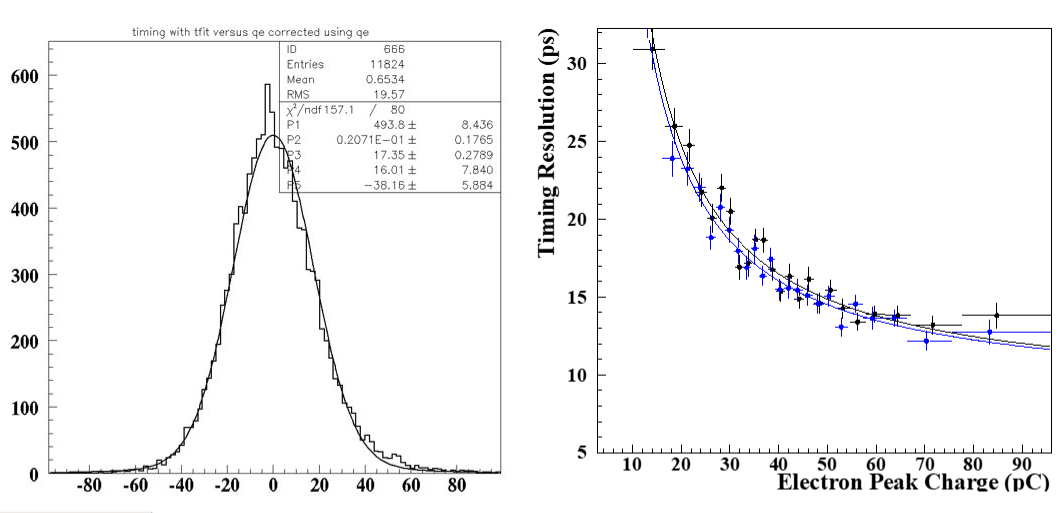}
\caption{Test of the Neural Network giving less digitization points as input layer. Resulting to a timing resolution of 19.02 $\pm$ 0.5\,ps. On the left the Signal Arrival Time of the Neural Network timing difference of the reference time. On the right the timing resolution curve as a function of electron peak charge both for the Constant Fraction Discrimination Technique (blue colored points) and for the Neural Network (black colored points).}\label{fig:les_digit}
\end{figure}

Having passed successfully all the above tests, the neural network proved to be a reliable tool in order to make precise timing. Another test, in which we submitted the neural network, is to give as a starting point of the input digits 1.25\,ns before the crossing point on the constant timing threshold applied, instead of the 1.5\,ns, but with the same 50\,ps digitization spacing. Apart from that we change also the random way we define the starting point this time with a Gaussian distribution with mean value to zero and sigma to 100\,ps. This is chosen in order not to be outside of the timing interval of the training set waveforms. Neural Network results to a distribution, which can be seen in Figure \ref{fig:nn_new_start}, which RMS is 18.4 $\pm$ 0.6\,ps, giving the final timing resolution. This final result is similar to our reference result from the Constant Fraction Discrimination Technique, giving strong evidence that our network works independently from biases on noisy data, or the definition of starting point. The last test has to do with the input parameters, i.e with the digitizations given on the input layer of the Neural Network. Using different spacing from 50\,ps to 200\,ps spacing results to 16 digits. That demands also a different training set, for this reason we create multi-photoelectron pulses with wider digitization step, and then train the Network. Full offline analysis on those data result to a timing resolution of 19 $\pm$ 0.5 \,ps. Subsequently, tested with the EXP-set, Neural Network results to 19.02 $\pm$ 0.5\,ps. The behavior of Timing Resolution as a function of electron peak charge can be seen in Figure \ref{fig:les_digit}.

\chapter{Summary and Concluding Remarks }\label{cha:conclusion}
\paragraph{} Throughout this study, the PICOSEC-MicroMegas detector potential for precise timing, at a picosecond level, is demonstrated. This work focuses on the development of signal processing algorithms that explore the properties of the detector and offer the ability for online, precise timing.

We have used data, collected in Laser Beam Tests, where a PICOSEC-MicroMegas detector was illuminated by femtosecond laser pulses, producing, on average, 7.8 photoelectrons. When the detector signals were fully analyzed (by fitting the leading edge of the digitized waveform and applying CFD timing) a timing resolution of 18.3 $\pm$ 0.2 \,ps was found. In the following of this study, this value was used as the reference resolution to be compared with the results of the timing algorithm developed in this work.

We have developed a signal processing algorithm based on  Constant Threshold Discrimination timing and a slewing correction procedure, based on a charge above threshold variable. We prove that such a technique achieves the same precision as the reference resolution of the full off-line analysis. Notice that, this technique, is easily integrated into electronics, by utilizing the technological advantage of novel timing devices, like the NINO chip \cite{NINO}, complemented with ADCs measurements.

Recently, new fast digitizers, e.g. the SAMPIC with up to $\sim$9GSamples/s, became available offering the possibility of digitizing the leading edge of the PICOSEC-MicroMegas waveform. In this work, we propose an Artificial Neural Network for real-time signal processing, able to provide very precise timing, which can be used for very fast event selection. The main demand for such a technique is the availability of adequate training samples. We have developed a simulation model that is utilized to generate PICOSEC-MicroMegas type of waveforms. After proving that our simulation model describes honestly the detector's response to many photoelectrons, we have used the aforementioned emulated waveforms to train an ANN in order to process and time PICOSEC-MicroMegas signals. The ANN timing performance was extensively tested by using real data. A very high precision timing performance (the same as the reference resolution) was found. In parallel, we executed several other tests for systematic biases and we demonstrated that the ANN performance is unbiased and consistent.

Lastly, it is worth noticing that the simulation model, employed in this analysis, generates waveforms that are not completely, mutually uncorrelated. Also, the emulated waveforms differ slightly from the real pulses concerning the random electronic noise and the primordial time jitter. However, as it is proven in this work, the emulated data comprise the adequate information needed by the ANN to learn a signal analysis procedure for precise timing.

\addcontentsline{toc}{chapter}{Bibliography}
\printbibliography
\appendix
\chapter{Introduction to Artificial Neural Networks } \label{appendix:A}
\begin{appendices}
 
\begin{figure}[hbt!]
 \centering 
 \includegraphics[width=0.8\textwidth]{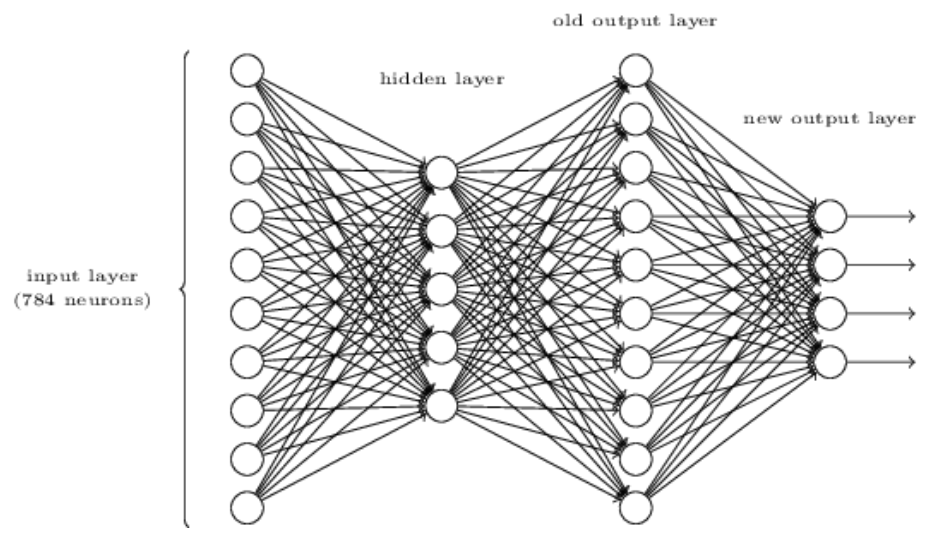}
\caption{Sketch of an Artificial Neural Network Architecture \cite{neural}.}\label{fig:neural_network}
\end{figure}

It is convenient to reference neural networks within an example. Imagine that you have a photo of $28\times28=784$ pixels that shows you a number (let's say 4), only the digits that present the number are allowed to be white(defining as 1), all other shades from black to white have lower number id than 1. All 784 pixels called neurons, make up the first layer of our network, like the representation in Figure \ref{fig:neural_network}. The number we want to recognize let's assume belongs to the range of 0 to 9, so these numbers will construct our last layer of the network.

Between the first and the last layer, there might be several layers (called the \emph{hidden layers}), which are used to encode abstract representations of the features/patterns that create differences between digit shapes and connect all the 784 digits from the first layer with that of the second layer. Only the neurons that fit a pattern will be activated on the second layer and so on, on every layer, it is defined. Note that the number of layers and the number of neurons on each layer are arbitrarily chosen, but on each step, when all neurons are inter-connected, the structure is called densely connected and the layers are called dense . From the first layer, we have to define an activation function reprocess the output of one layer to transform it into the input to the next layer, with every neuron having a different weight into that function. The learning process of the network is to find the right weights and biases to every one step of the previous process. Jumping to the last layer, the most intuitive approach is to have ten neurons, each representing one of the digits. 

The activation function of the last layer can be, in this case, a function that maps the weights and biases of the last layer to a simple number between 0 and 9. In this way, it represents how much the confidence of the algorithm that a given image corresponds to a given digit.

When it comes to this image, the information that is fed into this structure consists of the brightness of the 784 input image pixels. As the computation propagates forward into the network of nodes, the brightness pattern of pixels is transferred to the weights and biases of the nodes through layer activations.
\begin{figure}[hbt!]
 \centering 
 \includegraphics[width=0.8\textwidth]{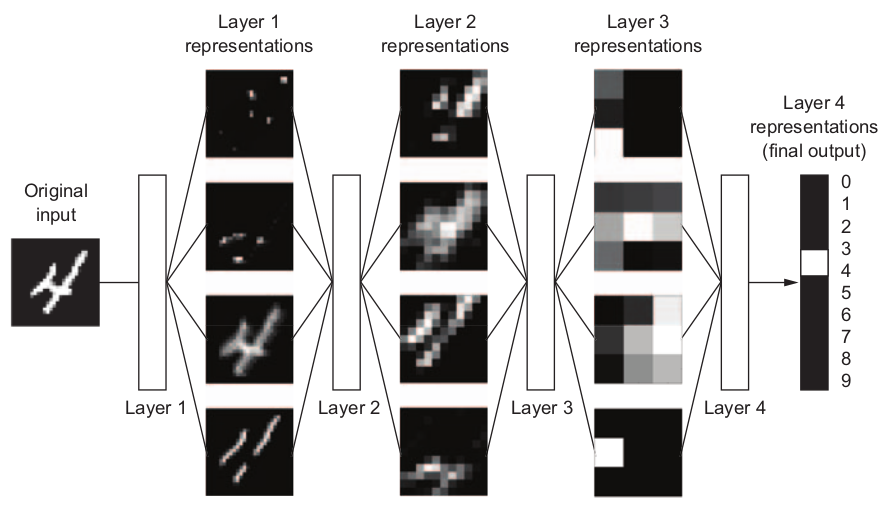}
\caption{Neural Network representations learned by a model using digit-classification. Figure adopted from \cite{francois}.}\label{fig:number_example_network}
\end{figure} 

But how exactly the activations in one layer might determine the activations in the next? The goal is to have some mechanism that could conceivably combine pixels into edges, or edges into patterns or patterns into digits. As an example let's consider the capability of a specific neuron in the second layer to pick up on whether or not the image has an edge in a specific region of the image, but what parameters should the network have, so that it's specific enough to potentially capture the specific pattern or the pattern that several edges can make a loop and other such things, a behavior well presented in the Figure \ref{fig:number_example_network}. So, let's assume our neural network is just as a mathematical framework to learn representations from given data \cite{francois}.

The specifications of how to reprocess the information to the next layer, it is convenient to assign a weight to each one of the connections between our neurons, meaning that the transformation is implemented by the parameterization of its weights. These weights are real numbers, which with activation functions of the first layer connected, can result in the computation of their weighted sum according to these weights, which in mathematical expression can be written in:
\begin{equation}
    \sum_{i=1}^{i=n}w_i\alpha_i = w_1\alpha_1 + w_2\alpha_2 + w_3\alpha_3 + \dots + w_n\alpha_n 
\end{equation}

When you compute a weighted sum like this, you might come out with any number, but the purpose of this type of recognition we want for activations to be some value between zero and one, which means we have to change this weighted sum, in a function that limits our output result into the range of $[0,1]$. A common function that does this is called the sigmoid function, well known to the readers of this thesis as a logistic curve defined as: 
\begin{equation}
    \sigma(x) = \frac{1}{1+e^{-x}}
\end{equation} 
resulting to the full expression for one activation to be:
\begin{equation}
    \alpha_0^{(1)} = \sigma \Bigg( w_1\alpha_1 + w_2\alpha_2 + w_3\alpha_3 + \dots + w_n\alpha_n + b_0 \Bigg)
\end{equation}
So the activation of the neuron here is a measure of how positive is the relevant weighted sum. But maybe you need another constraint, for the network to be activated when the sum is greater than a constant value, called bias. Where the weights represent what pixel pattern, in this neuron in the second layer, is picking up on, while the bias represents how high the weighted sum needs to be before the neuron starts getting meaningfully active. And that is just the process followed by one neuron, every other neuron in this layer is going to be connected to all 784-pixel neurons from the first layer and each one of those 784 connections has its own weight associated with it. Moreover, each one has some bias, some other number that is added to the weighted sum before squishing it with the sigmoid function. This is followed by the connections to other layers, which also have a bunch of weights and biases associated with them. In other words, this is the work done in the learning process, which means that the network has to find the suitable weights and biases for all the layers, such that it can correctly map inputs to their targets. A more notationally compact way that these connections can be represented is to organize all the activations from one layer into a column as a vector, then all the weights as a matrix where each row of that matrix corresponds to the connections between one layer and a particular neuron in the next layer. What that means is that taking the weighted sum of the activations in the first layer according to these weights, corresponds to one of the terms in the matrix-vector product of everything on the following equation. Instead of talking about adding the bias to each one of these values independently we represent it by a relevant vector added to the matrix-vector product. Then as a final step, we wrap a sigmoid around this matrix structure :
\begin{equation}
\sigma \begin{pmatrix} \begin{bmatrix}
w_{0,0} &  w_{0,1} & \ldots & w_{0,n}\\
w_{1,0} &  w_{1,1} & \ldots & w_{1,n}\\
\vdots & \vdots & \ddots & \vdots\\
w_{k,0}& w_{k,1} &\ldots & w_{k,n}
\end{bmatrix} \cdot
\begin{bmatrix}
\alpha_0^0\\
\alpha_1^0\\
\vdots \\
\alpha_n^0
\end{bmatrix} + 
\begin{bmatrix}
b_0\\
b_1\\
\vdots \\
b_n
\end{bmatrix} \end{pmatrix}
\end{equation}
What that's supposed to represent is that the network will apply the sigmoid function to each specific component of the resulting vector inside. Once we have written down this weight matrix and these vectors, as their own symbols it can communicate the full transition of activations from one layer to the next in an extremely tight and neat expression and this will make the relevant code both a lot simpler and a lot faster since many existing computational libraries are optimized for matrix multiplication.
\begin{equation}
    \vec{\alpha}^{(1)} = \sigma(\vec{W}\vec{\alpha}^{0} + \vec{b}) 
\end{equation}

\subsubsection{Gradient descent method of learning}
\paragraph{} So far it has been discussed how the network learns, what is needed is an algorithm where you can show this network a bunch of training data, which comes in the form of a collection of different images of handwritten digits along with labels for what they actually represent and it will adjust the necessary weight and biases as to improve its performance on the training data. Hopefully, this layered structure will mean that what it learns generalizes to images beyond that training data. The way we test that is that after you train the network, you show it more labeled data that it is never seen before and you see how accurately it classifies those new images. 

Remember that conceptually, we are thinking of each neuron as being connected to all of the neurons in the previous layer, and the weights in the weighted sum defining its activation are kind of like the strengths of those connections, while the bias is some indication of whether that neuron tends to be active or inactive. To start things off, we are just gonna initialize all of those weights and biases totally randomly. Immediately after such an initialization, the network naturally performs very poorly on training examples on a given training example since it is just going something random, since there are no encoded patterns in the hidden layers. For example, you fit with an image of number 3 and the output layer it just looks like a mess. So what you do is you define a cost function, to tell the network that the output should have activations that are zero for most neurons, but one for this neuron representing the desired recognized number should be different, like $y(x) = (0,0,0,1,0,0,0,0,0,0)^T$ to be the desired output. From a mathematical point of view, that means, that you have to add up the squares of the differences between the resulting output of the network activations and the value that we want them to have and this is what we will call the cost of a single training set. The sum should be small when the network confidently classifies the image correctly, but large when the opposite is true. To quantify how well this goal is being achieved, the analytic formula for the cost function is :
\begin{equation}
    C(w,b) = \frac{1}{2n}\sum_{x}||y(x) - \alpha||^2
\end{equation}
where w is the weight collection of the network, b is the bias collection, n is the total number of training inputs, a is the vector of outputs with x inputs. In this way, we are trying to inform the network how to change those weights and biases so that the cost function will be minimized. In other words, we are searching for sets of weights and biases which can make the cost minimum. 

\begin{figure}[hbt!]
 \centering 
 \includegraphics[width=0.6\textwidth]{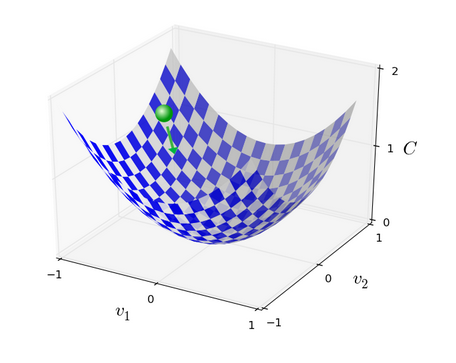}
\caption{Gradient descent representation as a ball rolling down a valley. Figure adopted from \cite{francois}.}\label{fig:gradient_descent}
\end{figure}

An algorithm suitable for this purpose is known as \emph{gradient descent}.  We try to minimize $C(u) = C(w,b)$ and we would like to find a global minimum of C. The well-known technique, of computing derivatives with respect of the variables of the cost function, and then use them to find places where C is an extreme value. When this cost function is a multi-variable function, as in the case of a neural network, this job turns to be very difficult. Eventually there is an algorithm to do the same work. Imagining a ball strolling down the slope of a valley. The rolling stops at the bottom of the valley and that is a minimum of the function that describes the curve of the valley. So that is the analogy, we imagine a ball that rolls down the valley and we simulate its motion down to the bottom, from ball's point of view, not taking actual the neuton's equations of motion. Moving the ball at small amount both in two directions, $u_1, u_2$, then the cost function changes as: 
\begin{equation}
    \Delta C \approx \frac{\partial C}{\partial u_1}\Delta u_1 + \frac{\partial C}{\partial u_2}\Delta u_2 = \nabla C\cdot \Delta u
\end{equation}
The idea is to find a way of choosing, $\Delta u_1, \Delta u_2$ so that $\Delta C <0$, i.e. that the ball is rolling down. Setting $\Delta u$ to be the vector of changes $\Delta u \equiv (\Delta u_1, \Delta u_2 )^T$, then the $\nabla C$ will be the vector of partial derivatives:
\begin{equation}
    \Delta C = (\frac{\partial C}{\partial u_1}, \frac{\partial C}{\partial u_2})^2 \approx \nabla C \cdot \Delta u 
\end{equation}
where $\nabla C$ relates changes in u to changes in C and that gives us a way of knowing how to choose $\Delta u$, in order to make $\Delta C$ a negative value, which denotes the gradient reduction like :
\begin{equation}
    \Delta u = - \eta \nabla C
\end{equation}
where $\eta$ is a small non negative parameter (called \emph{learning rate}), and then 
\begin{equation}
    \Delta C = - \eta |\nabla C|^2
\end{equation}
while we are changing u within the limits, C will always decrease. Our law of motion is : 
\begin{equation}
    u \xrightarrow[]{} u' = u -\eta \nabla C
\end{equation}
and use it over and over again for the next move, and additionally decreasing C, until we reach a global minimum as we have intended. This algorithm repeatedly computes the cost gradient $\nabla C = - \Delta u /\eta$, and then moving in the opposite direction, like a ball rolling down of a valley, visualized in Figure \ref{fig:gradient_descent}.

Another parameter that needs to be clarified, is the learning rate $\eta$, which needs to be taken into account to make the gradient descent work correctly. As we have mentioned this should be a small positive value in order to accept $\Delta \equiv \nabla C \cdot \Delta u$ as a well-defined approximation. If $\eta$ is chosen to be very small $\Delta u$ will be very small too, and then the algorithm will be inefficient .

Averaging all above assumptions, to a cost function that is a multi-variable function, 
\begin{align}
\Delta u &= (\Delta u_1,\dots,\Delta u_m )^T \\
\Delta C &= (\frac{\partial C}{\partial u_1},\dots, \frac{\partial C}{\partial u_m} )^T \\
\Delta C &\approx \nabla C \cdot \Delta u\\
u \xrightarrow[]{} u' &= u -\eta \nabla C
\end{align}
Applying that to a neural network, the idea depends on finding the weights $w_k$ and biases be those which minimize the cost function. Naming our position with a variable with two parameters $w_k, b_l$, and gradient vector $\nabla C$ has its related derivatives $\frac{\partial C}{\partial w_k}, \frac{\partial C}{\partial b_l}$. So in addition to the u change, we now have: 
\begin{align}
w_k \xrightarrow[]{} w_k' &= w_k -\eta \frac{\partial C}{\partial w_k}\\
b_l \xrightarrow[]{} b_l' &= b_l -\eta \frac{\partial C}{\partial b_l}
\end{align}
So repeatedly applying the rule we are "rolling down the valley" and the only think left is to hope finding a minimum of cost function, in other words to successfully learn our neural network. 

One major problem is that the cost function is expressed as $C = \frac{1}{\eta} \sum_{x} C_x $ where $C_x = \frac{|y(x) - \alpha|^2}{2}$, which denoted an average over costs for every one training set. So to compute $\nabla C$ it is needed to be computed all $\nabla C_x$ for every training set and then to average them. 
\begin{equation}
    \nabla C = \frac{1}{\eta} \sum_x \nabla C_x
\end{equation}
but when the number of training inputs is very large this can be a very costly procedure on time and learning will be slowly. In order to speed up learning we may use the method of stochastic gradient descent. The difference is that by this method we can compute $\nabla C_x$ for a small sample of training inputs that are chosen randomly, and averaging to this sample. In this way the training method becomes quicker. It is common to have those random training inputs as mini-batch. So that : 
\begin{align}
\nabla C &\approx \frac{1}{\eta}\sum_j^{m}\nabla C_{xj}\\
w_k \xrightarrow[]{} w_k' &= w_k - \frac{\eta}{m} \sum_j \frac{\partial C_{xj}}{\partial w_k}\\
b_l \xrightarrow[]{} b_l' &= b_l - \frac{\eta}{m} \sum_j \frac{\partial C_{xj}}{\partial b_l}
\end{align}
After that we pick another randomly chosen mini-batch and train the network again, and so on, until we have used all the training inputs. This procedure is said to be a completeness of an \emph{epoch} of training. At that point we may start with a new training epoch. 

This technique seems to be very useful since it is easier to sample mini-batches than it is to apply gradient descent to the full batch, and we do not care that the estimation will be affected by serious statistical fluctuation, because the only think we do really care is to find the direction to move, in order to decrease the cost function, in fact we do not need an exact computation of $\nabla C$. 

The algorithm for computing this gradient efficiently which is effectively the heart of how a neural network learns is called back-propagation. More information you can find in the \cite{francois}.
\end{appendices}

The standard algorithm of gradient descent uses the whole set of data to compute the gradient, hence the updates of the weights and biases. This strategy though is time consuming, and most of the times is intractable on a single machine, if the data-set is too big to fit in main memory. Another issue with batch optimization methods is that they don't give an easy way to incorporate new data in an \say{online} setting. As a result different kind of methods are used to approximate the gradient of the loss function with some of them being: 
\begin{itemize}
    \item Stochastic Gradient Descent, is an iterative method for optimizing a differentiable objective function, a stochastic approximation of gradient descent optimization.
    \item Adaptive Gradient Descent, is a modified stochastic gradient descent with per-parameter learning rate. Informally, this increases the learning rate for more sparse parameters and decreases the learning rate for less sparse ones. 
    \item The Root Mean Square Propagation, is a method which learning rate is adapted for each of the parameters. The idea is to divide the learning rate for a weight by a running average of the magnitudes of recent gradients for that weight. 
    \item Adaptive Moment Estimation is an update to the RMSProp optimizer. In this optimization algorithm, running averages of both the gradients and the second moments of the gradients are used.
\end{itemize}

\end{sloppypar}

\end{document}